\documentclass[manuscript]{aastex}

\usepackage{morefloats}
\usepackage{graphicx}

\shorttitle{ISM In Solar Rearview Mirror}
\shortauthors{Wyman \& Redfield}

\begin{document}

\title{Probing Our Heliospheric History I: High-Resolution Observations of \ion{Na}{1} and \ion{Ca}{2} Along the Solar Historical Trajectory}

\author{Katherine Wyman$^{1,2}$ and Seth Redfield$^{2}$}
\affil{$^{1}$Harvard-Smithsonian Center for Astrophysics, 60 Garden Street, Cambridge, MA 02138, USA kwyman@cfa.harvard.edu\\ $^{2}$Astronomy Department, Van Vleck Observatory, Wesleyan University, Middletown, CT 06459, USA sredfield@wesleyan.edu}

\begin{abstract}
Over the course of its motion through the Galaxy, our solar system has encountered many interstellar environments of varying characteristics.  Interstellar medium (ISM) density variations spanning seven orders of magnitude are commonly seen throughout the general Galactic environment, and a sufficiently dense cloud within this range has the potential to dramatically alter the structure of the heliosphere.  We present observations of the ISM environments the Sun has most recently encountered based on high-resolution optical spectra toward nearby stars in the direction of the historical solar trajectory. The data were obtained with the highest-resolution spectrographs available, including the Tull Spectrograph on the Harlan J. Smith Telescope at McDonald Observatory and the Ultra-High-Resolution Facility on the Anglo-Australian Telescope at the Anglo-Australian Observatory.  Observations were made of interstellar \ion{Na}{1} and \ion{Ca}{2} doublet absorption toward 43 bright stars within $\sim$500 pc.  No absorption is seen out to a distance of 120 pc (consistent with the outer boundary of the Local Bubble), but a complex collection of absorbers is seen in stars beyond 120 pc.  While common absorbers are consistently seen in most sight lines, significant spatial variation is also detected, even between closely spaced sight lines.  This pervasive evidence of small-scale structure not only speaks to the complexity of the morphology or physical properties of the gas in the ISM, but also emphasizes that dramatic structural changes to the heliosphere are common and it is important to understand the implications of such changes, such as the modulation in the cosmic ray flux, on planets.

\end{abstract}


\section{Introduction}

As stars hurtle along their orbits around the Galaxy, they pass through a diverse array of
interstellar medium (ISM) environments, ranging from dense molecular clouds
(e.g., molecular knots in Orion, $n \gtrsim 10^5$ cm$^{-3}$; \citealt{odell01}) to rarefied regions, heated and
largely vacated by strong supernovae or stellar winds (as seems to be
the case with the Local Bubble, $n \lesssim 10^{-2}$ cm$^{-3}$; \citealt{welsh09}).  Such a dramatic variation in the characteristics of the local interstellar environment of stars can have a profound impact on the conditions in the volume occupied by planets.  A balance of the
pressures, between an outward moving stellar wind and the inward
confinement of the surrounding interstellar material, defines an
interface, the boundary between the star and the rest of the universe.
In the case of our Sun, this interface, or heliopause, is located at
approximately 300 AU \citep[e.g.,][]{muller06}, a direct consequence of the properties
of the local interstellar medium (LISM) and the solar wind (i.e.,
their density and velocity).

Changes in the extent of the heliosphere occur over a multitude of time scales.  In the short term, the solar wind strength fluctuates over its 22-year magnetic cycle while solar flares and coronal mass ejections can drive variability on even shorter time scales.  Coronal activity has the potential to asymmetrically modify the morphology of the heliosphere.  An event like this occurred in 2003  and was recorded by instruments aboard Voyager 1 \citep{decker05}.  On longer time scales, a gradual change in the solar wind strength will occur over the 10-billion-year life span of the Sun.  Observations of other solar type stars suggest that the historical solar wind may have been $\sim$50$\times$ stronger than it is today \citep{wood05let}.  However, the most dramatic driver of the structure of the heliosphere originates outside its boundaries (i.e., the properties of the surrounding ISM).

Currently the Sun is moving through a large ($R$ $\sim$ 100 pc), hot ($T \sim 10^6$ K), and low density ($n \sim 0.005$ cm$^{-3}$) cavity known as the Local Bubble.  Found within the Local Bubble are a collection of smaller (1--10 pc), higher density ($n \sim 0.1$ cm$^{-3}$), cooler ($T \sim 7000$ K), partially ionized clouds \citep{redfield08}.  Given current solar velocity measurements \citep[e.g.,][]{dehnen98}, a cloud 1 pc in diameter would pass over the solar system in just 70,000 years.  Currently, the Sun resides very near to the inside edge of one of these clouds referred to as the Local Interstellar Cloud (LIC).  Given the relative velocity vectors of the Sun and the LIC, the Sun will be transitioning out of the LIC sometime in the next few thousand years \citep{redfield00}.  In fact, the farthest Oort Cloud objects might have already breached this boundary and moved into hot Local Bubble material.  The Local Bubble is also home to clouds of much higher densities and cooler temperatures.  The Local Leo Cold Cloud, a thin cloud first identified in \ion{H}{1} 21~cm emission has recently been identified to be located between 11.3 and 24.3 pc away \citep{meyer06,peek11}.  This cloud appears to have a temperature of $\sim$20 K and a \ion{H}{1} density of $\sim$3000 cm$^{-3}$ \citep{meyer12}.  In addition, several of the MBM clouds \citep{magnani85}, originally identified by 2.6 mm CO emission, may be at or within the Local Bubble boundary.  In particular, in roughly the same direction as the survey presented here, high-resolution observations of nearby stars were used to pinpoint the distance of MBM 20 \citep{penprase93,hearty00}.  A comprehensive review of the LISM is provided by \citet{frisch11}.

Given the pliable nature of the heliosphere and the configuration of our local galactic environment, several questions arise: How will the heliosphere respond when the Sun exits the LIC?  What would happen if the Sun were to traverse a dense interstellar cloud, such as the Local Leo Cold Cloud?  The current paper develops the foundation for such an investigation by asking: What were the interstellar environments of the Sun in its most recent past?    

The implications of a dynamic heliosphere go well beyond structural
changes and could be important for planets and their atmospheres.  The
relationship between our interstellar environment and the Earth has
interested researchers for almost a century.  \citet{shapley21} first
formally suggested a link between the passage of the solar system
through dense ISM clouds and quasi-periodic
climate catastrophes and mass extinctions of organisms on Earth.  As a
testament to its broad appeal, theoretical speculations on this
relationship have continued to this day \citep{hoyle39,fahr68,begelman76,talbot76,frisch86,thaddeus86,zankfrisch99,shaviv03}.
A recent compendium of articles related to the solar journey through the Galaxy is given in \citet{frisch06book}.  
Mechanisms
invoked to link the ISM and planetary atmospheres include: (1)
fluctuations of cosmic ray flux modulated through the heliosphere, given that the solar magnetic field extends only as far as the termination shock \citep[e.g.,][]{giacalone99,usoskin05,smith09}; and (2)
dust accretion directly onto planetary atmospheres \citep[e.g.,][]{pavlov05}.  These processes
have potential consequences for climate, ozone chemistry, and DNA
mutation rates for surface organisms \citep[see review by][]{scherer06}.

This work builds on a legacy of high-resolution ISM observations in
the optical.  The strength of the \ion{Na}{1} and \ion{Ca}{2}
transitions has made them invaluable tracers of the ISM, even when
they are trace ions in interstellar gas (as is the case in the LISM,
where for example, \ion{Ca}{3} is the dominant form of calcium and \ion{Na}{2} is the dominant form of sodium;
\citealt{slavin08}).  Surveys have been made along sight lines toward most of the bright stars across the entire sky \citep{vallerga93,welty94,welty96}, of star-forming
regions (e.g., Orion; \citealt{frisch90b,welsh05b}), and of wide binaries used to
search for small-scale structure in the ISM along almost identical
sight lines \citep[e.g.,][]{meyer96}.  However, as far as we know, this is the densest high-resolution absorption line survey of any particular direction through the ISM.  Not only is this particular sight line of interest to understand the history of solar interactions with the surrounding interstellar gas, but also this survey provides a detailed and densely sampled look at the physical properties of the galactic ISM.

A companion paper \citep[][hereafter Paper II]{wyman13} explores the potential consequences an encounter with the ISM environments presented in this paper could have on the solar system.  Models of the heliosphere have been made to explore a wide range of different interstellar environments by \citet{muller06}.  In particular, models were made using characteristics of the clouds in the LISM, as well as the hot Local Bubble and cold dense molecular clouds.  These models provide relations between the interstellar properties and heliospheric properties.  Here, we present observations of the interstellar clouds traversed by the Sun and plan to use these relationships to make estimates of the historical heliospheric response (and thereby the cosmic ray flux history on the Earth's atmosphere) in our companion paper.  We defer additional discussion to Paper II, including the likelihood that the current interstellar column in the direction of the historical solar motion is identical with that of the actual interstellar environments traversed by the Sun over that time.  In this paper, we focus primarily on presenting the rich dataset and providing an analysis of the ISM in this unique direction.

\section{Observations}
\label{observations.s}

Stars were observed to measure ISM absorption in the direction of the
Sun's path through the Galaxy, out to a distance of $\sim$500 pc.  The
direction and velocity of the solar peculiar velocity were derived from
\citet{dehnen98}, who measure the solar peculiar velocity from the
      {\it Hipparcos} dataset (see \citealt{schonrich10} for an
      alternative analysis).  The adopted values are
      $l_0=207.70^{\circ}$ and $b_0=-32.41^{\circ}$, in the
      constellation Eridanus.  The magnitude of the solar peculiar
      velocity is estimated to be $13.38 \pm 0.42$~km~s$^{-1}$, which
      corresponds to a time of $\sim$36.7 million years to traverse
      500\,pc.  This represents less than 1\% of the total path length
      the Sun traverses in a single orbit around the Galaxy.  This
      small percentage allows us to approximate the Sun's recent
      historical trajectory as a straight line toward $l_0$, $b_0$.

Stars for this study were chosen based on a close proximity ($\theta< 10^{\circ}$) to $l_0$, $b_0$, and properties that optimize ISM absorption analysis (e.g., bright [$V < 8$], rapidly rotating, early type [B, A, F] stars).  We use the revised {\it Hipparcos} parallaxes presented by \citet{vanleeuwen07} to calculate distances to our target stars, and restricted our sample to targets for which the relative error in distance is $\leq$0.5, whereas the median relative error is $\leq$0.1.  This restriction led us to remove three targets (HD30020, HD29173, and HD30076).  

\begin{figure}
\centering
\includegraphics[scale=1]{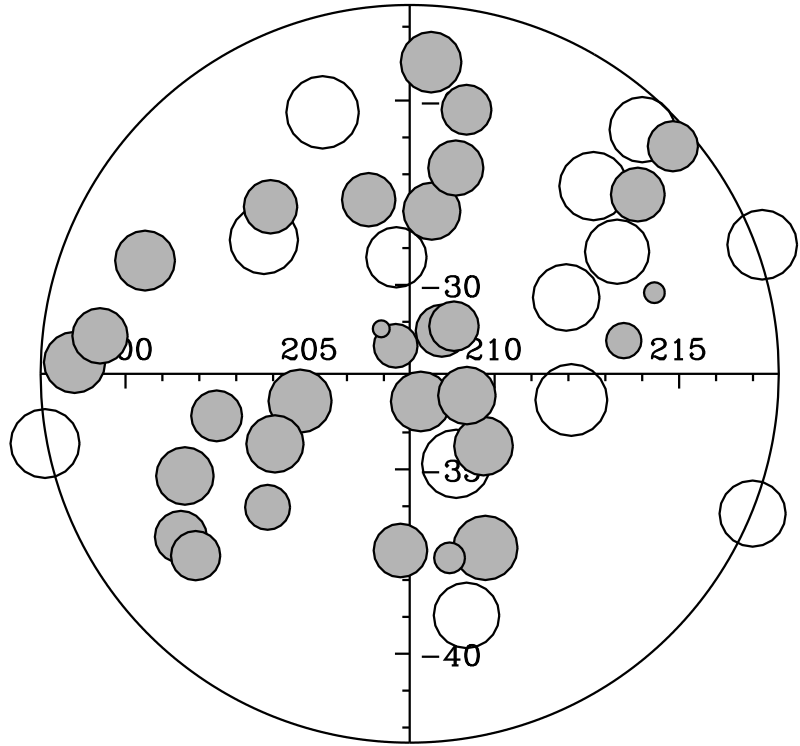}
\caption{Distribution of sight lines used in this survey and listed in Table~\ref{stars}.  We observed 43 stars within 9$^\circ$ of the historical solar trajectory at the origin \citep[$l_0=207.70^{\circ}$ and $b_0=-32.41^{\circ}$;][]{dehnen98}.  More than 50\% of the sample is within $\sim$5$^\circ$ of the solar trajectory.  In addition, 8 sight line pairs are separated from each other by $<$1$^\circ$.  Symbol size is inversely proportional to stellar distance, which extends from 27.40--610\,pc.  Open symbols indicate no ISM absorption was detected (almost exclusively targets within 120\,pc), while shaded symbols indicate absorption profiles that were fit in order to reconstruct the interstellar environments along the historical solar trajectory.}
\label{angplot}
\end{figure}

We observed two targets (HD26574, 37.31~pc; HD26994, 289~pc) that were more than 9 degrees from the direction of the historical solar trajectory.  HD26574, the closer target showed no interstellar absorption, while HD26994 had absorption that was dramatically different from what was observed for the rest of our sample.  For this reason, we decided to constrain our sample cone to $<$9$^\circ$ from the solar historical trajectory.  A third target, HD32964 (66 Eri), is a well-known short period binary ($P=5.5$ days; \citealt{makaganiuk11}) in which both stars displayed stellar absorption features at velocities at which we would expect to see LISM absorption.  Because this target is only 94.7~pc away, and given the results from our entire sample, we would not expect to see any interstellar absorption since the closest target with detected absorption is 125.9\,pc and 112.9\,pc for \ion{Ca}{2} and \ion{Na}{1}, respectively.  Therefore, we removed these targets from our sample.  The remaining 43 targets are listed in Table~\ref{stars} and their distribution on the sky is illustrated in Figure~\ref{angplot}.

Data were obtained throughout the time period of October 2003 through October 2005.  Absorption lines of neutral sodium (\ion{Na}{1} D$_1$ and D$_2$; 5895.9242 and 5889.9510 \AA) and singly ionized calcium (\ion{Ca}{2} H and K; 3968.4673 and 3933.6614 \AA) were chosen as the ISM component tracers since they both have strong doublet resonance lines in the optical that are sensitive to nearby interstellar clouds.  Observations were made with the highest-resolution spectrographs available in each hemisphere.  In the northern hemisphere, observations were taken with the Harlan J. Smith Telescope (HJST), a 2.7 meter telescope located at McDonald Observatory in west Texas.  Observations were made at three separate resolution ($R \equiv\frac{\lambda}{\Delta\lambda} $) settings: TS12: Tull Spectrograph ($R \sim$~400,000; \citealt{tull72}), TS21: Cross-Dispersed Echelle Spectrograph (2d coud\'{e}) focus 1 ($R \sim$~240,000; \citealt{tull95}), and TS23: Cross-Dispersed Echelle Spectrograph (2d coud\'{e}) focus 3 ($R \sim$~60,000; \citealt{tull95}).  In the southern hemisphere, observations were taken with the 3.9 meter Anglo-Australian Telescope (AAT), located at the Anglo-Australian Observatory in Australia.  Spectra were obtained with the Ultra-High-Resolution Facility (UHRF), an echelle spectrograph located at the coud\'{e} focus ($R \sim$~940,000; \citealt{diego95}).  A subset of targets were observed multiple times at different resolutions and with both facilities to provide consistency checks on measurements of absorption profile parameters (see Section~\ref{multiple.resolution.observations.ss} for a discussion of sight lines observed with multiple configurations).

Table \ref{obsParams} lists the observational parameters for all data included in this study and is ordered by distance from the Sun.  The approximate continuum signal-to-noise ($S/N$) ratio is given for each observation, as is the projected radial velocity of Earth's atmosphere at the time of the observation.  Telluric features, both water vapor and sodium, can contaminate this spectral region, and are easily identified by their radial velocity.

\subsection{Data Reduction}
\label{data.reduction.ss}

The data were reduced using Image Reduction and Analysis Facility (IRAF; \citealt{tody93}) and Interactive Data Language (IDL) routines to subtract the bias, flat-field the images, remove scattered light and cosmic-ray contamination, extract the echelle orders, calibrate the wavelength solution, and convert to heliocentric velocities. Wavelength calibration images were taken using a Th-Ar hollow cathode before and after each target.

The sodium region of the spectrum suffers from weak, but pervasive, telluric contamination from water in Earth's atmosphere that must be identified and removed from the data, particularly for observations toward nearby stars that may be expected to exhibit weak interstellar absorption.  A forward modeling technique demonstrated by \cite{lallement93} was used to remove telluric line contamination in the vicinity of the \ion{Na}{1} D lines based on a terrestrial atmosphere model (the Atmospheric Transmission [AT] program, from Airhead Software, Boulder, Colorado) developed by E. Grossman.  Figure~\ref{twores} shows a typical observed spectrum along with a fit to the telluric contamination.  Given that scores of telluric lines are involved in the fit, the telluric model is tightly constrained and highly successful at removing the water vapor contamination from our spectra.  Note that telluric lines are relatively absent in the vicinity of the \ion{Na}{1} D$_1$ line at 5895.9242~\AA, which further reduces any systematic influence of telluric features on the ISM profile fit.  Observing both transitions of the \ion{Na}{1} doublet is an important confirmation that the telluric subtraction is successful.  With two independent measurements of \ion{Na}{1} absorption at the same projected velocity, it is easy to identify contaminating telluric lines.  Broad features remaining in the spectrum after telluric absorption removal are typically stellar absorption, which can be confirmed given the known stellar radial velocity and removed with a low-order polynomial to reveal the ISM absorption.

\subsection{Spectral Analysis}
\label{spectral.analysis.ss}

All absorption profiles were fit using standard methods (see
\citealt{linsky96,redfield02}). This involves estimating the stellar
continuum, determining the fewest absorption components required to
obtain a satisfactory fit, convolving the absorption feature with an
instrumental line-spread function, and fitting all lines of the same
ion simultaneously.  Determining the stellar continuum is usually
straightforward when one is dealing with high-resolution data, and
systematic errors in this procedure are further reduced when multiple
resonance lines of the same ion are fit simultaneously.  Figure~\ref{twores} shows an example of a stellar continuum that is typically encountered.  The continuum is slowly varying across the narrow wavelength range and easily fit with a low ($\lesssim$5) order polynomial.  Voigt absorption profiles are
fit to \ion{Na}{1} and \ion{Ca}{2} lines simultaneously using atomic
data from \citet{morton03}.  While the continuum is not fit simultaneously with the ISM profile, in rare cases, the simultaneous ISM profile fits highlighted discrepancies between the two transitions that could be resolved with a better continuum model.  In these cases, an iterative approach to the continuum modeling was used.  Best-fit parameters are derived using a
Marquardt $\chi^2$ minimization technique \citep{press02}.  Each sight
line was initially fit with one component unless visual inspection of
the spectrum strongly suggested the presence of multiple components.
Additional components were added until an F-test \citep{bevington}
indicated that their inclusion was no longer statistically
warranted. The final model spectrum is a convolution between these intrinsic Voigt profiles and the instrumental line spread function. 
The
final fit parameters for each absorber are the central velocity ($v$),
the line width or Doppler parameter ($b$), and the column density
($N$) along the line of sight.  Errors on the fit parameters were derived from Monte Carlo simulations of the simultaneous fit.  In order to evaluate the contribution of systematic errors we compare our fit errors with the standard deviation of fit parameters of the two transitions fit independently and the simultaneous fit, and use the larger of the two.  Table~\ref{CAallcomps} lists the final
fit parameters for all individual \ion{Ca}{2} absorbers, while
Table~\ref{CAtotecomps} gives the total column density along each
sight line. Tables~\ref{NAallcomps} and \ref{NAtotecomps} are the same
for the \ion{Na}{1} observations.

Figures \ref{123} through \ref{470} are the flux-normalized, best-fit
results of the fitting process in the heliocentric rest frame, for each sight line in the sample with
detected interstellar absorption, ordered by distance to each
background star.  Absorption spectra for each line in both sodium and
calcium doublets are shown, as well as each individual absorption
component prior to convolution with the line-spread function.  The
hyperfine structure of \ion{Na}{1} is modeled, but only impacts the
fits of observations taken at the highest spectral resolution
\citep{welty94}.  The fit to the total absorption profile is plotted
after convolution with the instrumental line-spread function.

\subsection{Multiple-Resolution Observations}
\label{multiple.resolution.observations.ss}

\begin{figure}
\centering
\epsscale{0.75}
\plotone{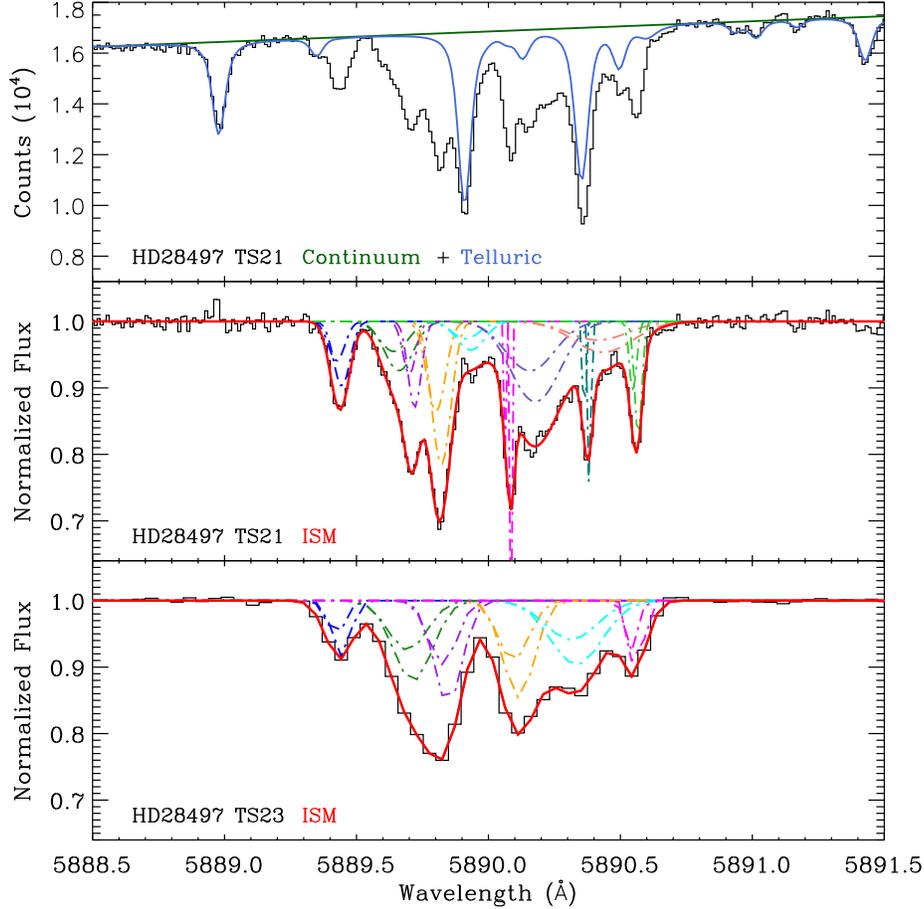}
\caption{HD28497 \ion{Na}{1} continuum normalization, telluric line fitting, and ISM absorption observed at two spectral resolutions.  The top panel is data showing the continuum modeled with a low-order polynomial along with a fit to the telluric contamination. The bottom two panels are the data after normalization and subtraction of the telluric features, where the middle panel is data taken with TS21 at a resolution of $R \sim$ 240,000, and the bottom panel was taken with TS23 at a resolution of $R \sim$ 60,000.  Strong absorbers, particularly if unblended (e.g., the most blueshifted component), are consistent between the two observations, but the narrowest absorbers (i.e., $b < 1$ km~s$^{-1}$) seen in the higher-resolution observation fail to be resolved in the lower-resolution observation.  However, both these observations yield the same measurement of total column density (within the errors) along this particular sight line (TS23: $N($\ion{Na}{1}$) = 11.948^{+0.014}_{-0.016}$, TS21: $N($\ion{Na}{1}$) = 11.936^{+0.036}_{-0.060}$).  Results of other sight lines observed at different spectral resolution are compared in Table~\ref{multres}}
\label{twores}
\end{figure}

\begin{figure}[p]
\centering
\figurenum{3a}
\plottwo{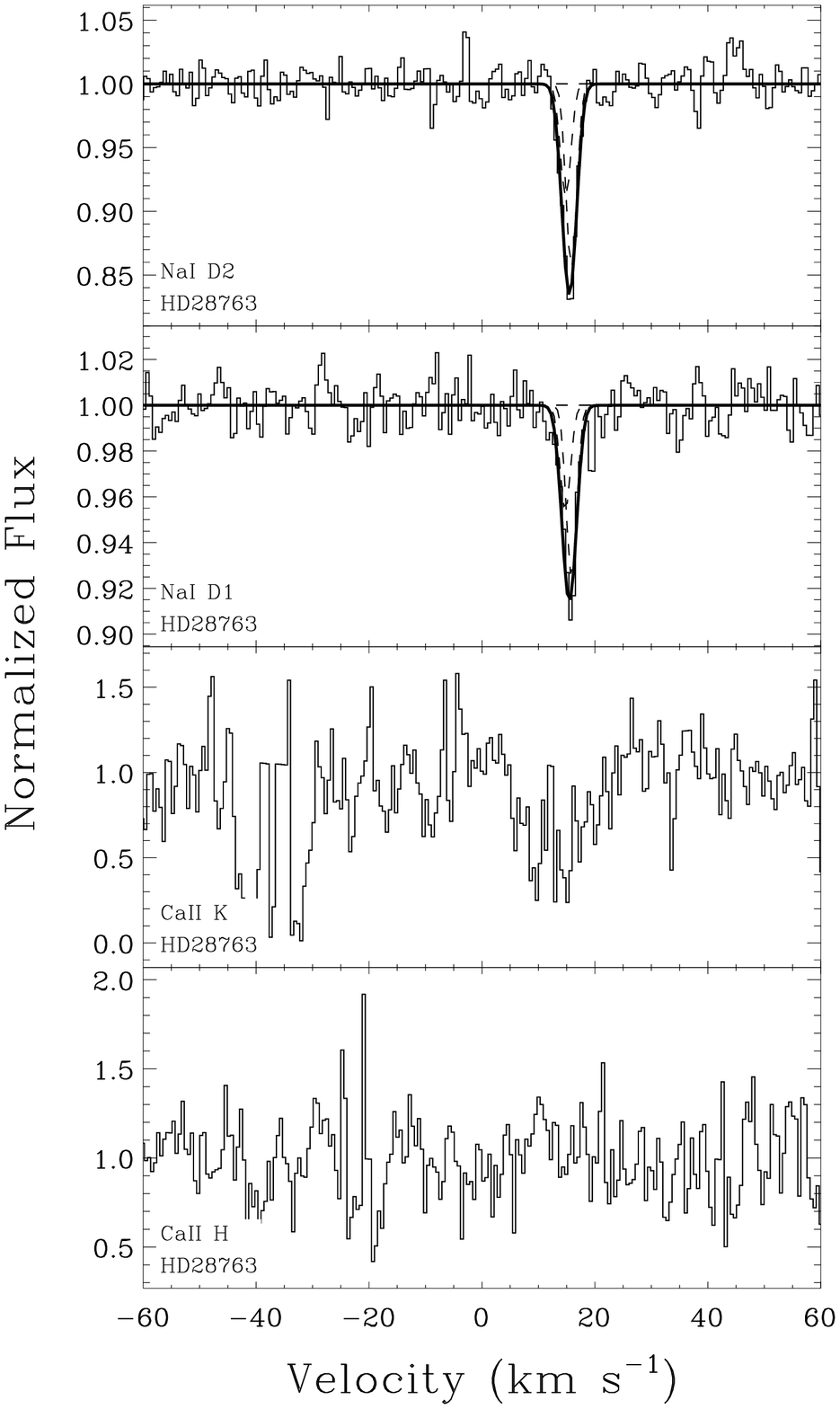}{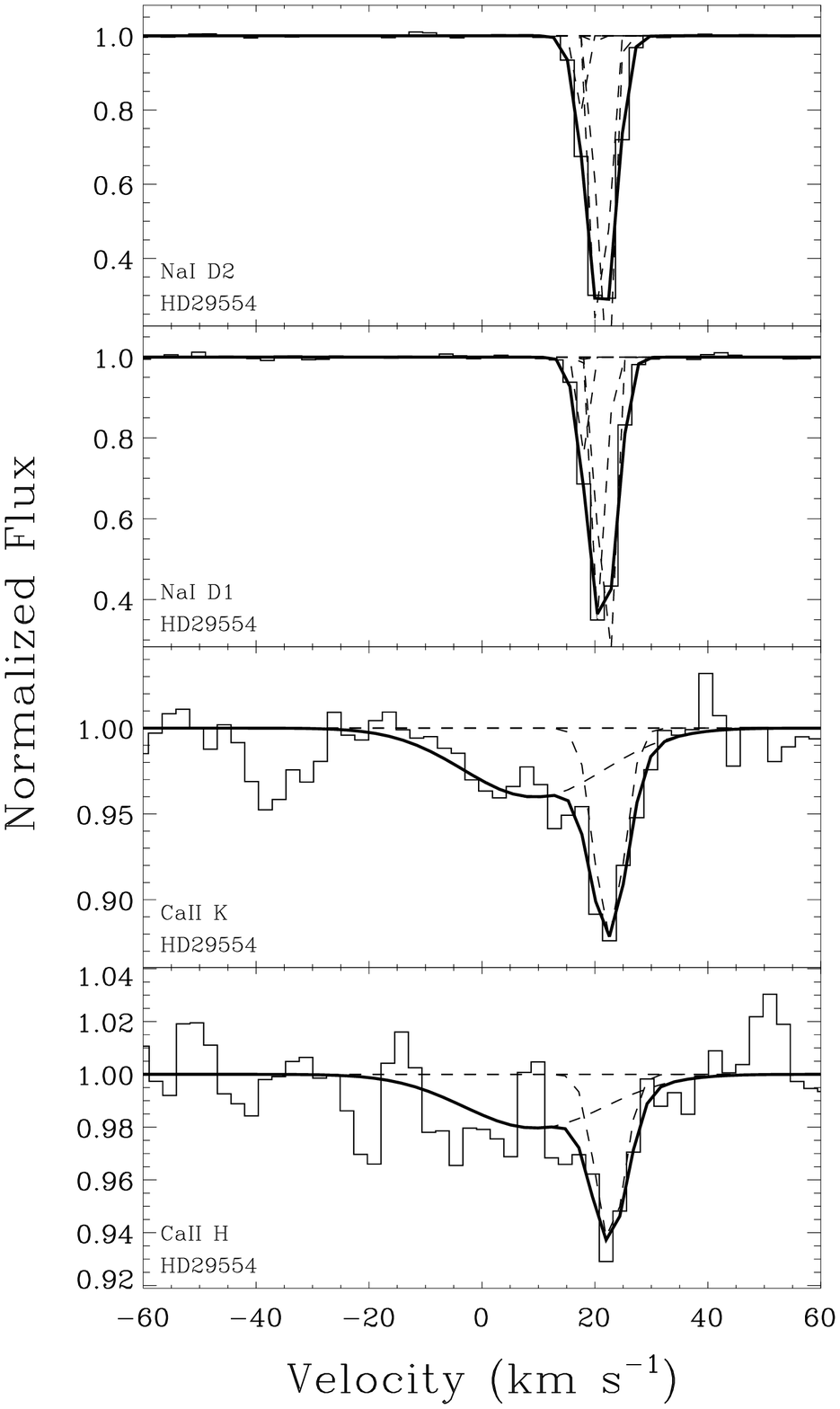}
\caption{\label{123} Fits to high-resolution \ion{Na}{1} and \ion{Ca}{2} spectra that show ISM absorption toward stars along the historical solar trajectory.  The spectra have been normalized (thin solid line).  If absorption is detected, dashed lines indicate the profile of individual absorbers (dashed lines), while the thick solid line is the total absorption of all components convolved with the instrumental line-spread function.  Fit parameters based on these fits are listed in Tables~\ref{CAallcomps} and \ref{NAallcomps}.}
\end{figure}

\begin{figure}[p]
\centering
\figurenum{3b}
\plottwo{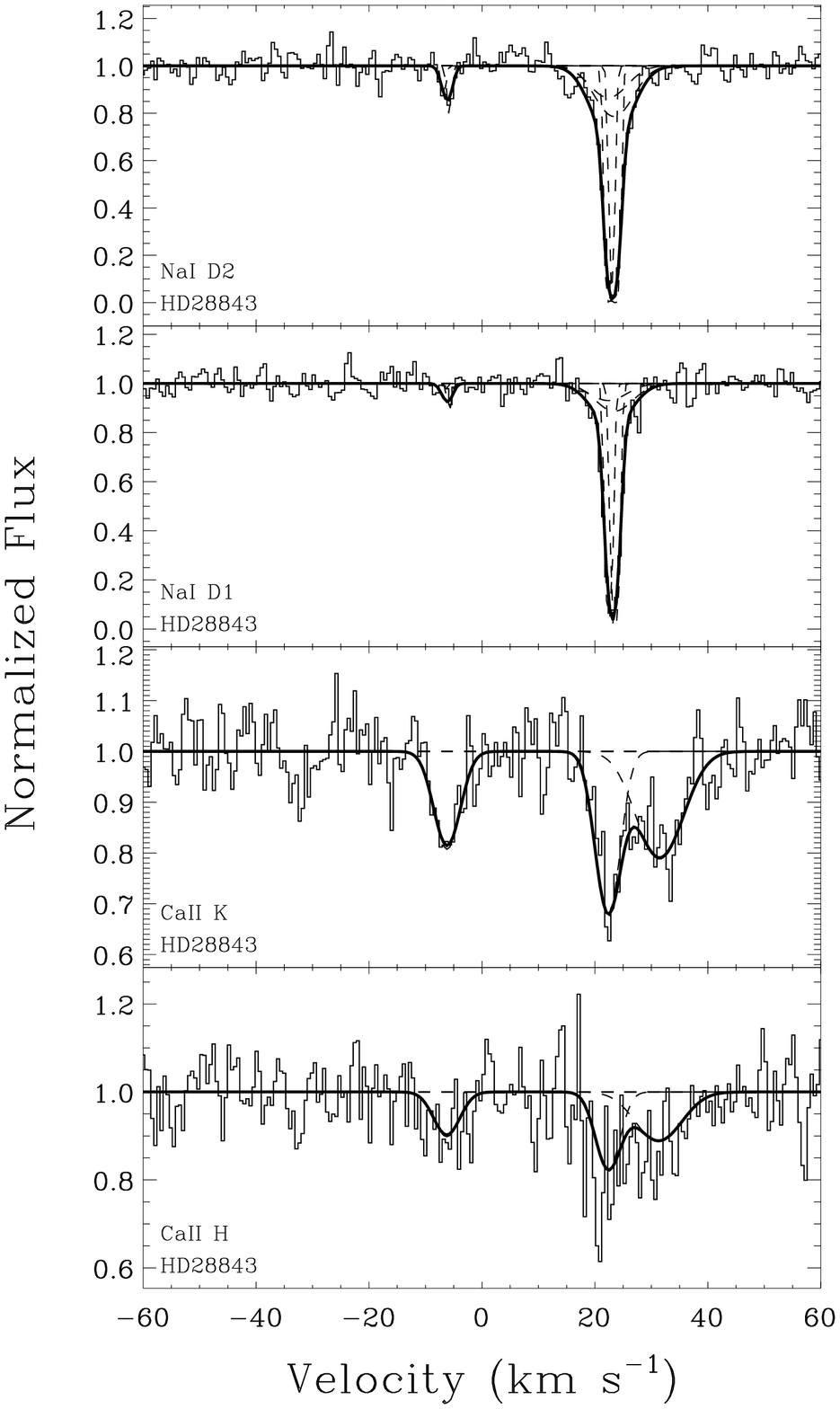}{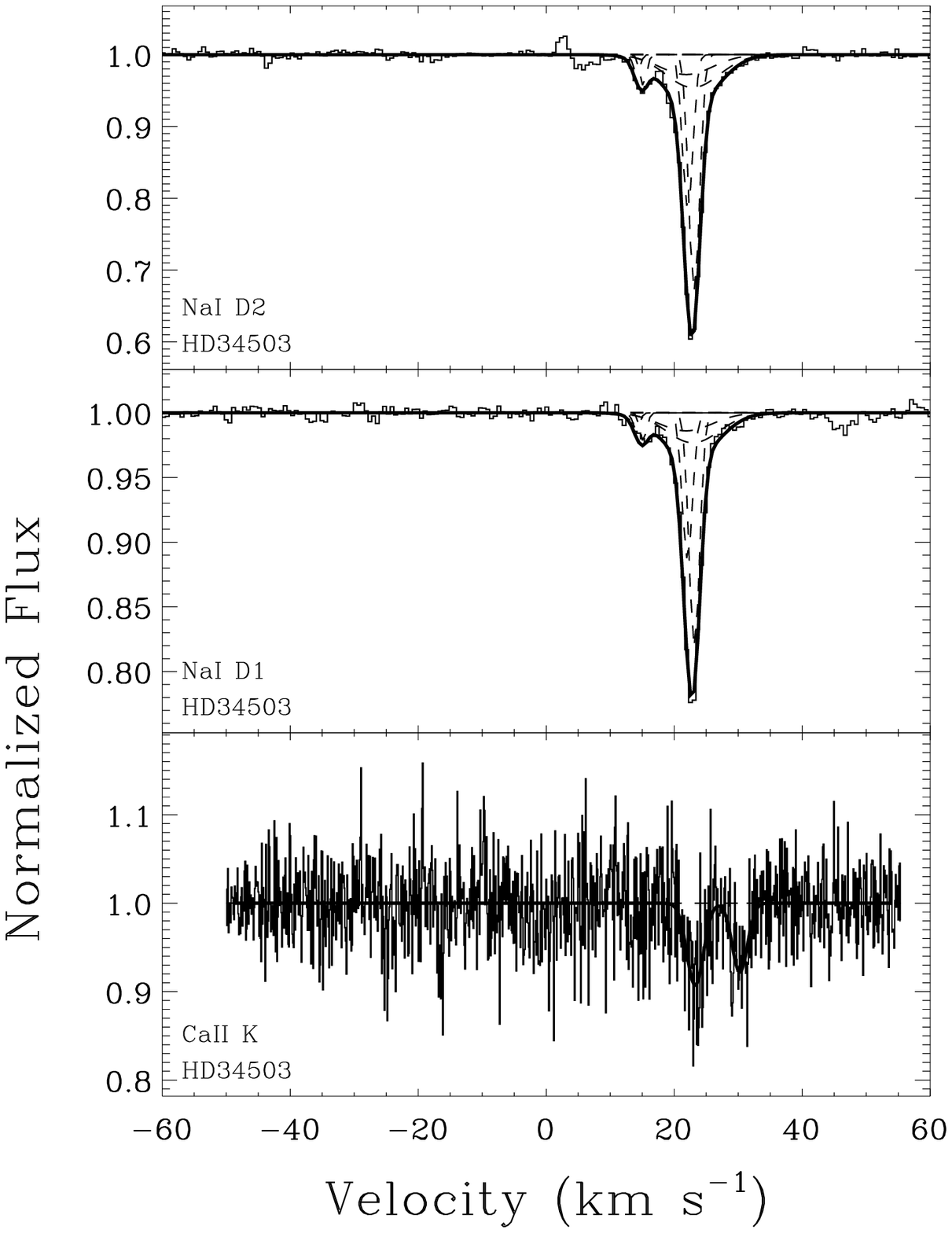}
\caption{\label{131} Same as Figure \ref{123}}
\end{figure}

\begin{figure}[p]
\centering
\figurenum{3c}
\plottwo{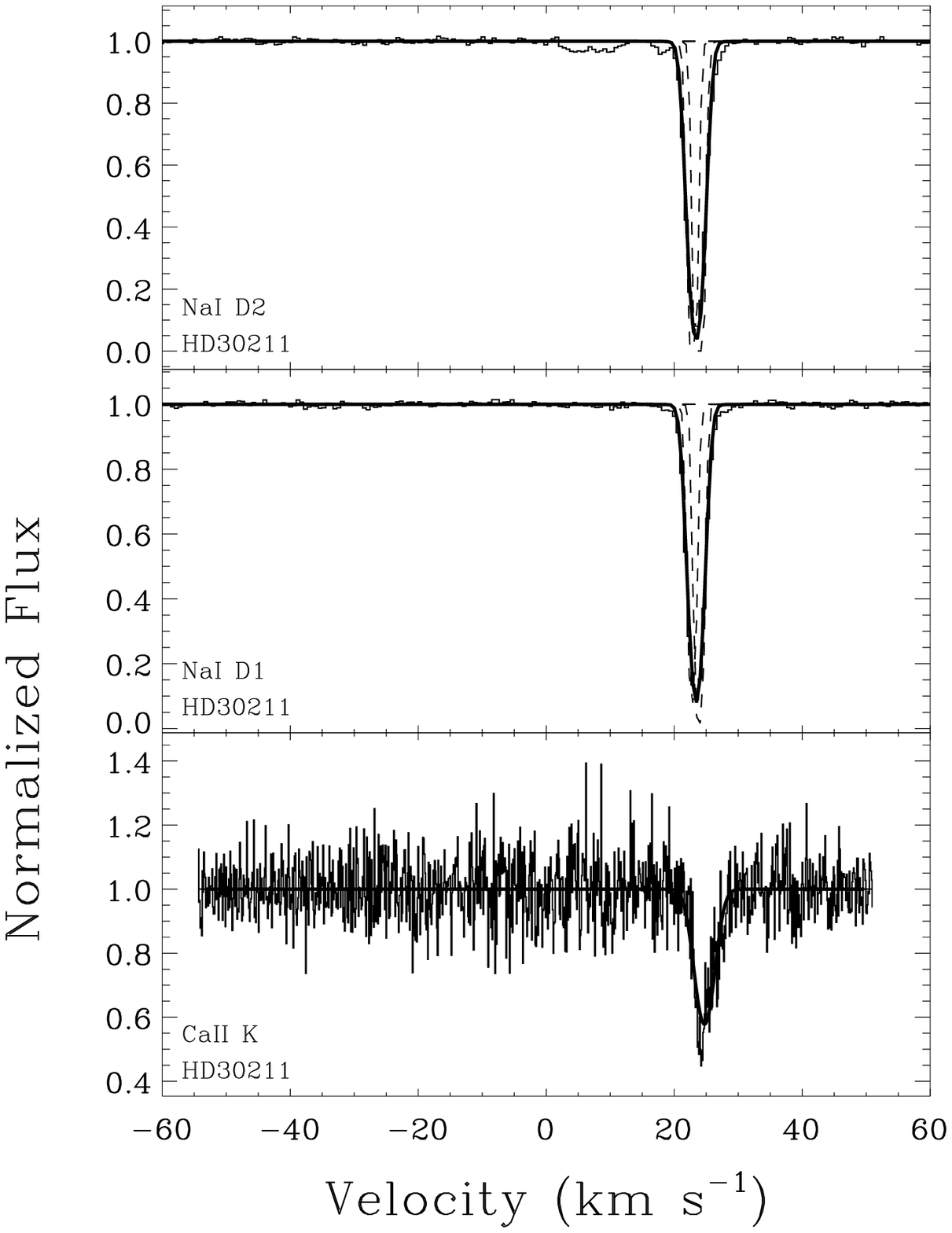}{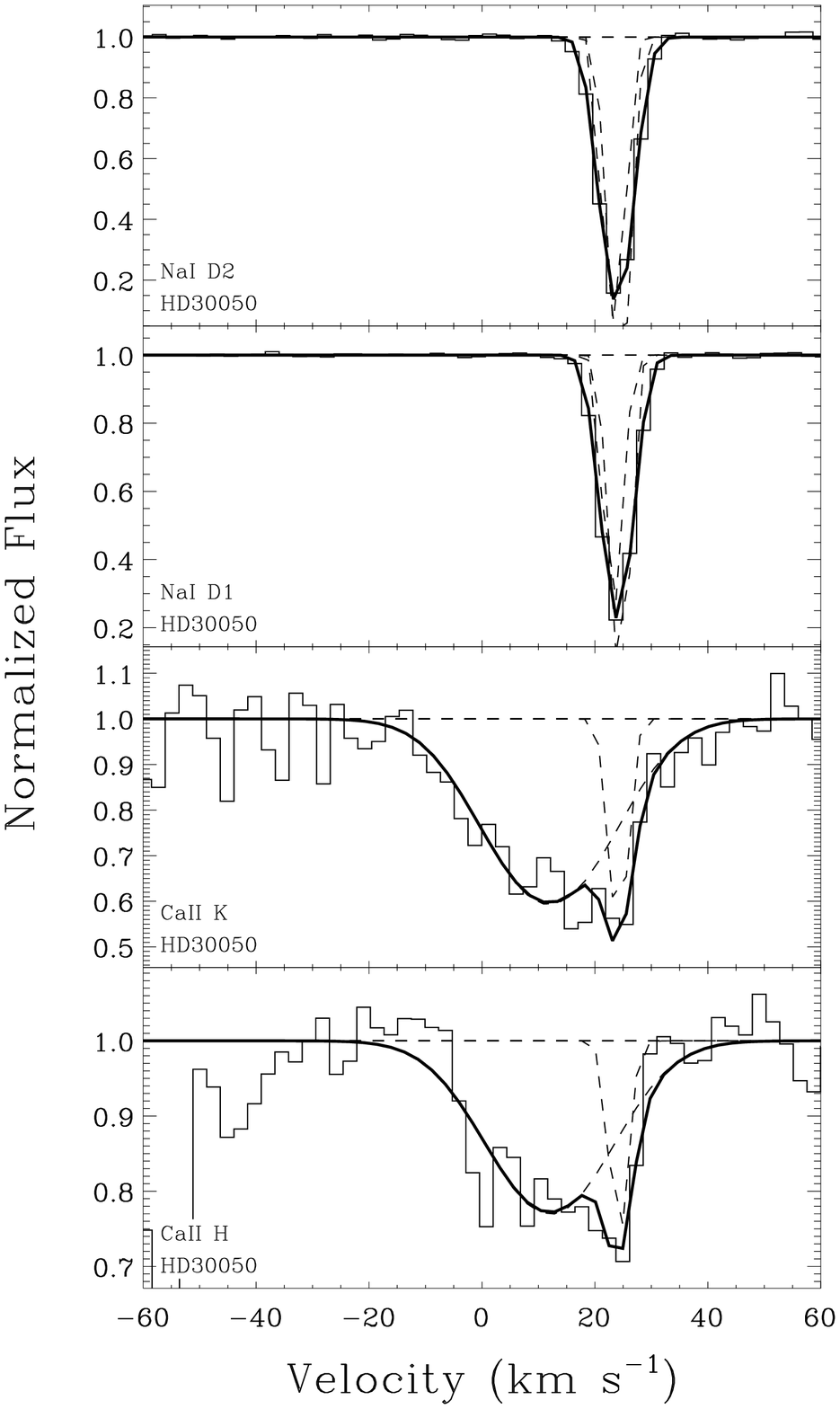}
\caption{\label{161}Same as Figure \ref{123}}
\end{figure}

\begin{figure}[p]
\centering
\figurenum{3d}
\plottwo{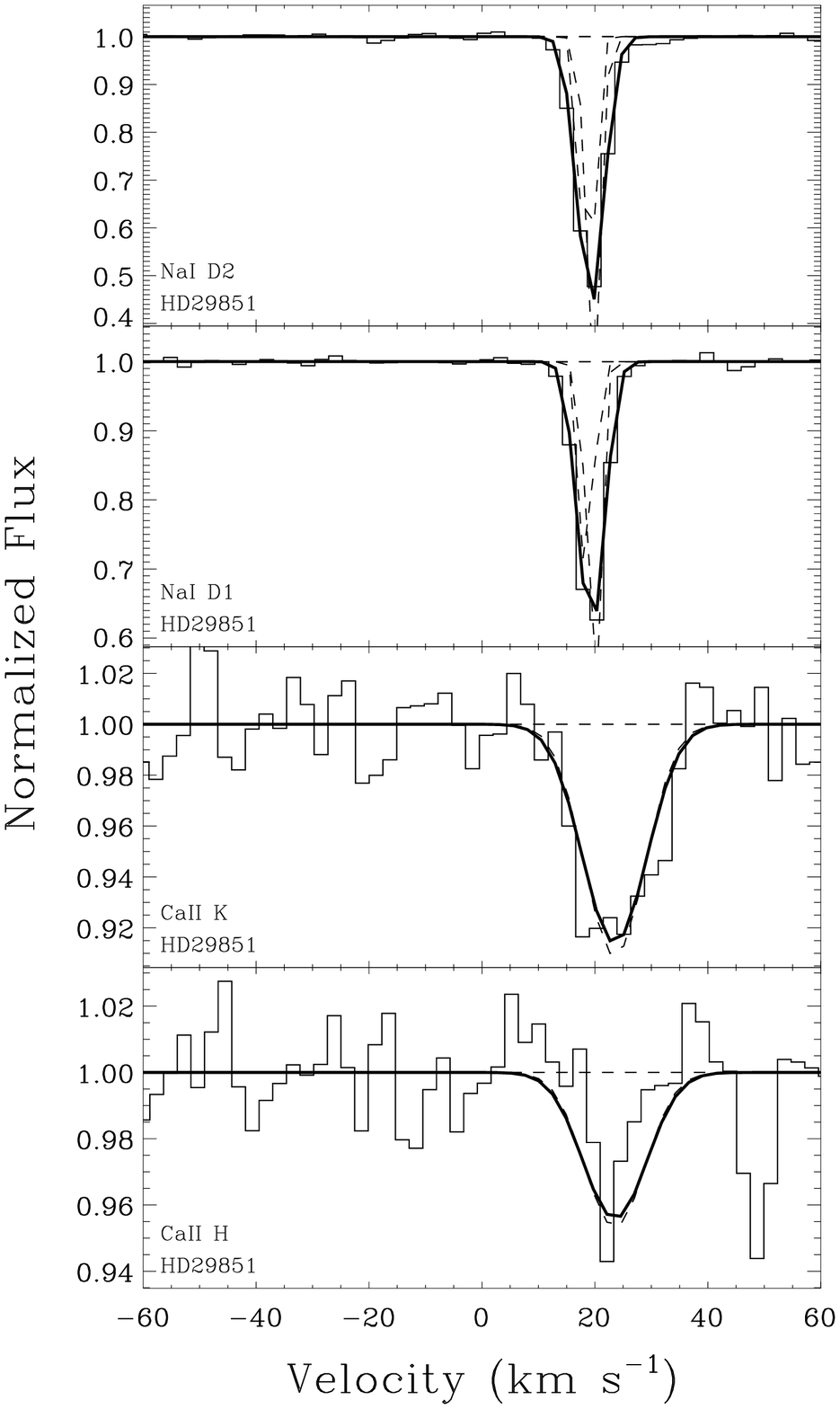}{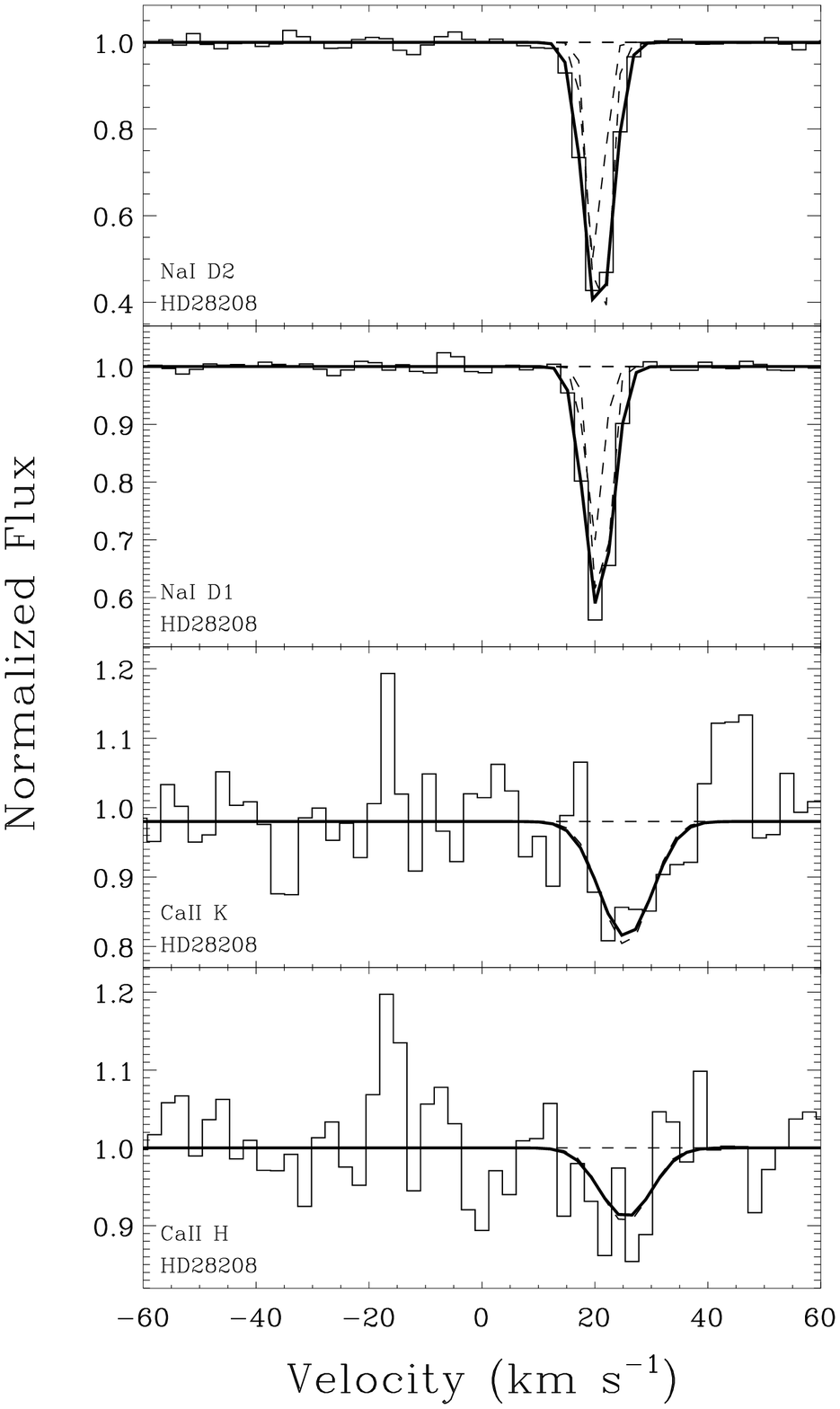}
\caption{\label{170}Same as Figure \ref{123}}
\end{figure}

\begin{figure}[p]
\centering
\figurenum{3e}
\plottwo{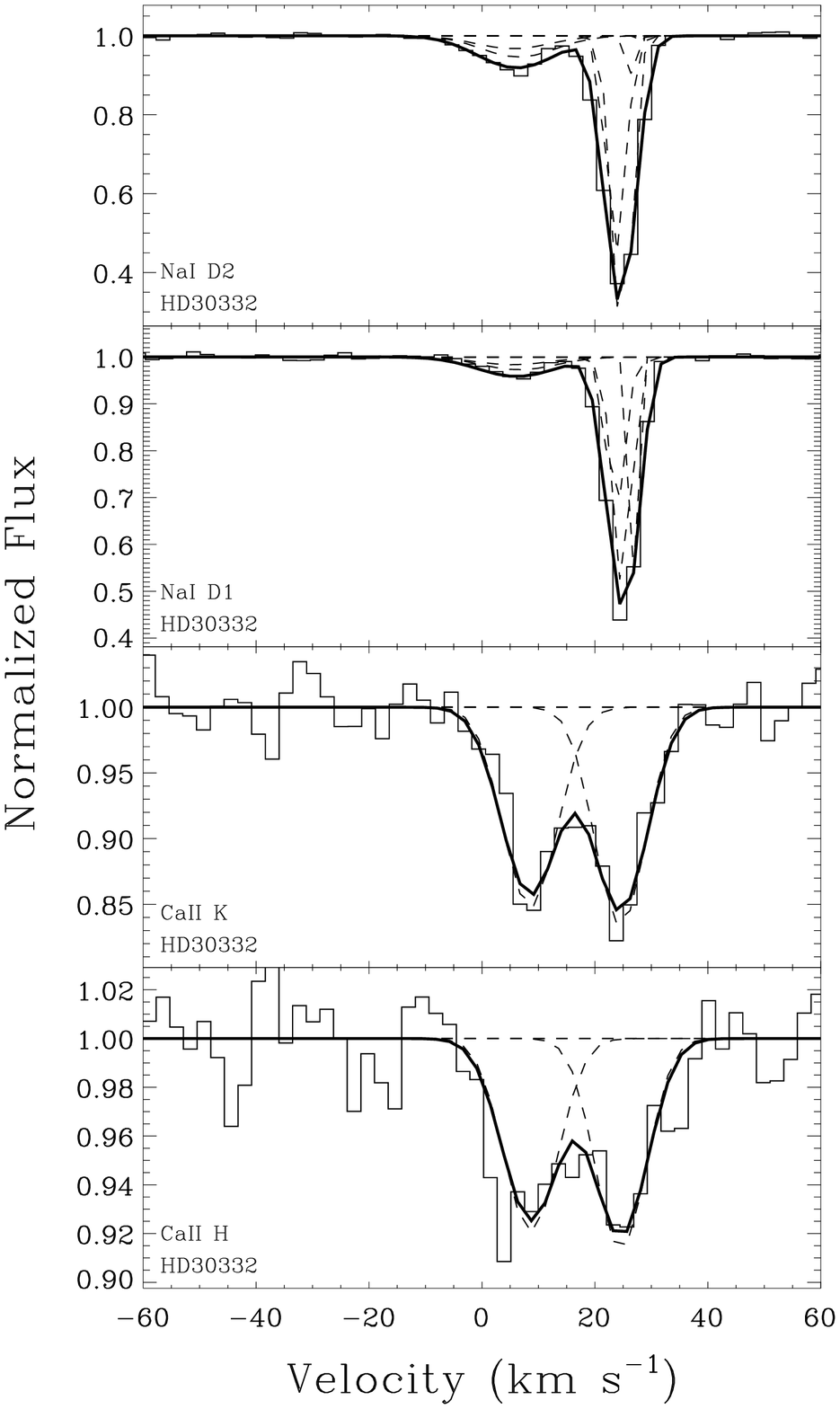}{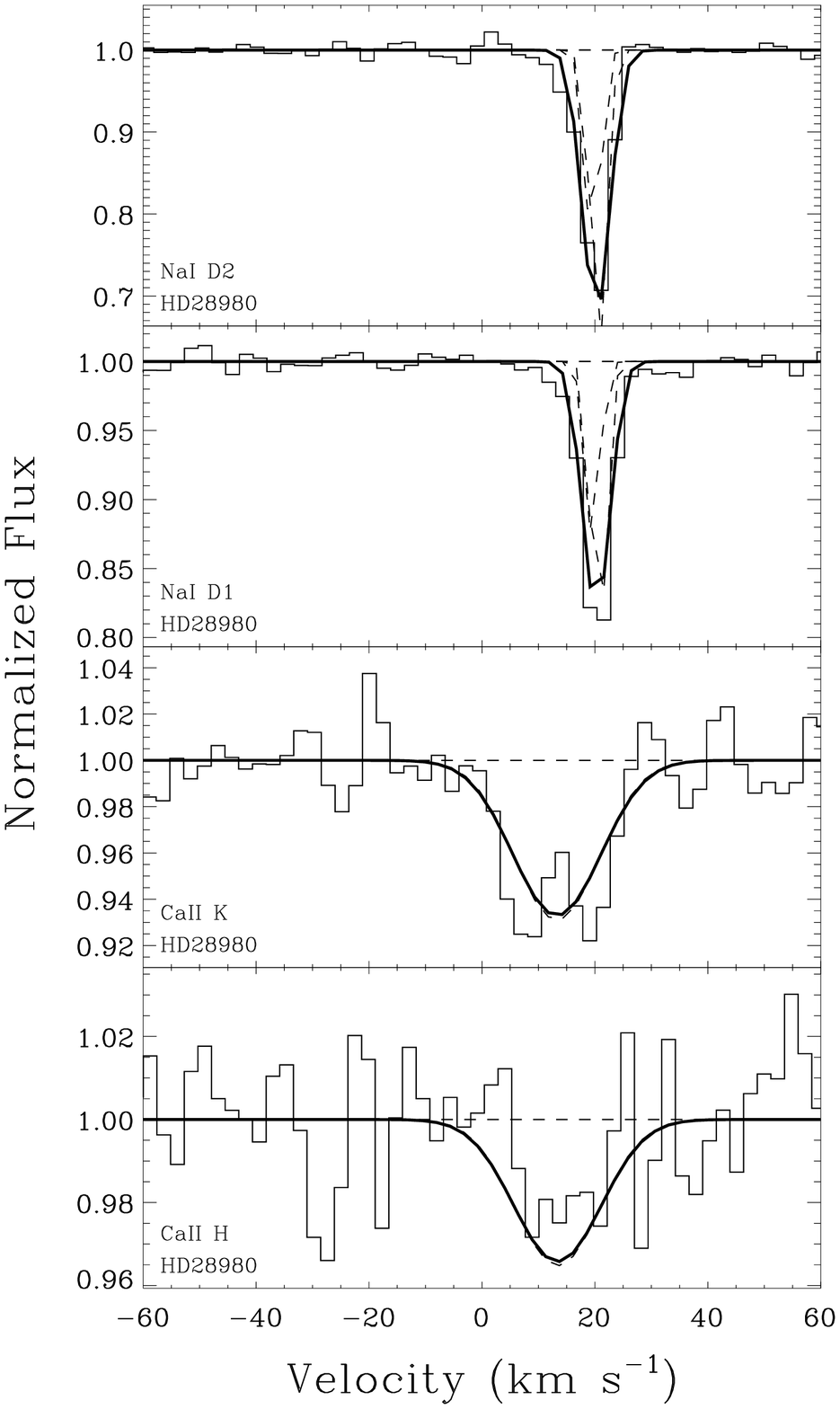}
\caption{\label{172}Same as Figure \ref{123}}
\end{figure}

\begin{figure}[p]
\centering
\figurenum{3f}
\plottwo{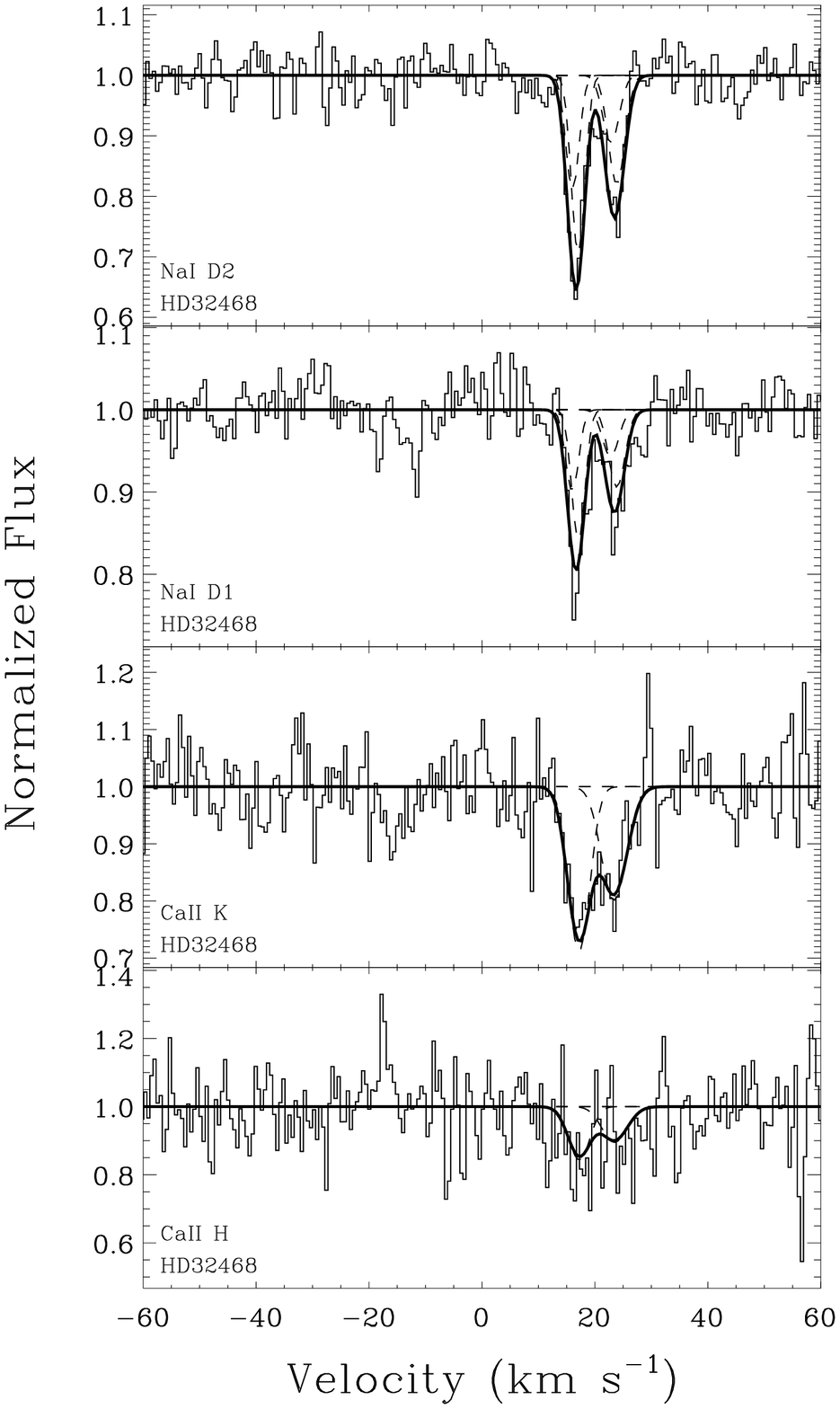}{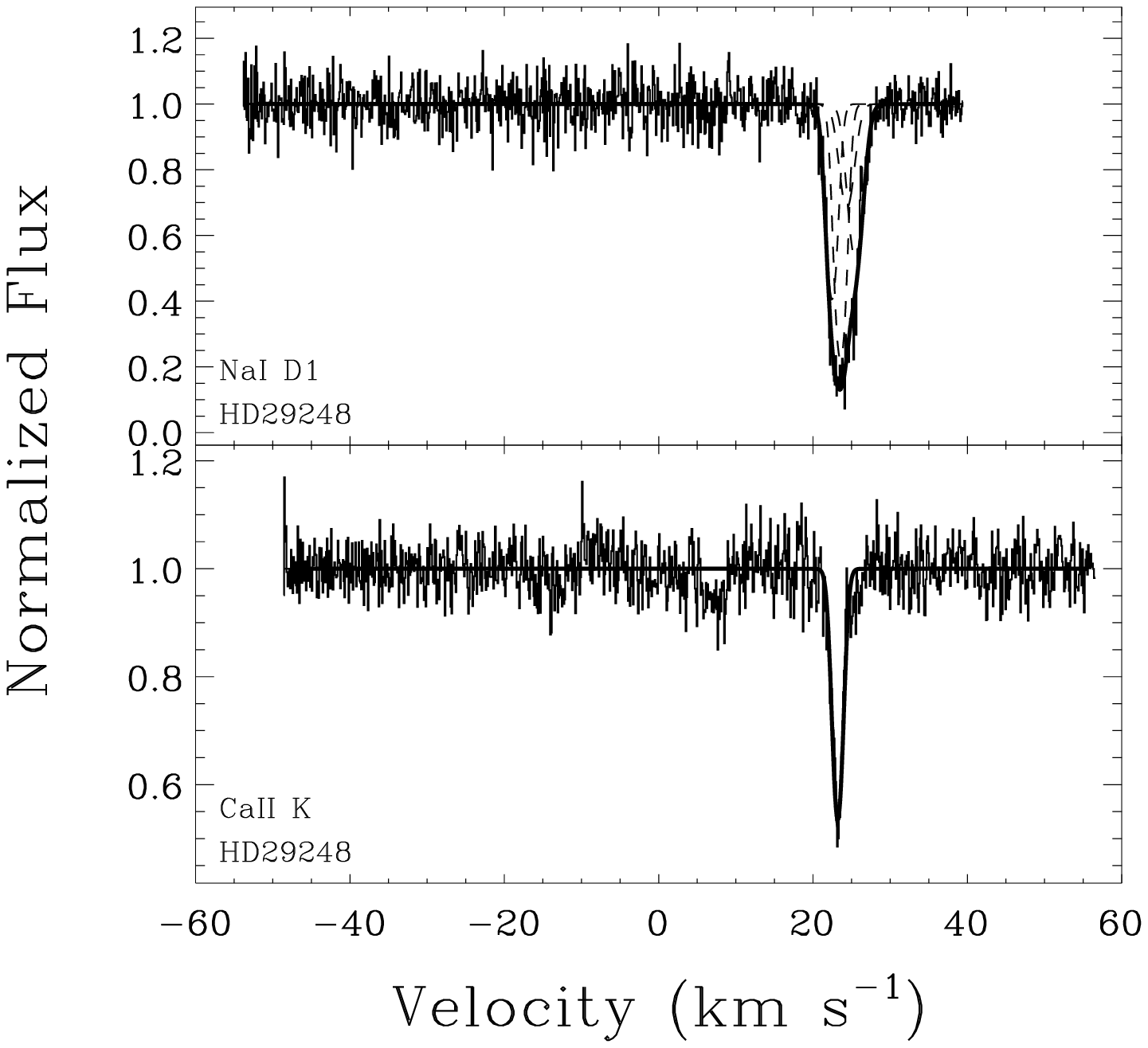}
\caption{\label{185}Same as Figure \ref{123}}
\end{figure}

\begin{figure}[p]
\centering
\figurenum{3g}
\plottwo{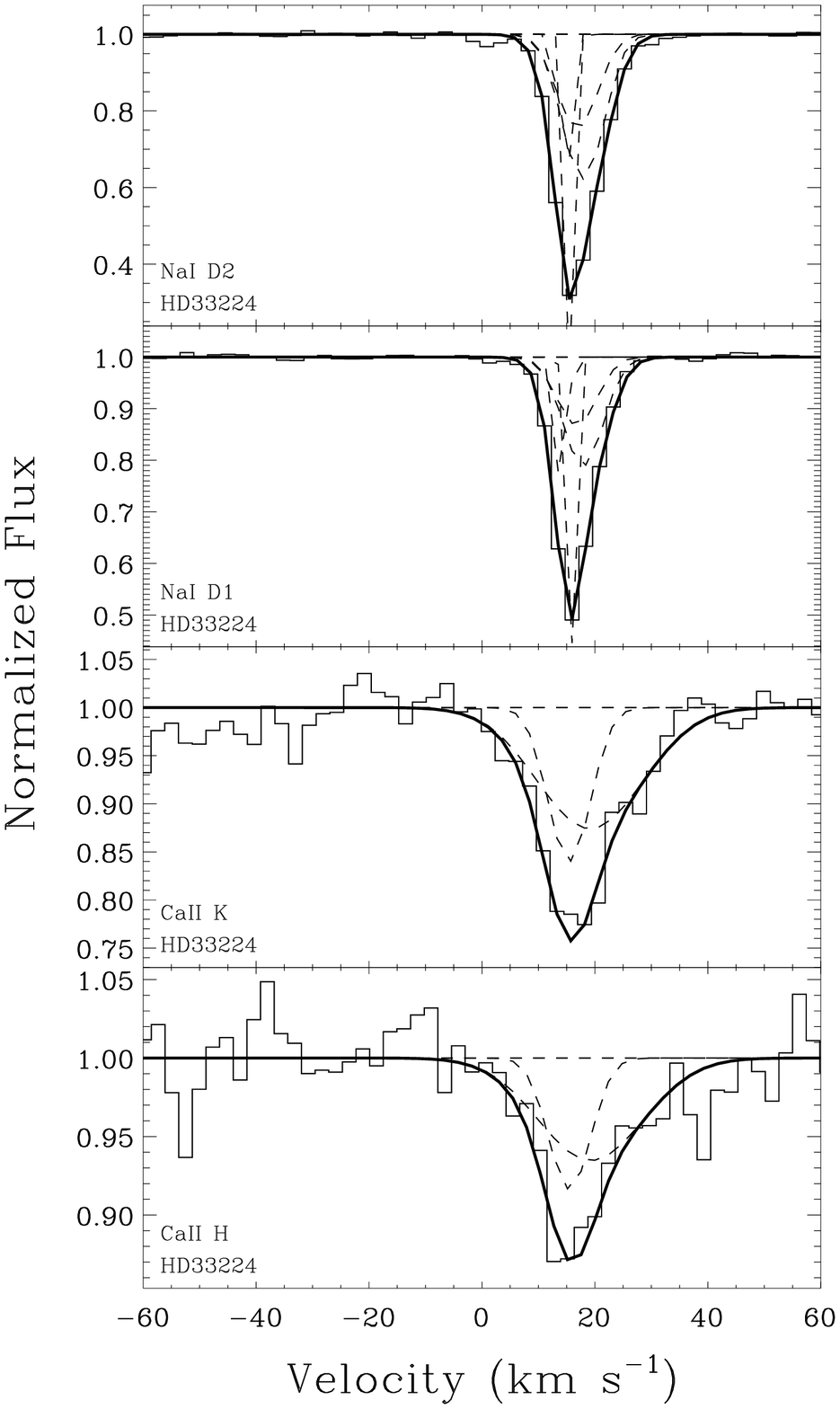}{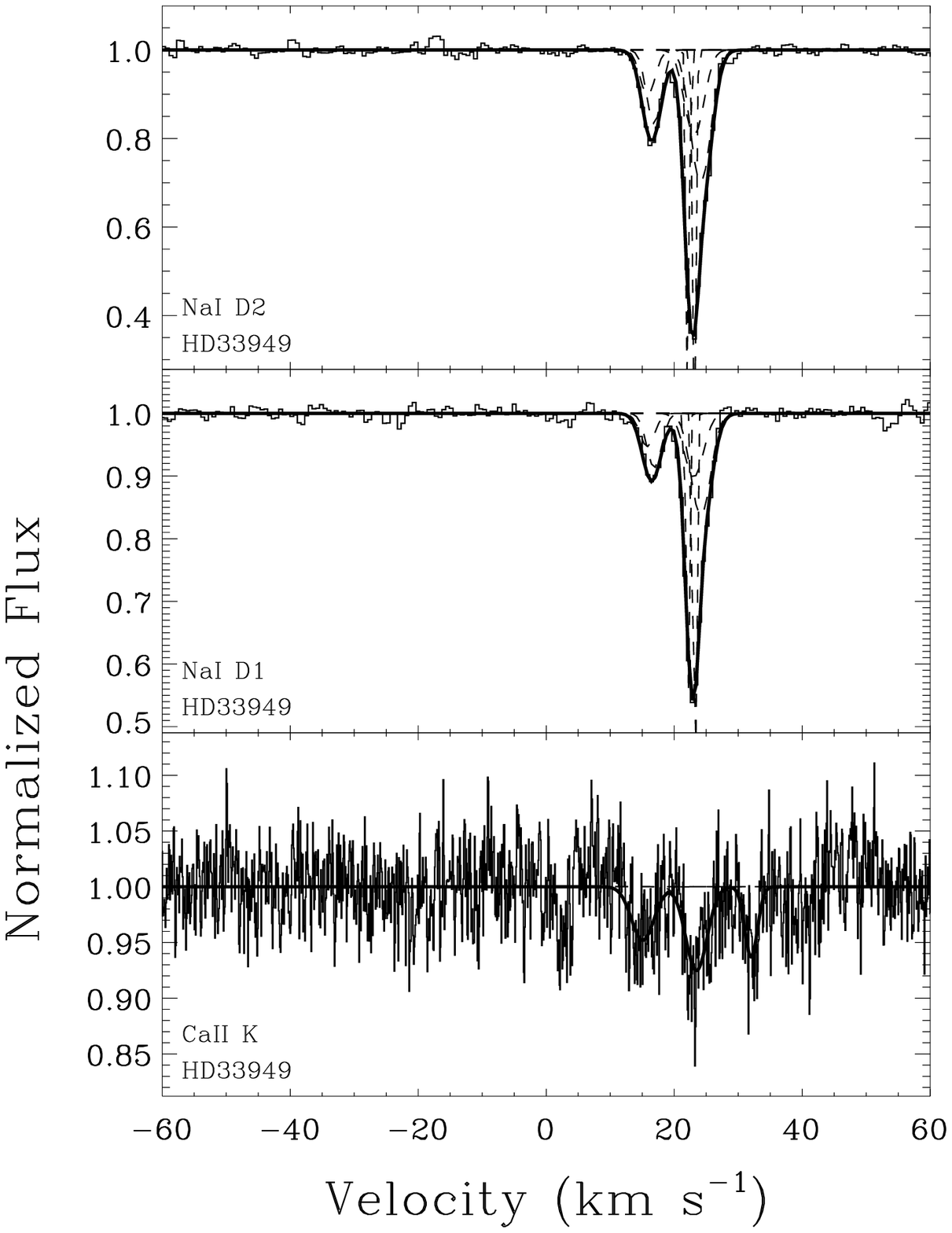}
\caption{\label{224}Same as Figure \ref{123}}
\end{figure}

\begin{figure}[p]
\centering
\figurenum{3h}
\plottwo{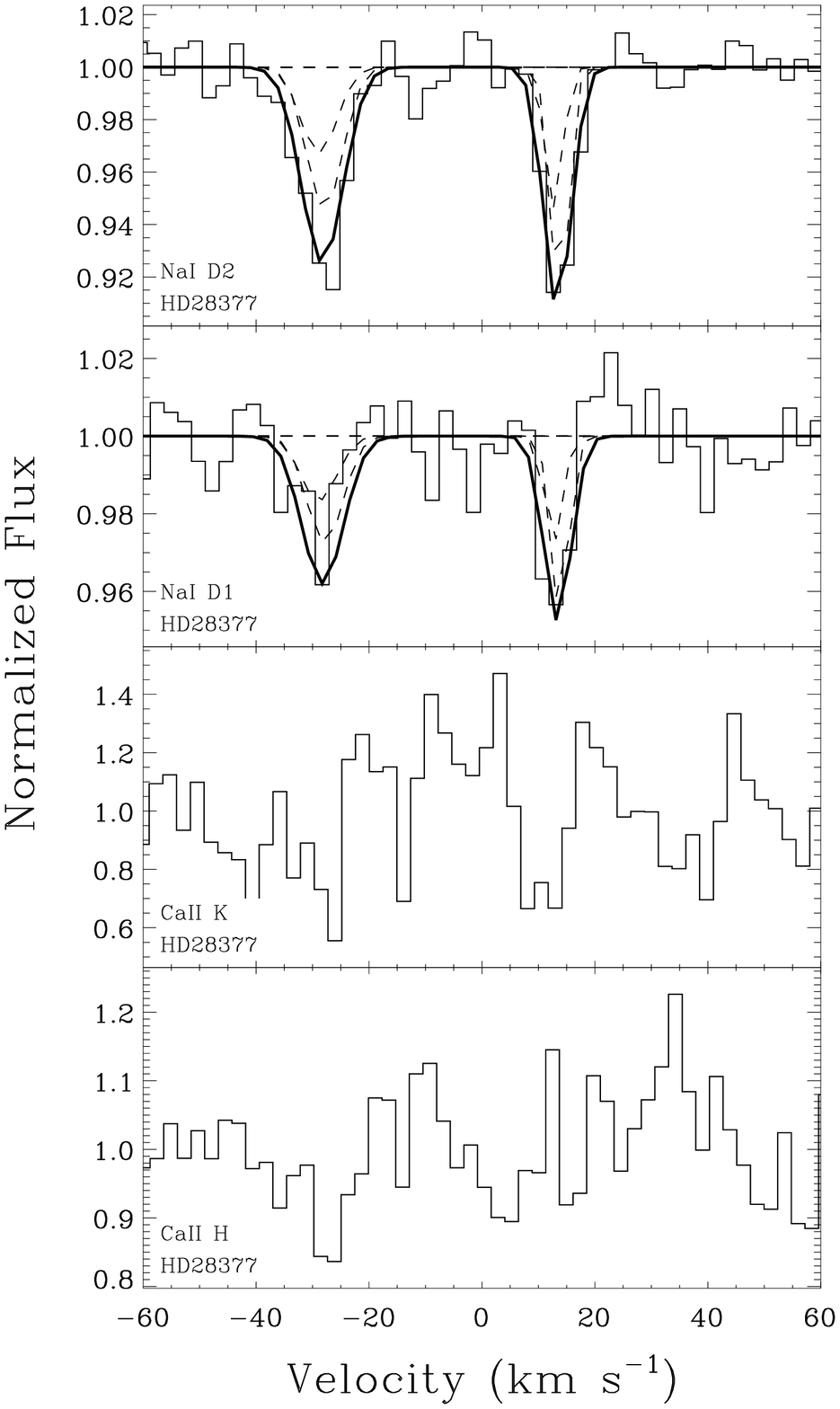}{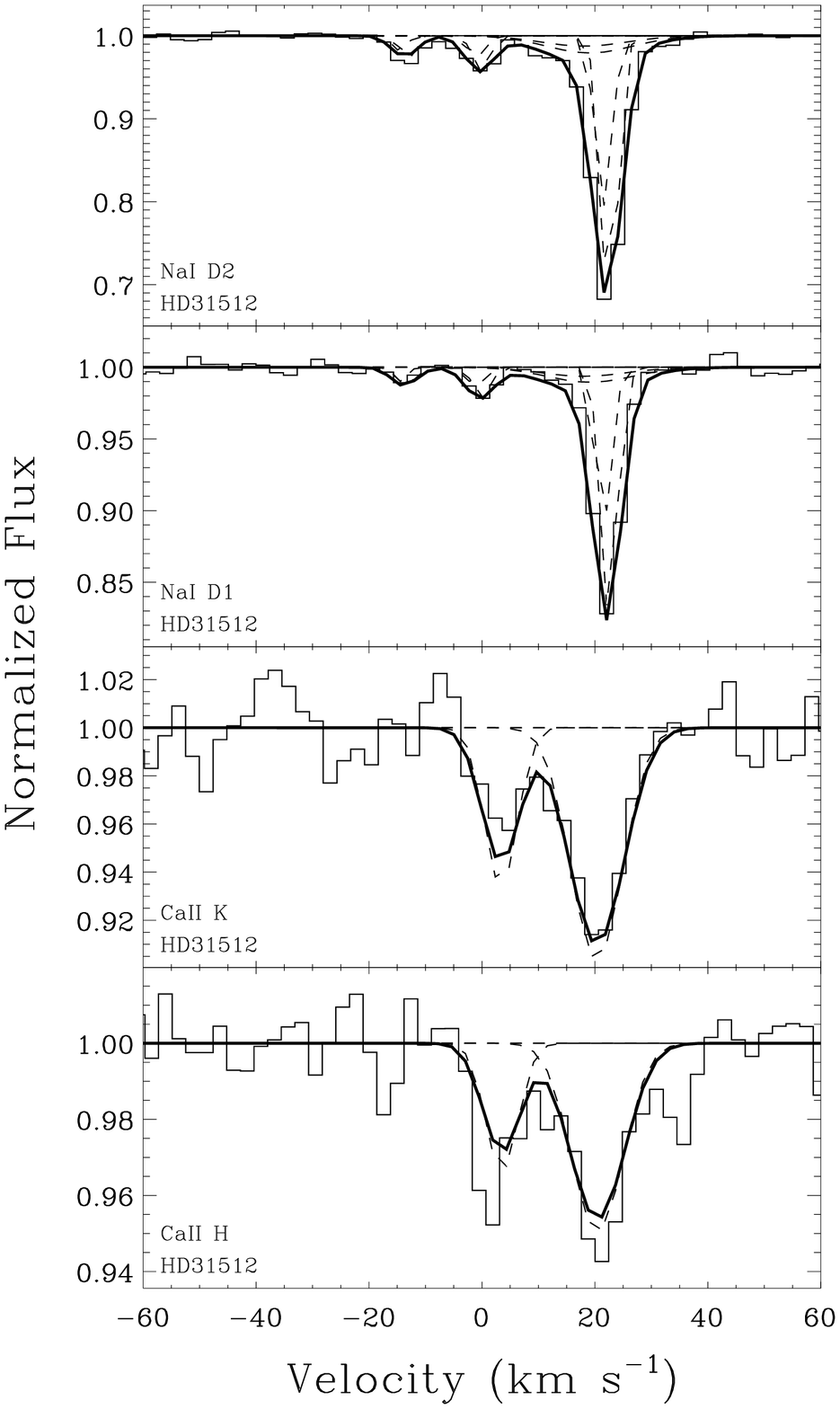}
\caption{\label{232}Same as Figure \ref{123}}
\end{figure}

\begin{figure}[p]
\centering
\figurenum{3i}
\plottwo{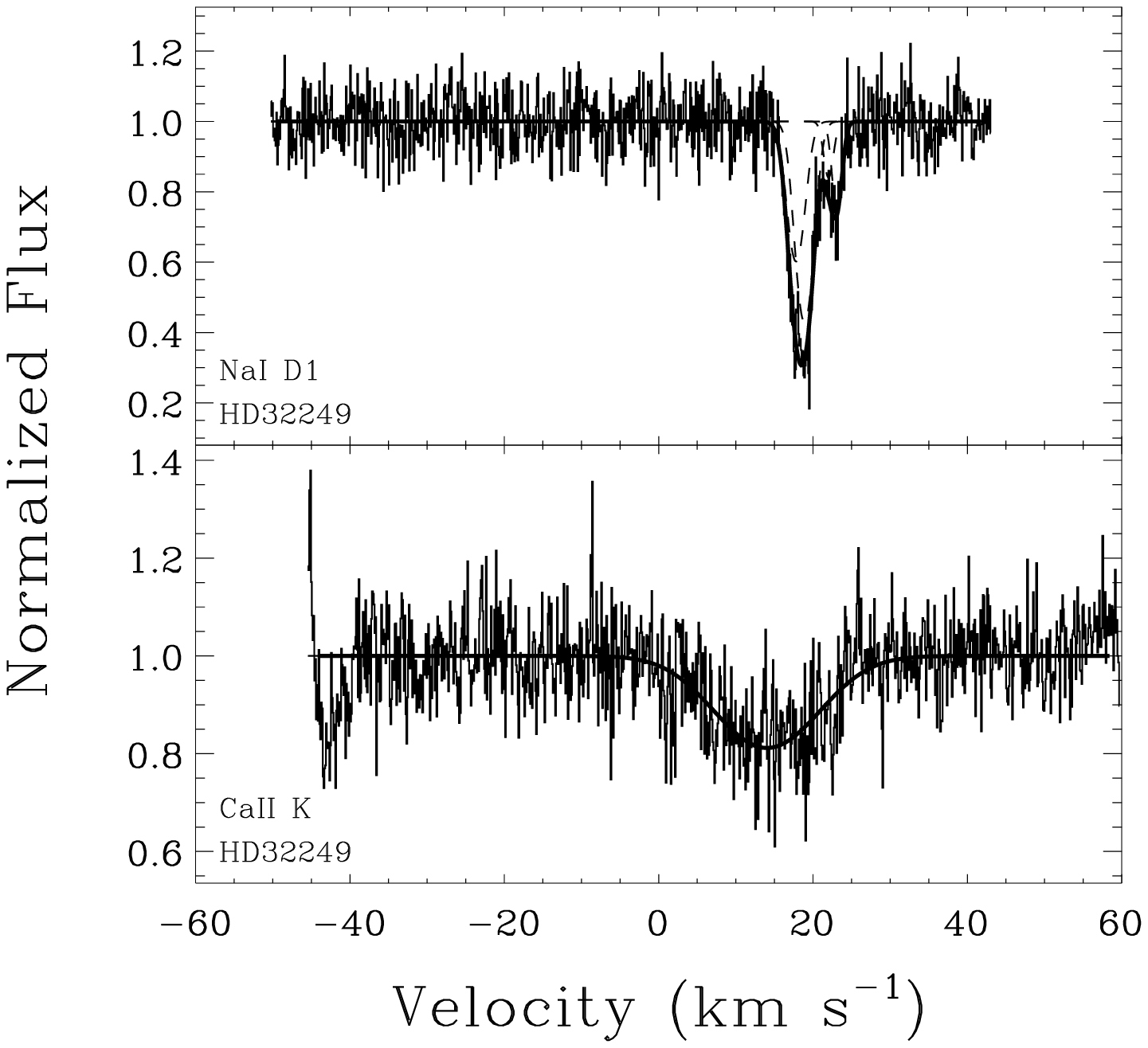}{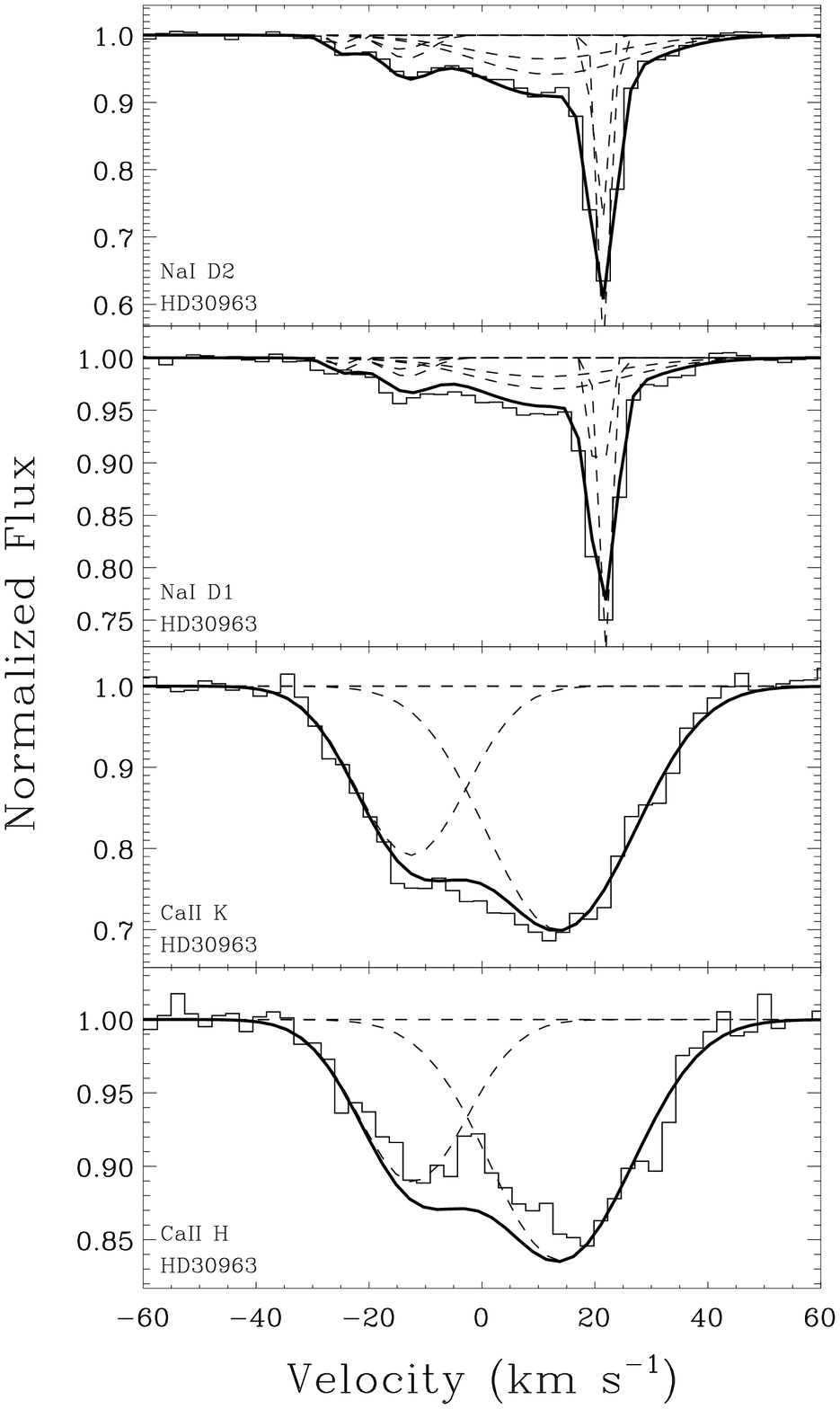}
\caption{\label{233}Same as Figure \ref{123}}
\end{figure}

\begin{figure}[p]
\centering
\figurenum{3j}
\plottwo{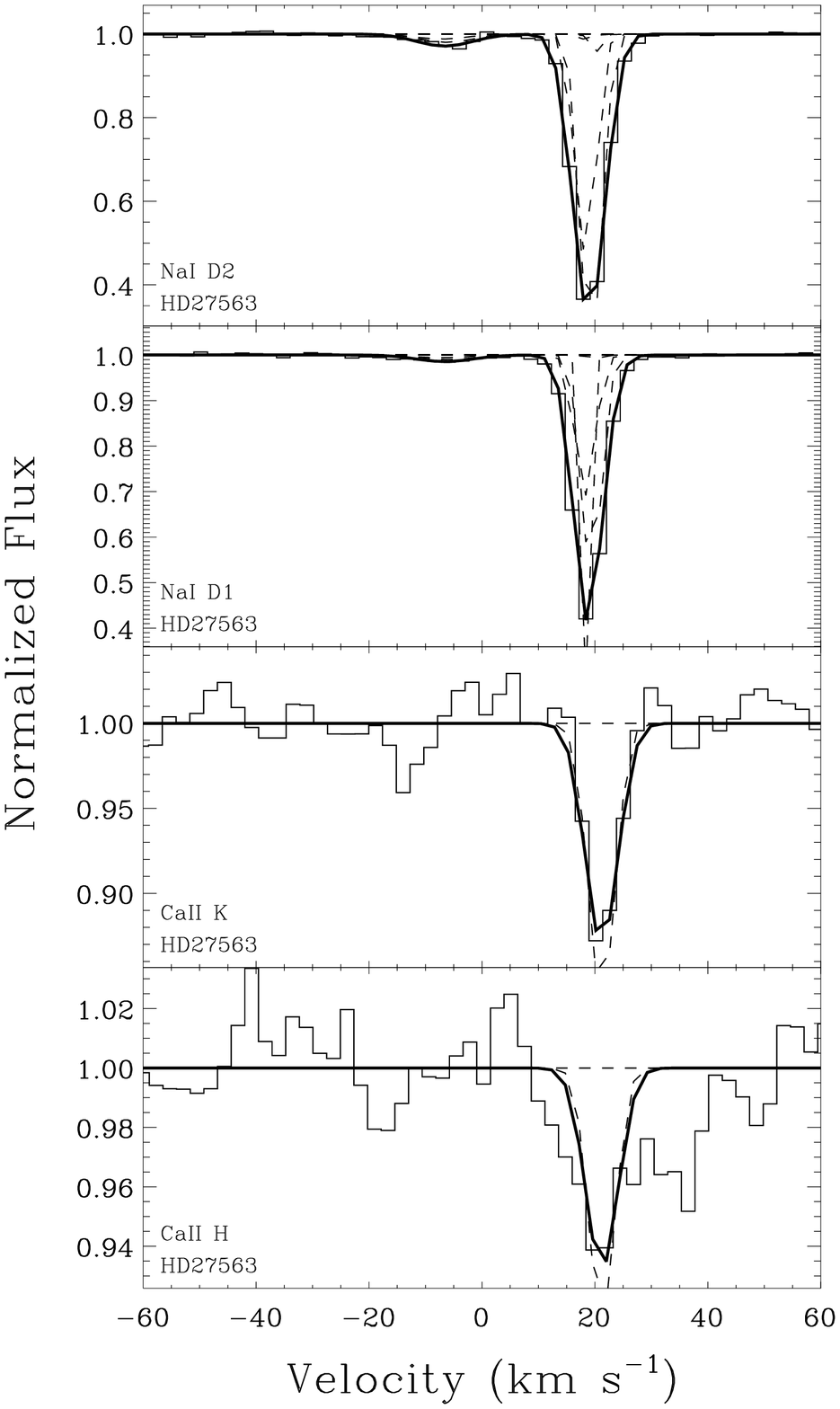}{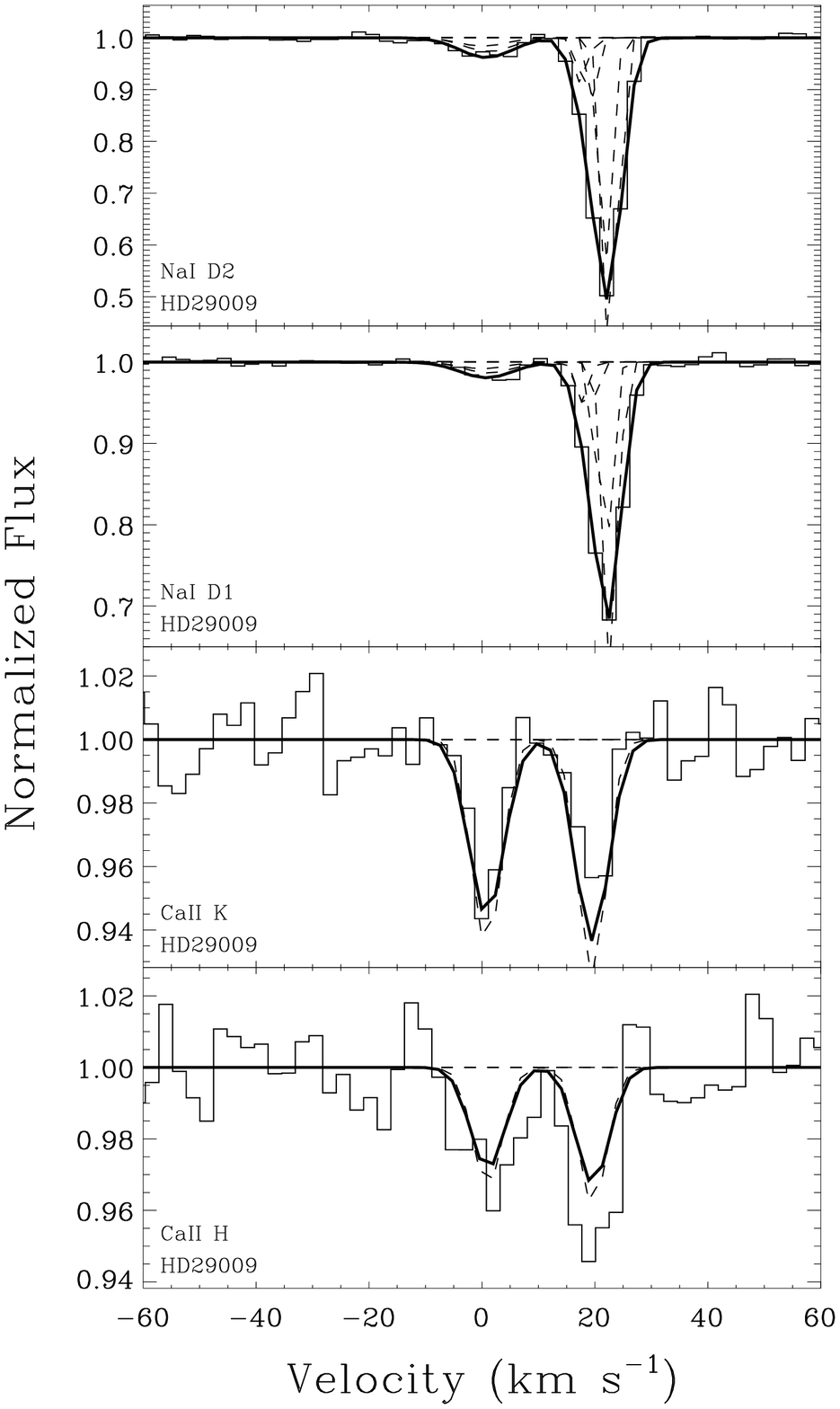}
\caption{\label{239}Same as Figure \ref{123}}
\end{figure}

\begin{figure}[p]
\centering
\figurenum{3k}
\plottwo{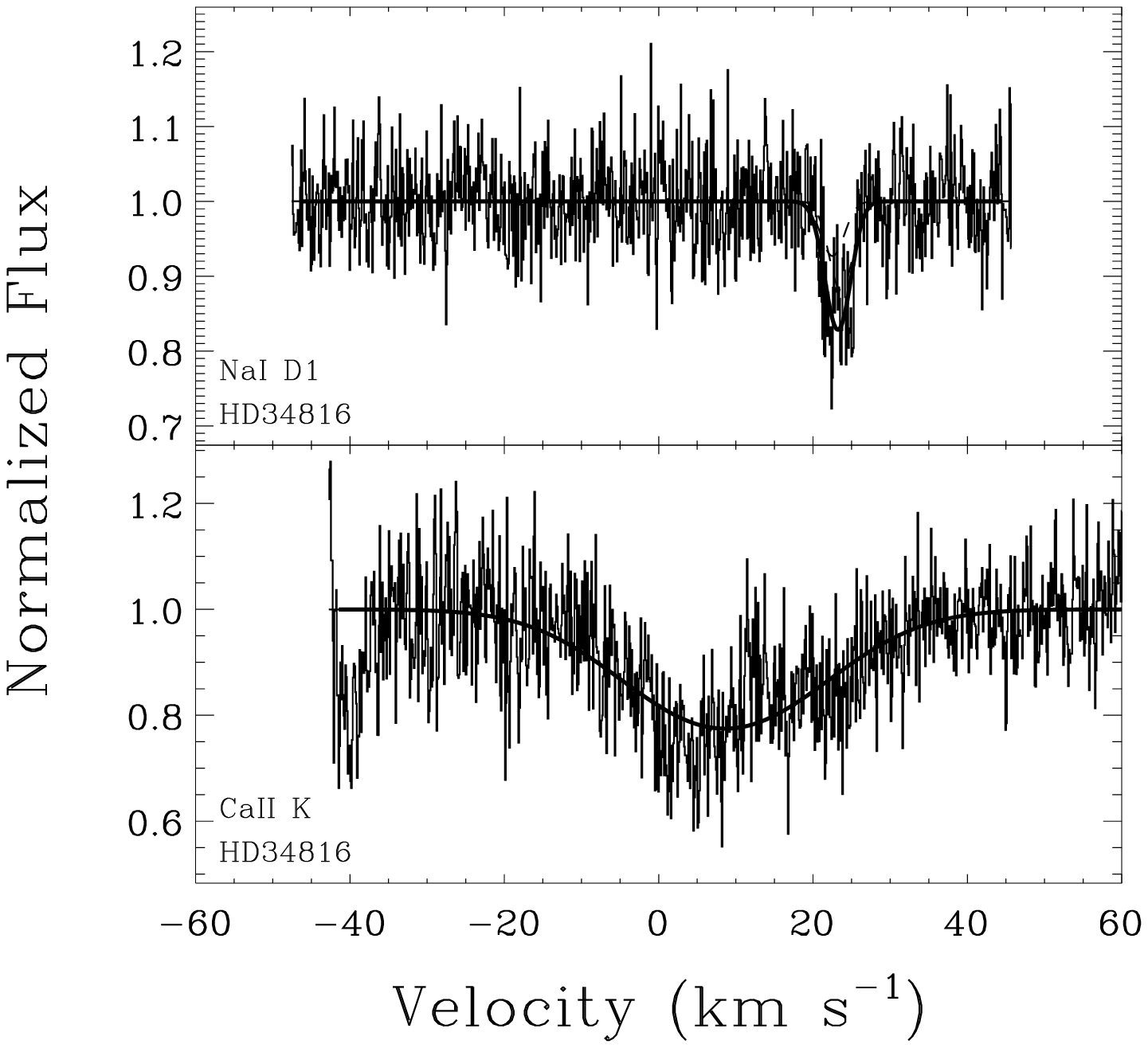}{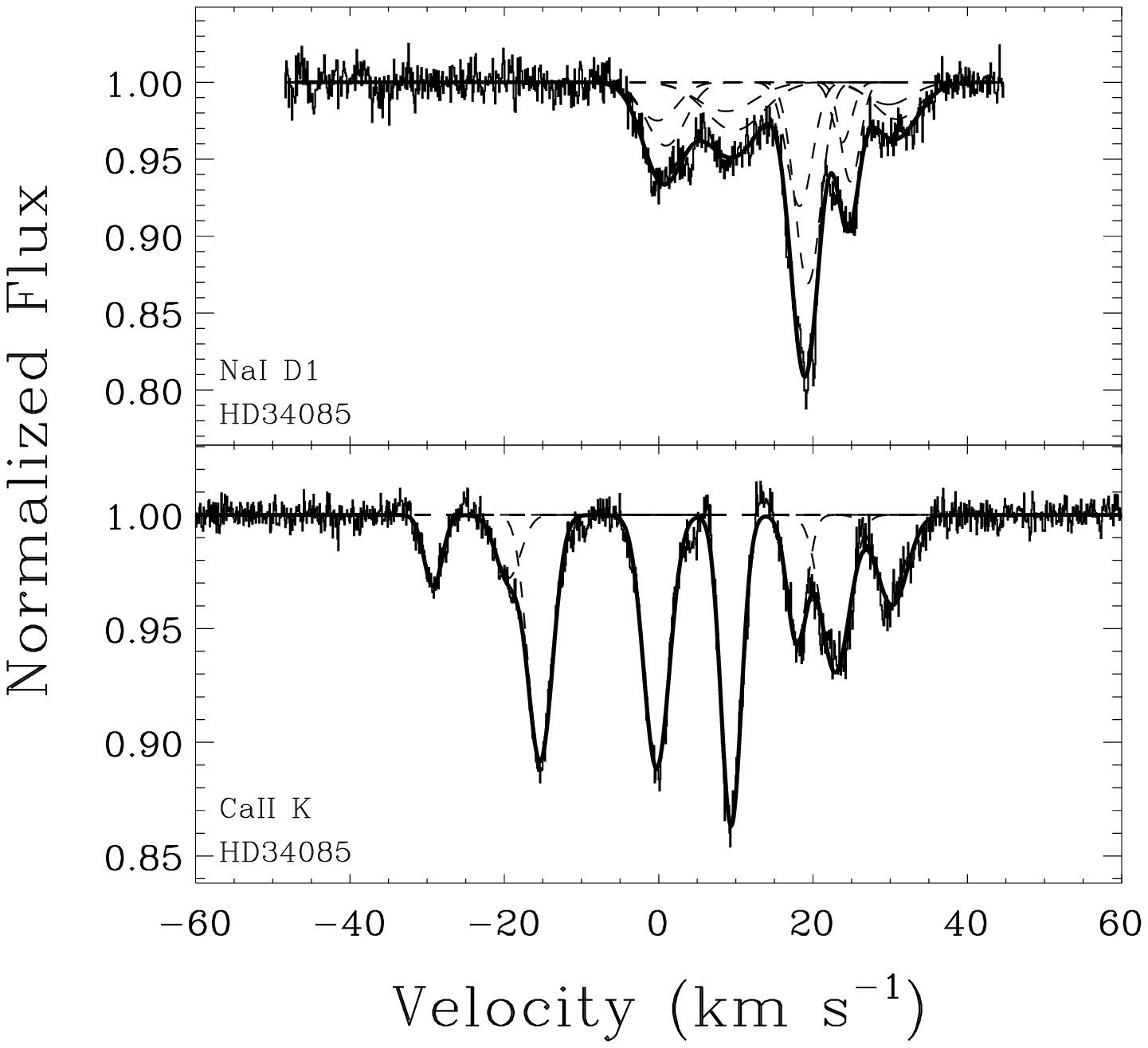}
\caption{\label{293}Same as Figure \ref{123}}
\end{figure}

\begin{figure}[p]
\centering
\figurenum{3l}
\plottwo{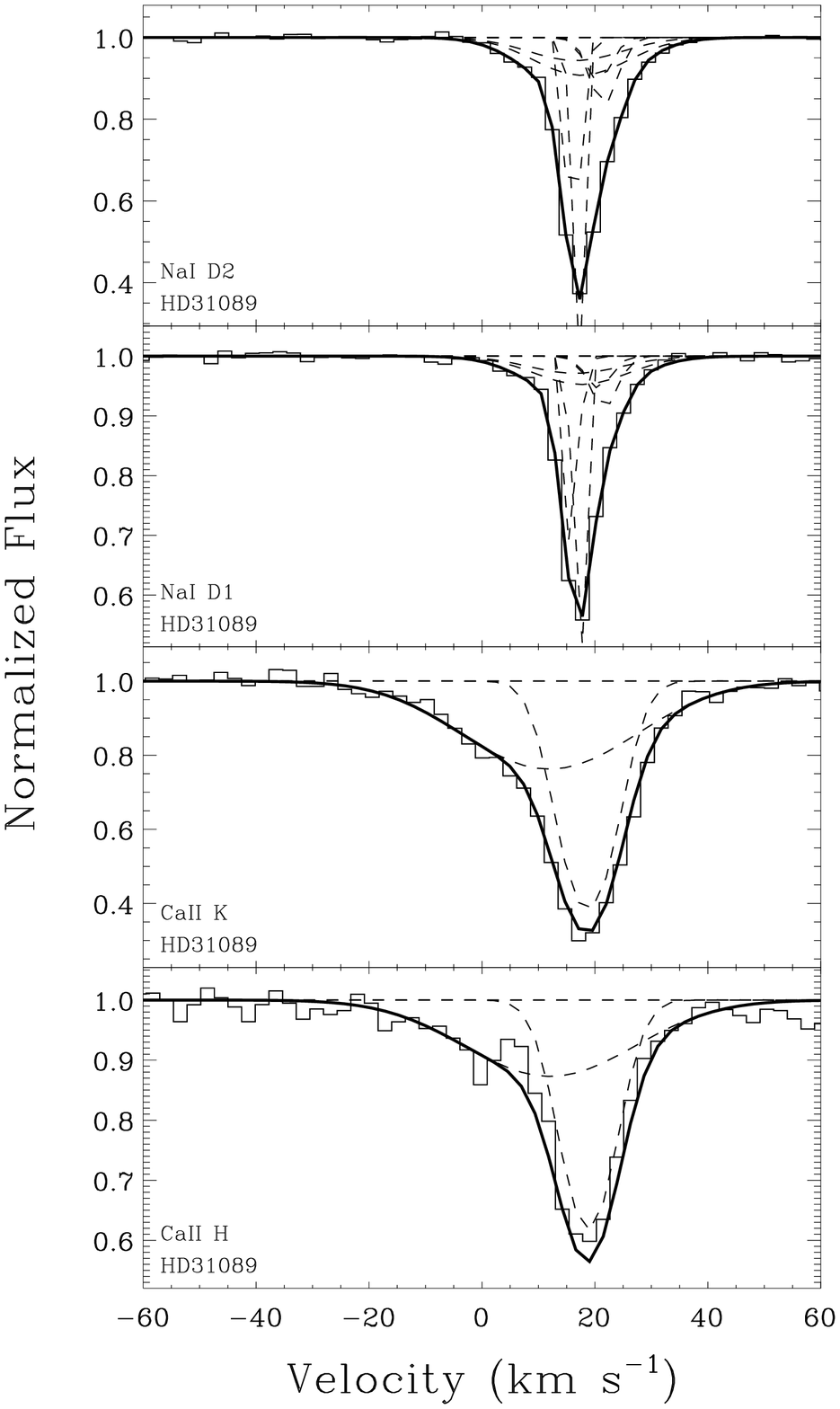}{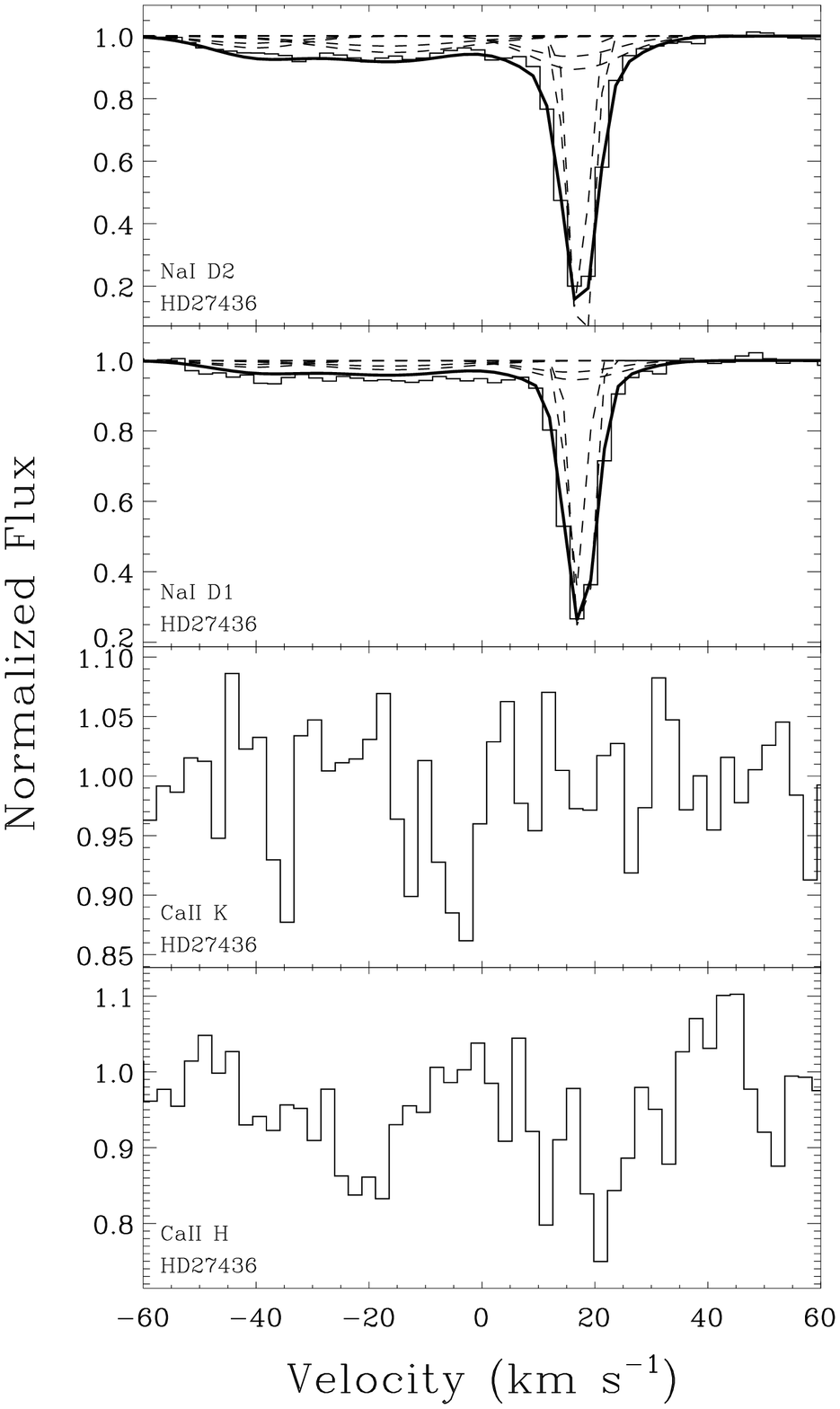}
\caption{\label{305}Same as Figure \ref{123}}
\end{figure}

\begin{figure}[p]
\centering
\figurenum{3m}
\plottwo{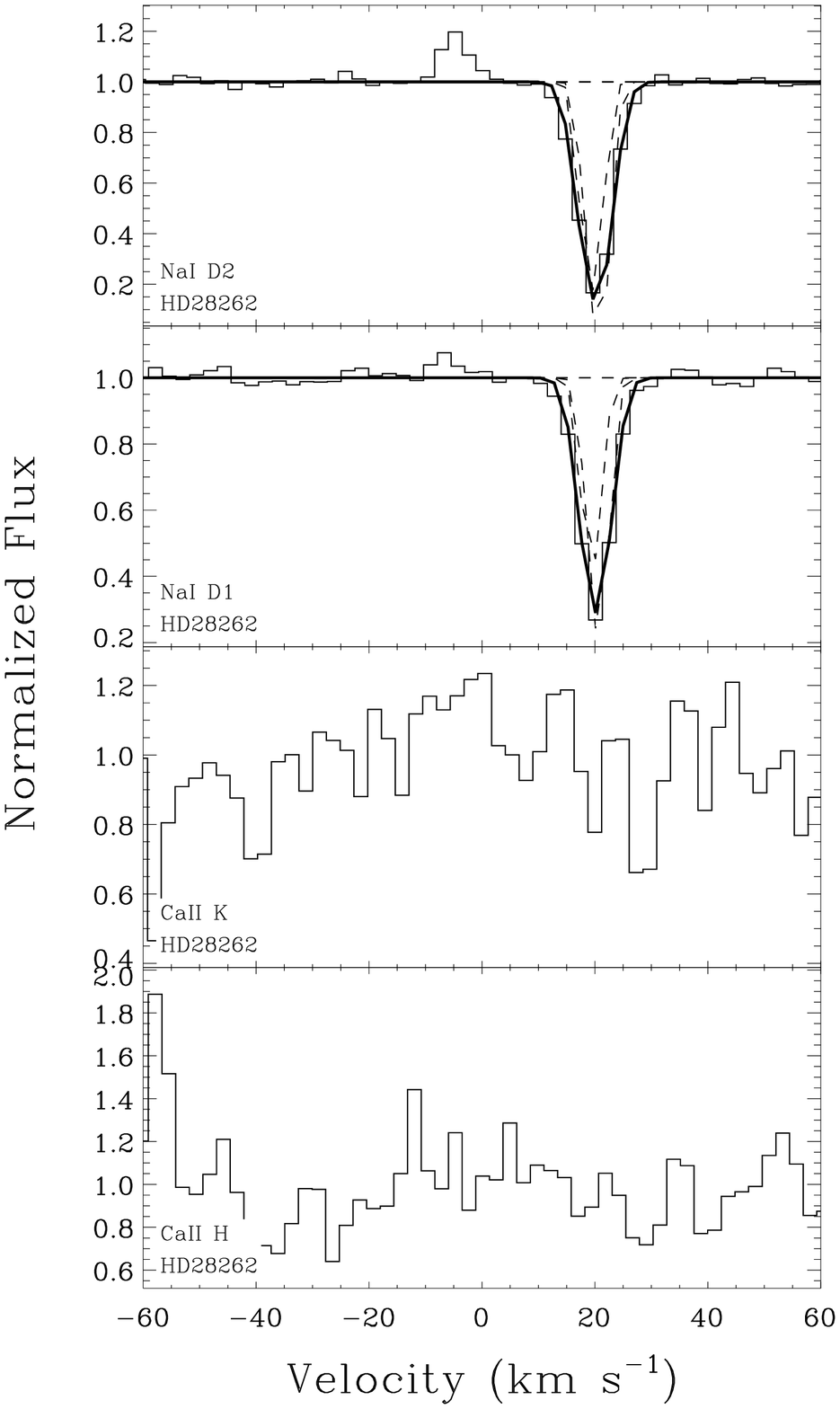}{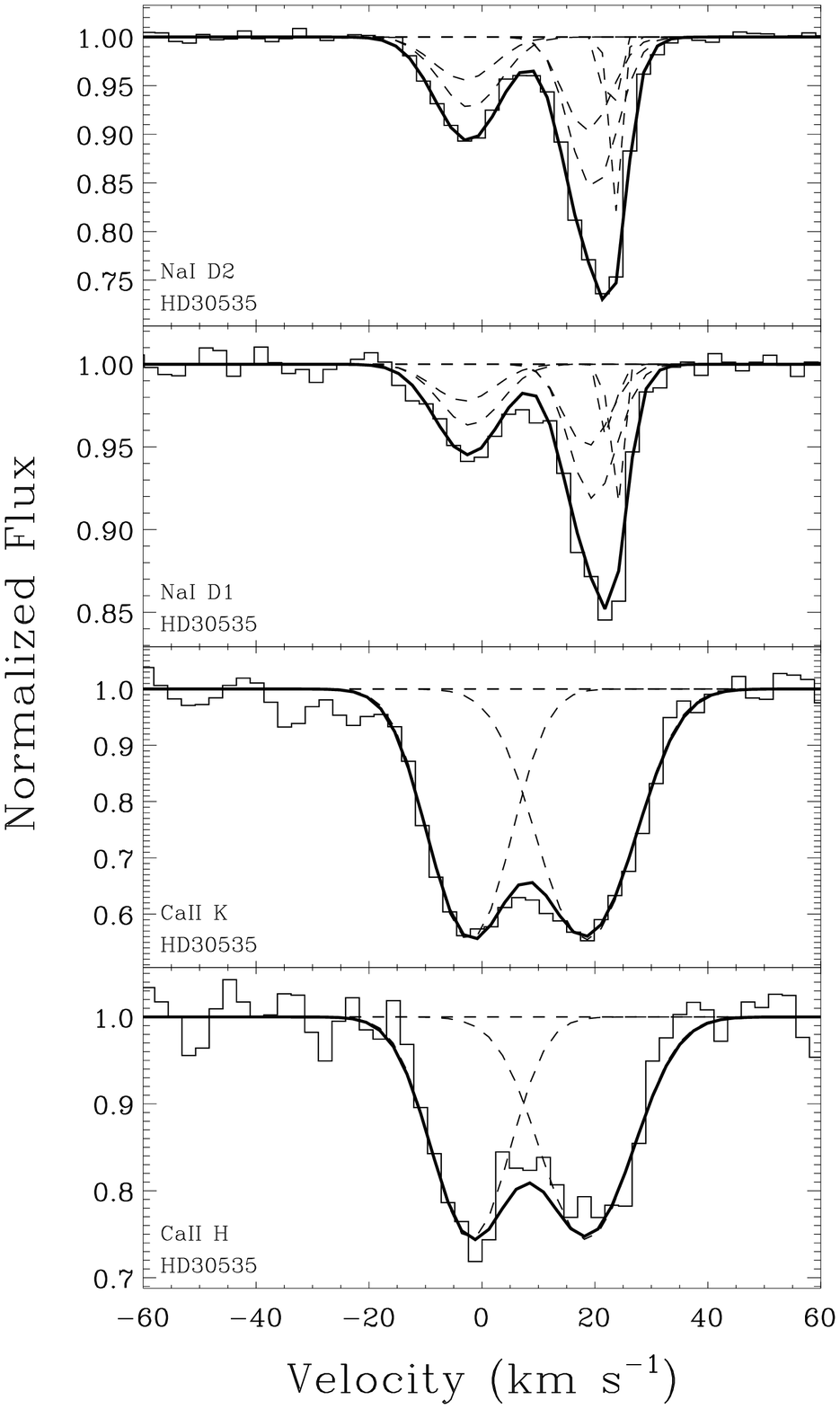}
\caption{\label{375}Same as Figure \ref{123}}
\end{figure}

\begin{figure}[p]
\centering
\figurenum{3n}
\plottwo{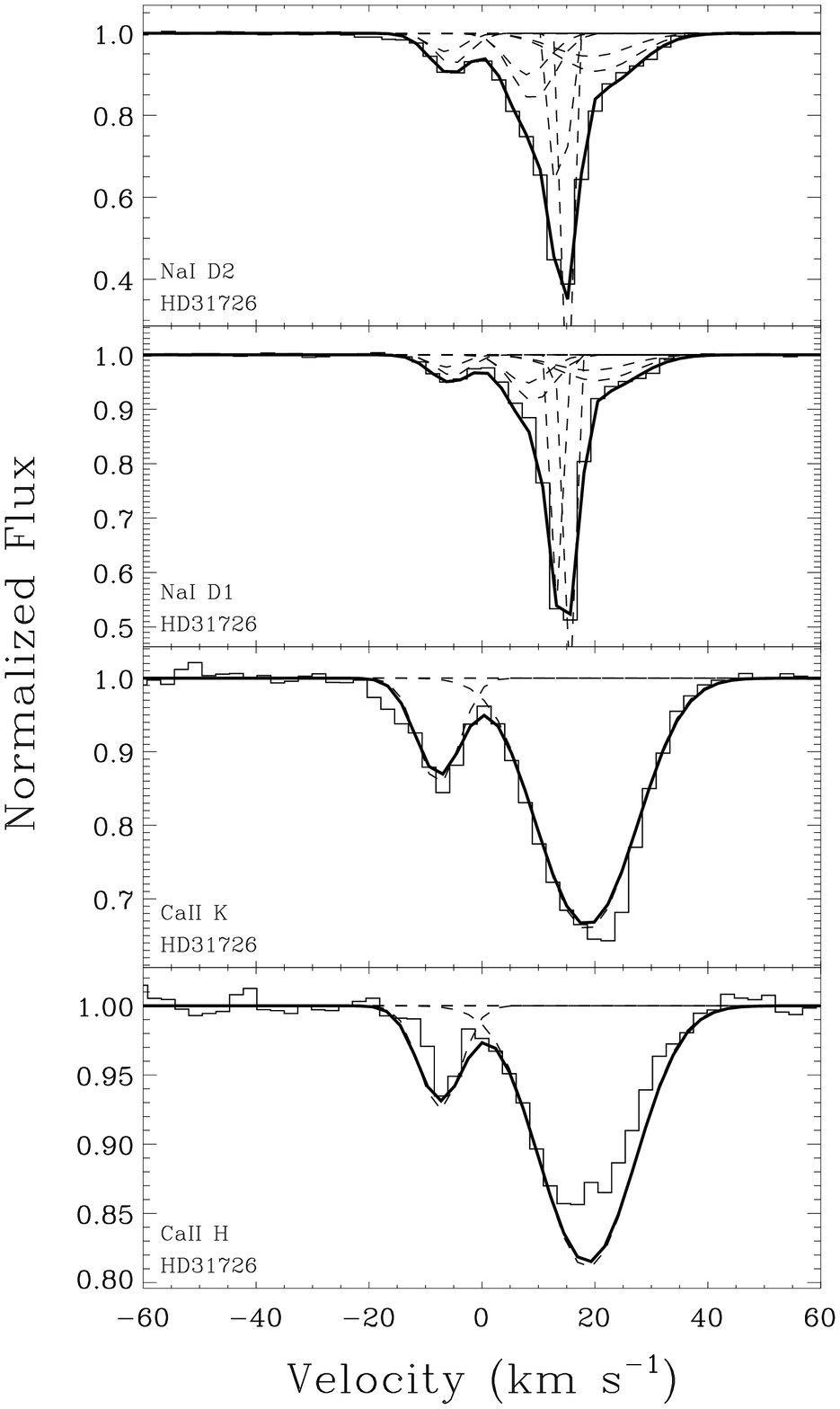}{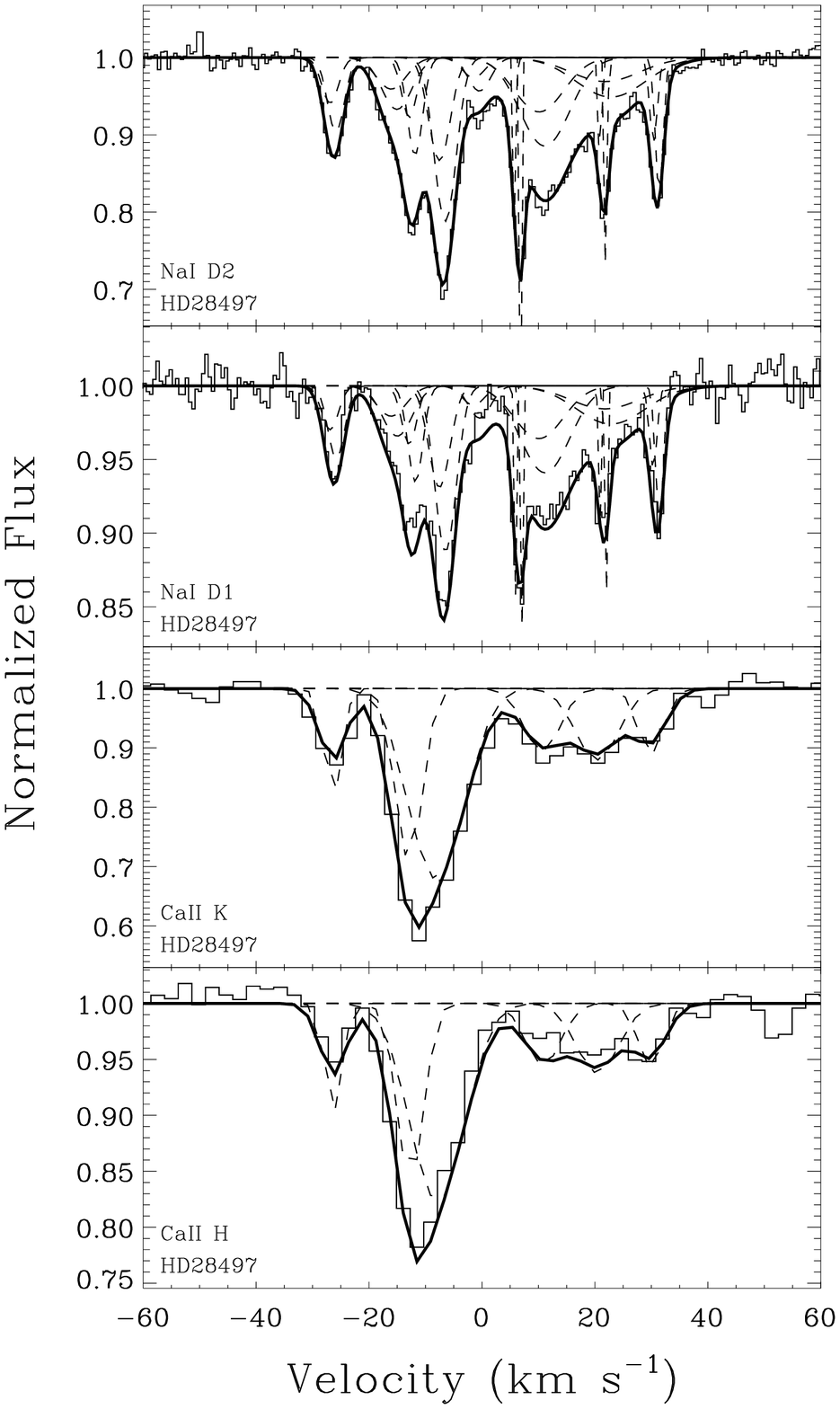}
\caption{\label{410}Same as Figure \ref{123}}
\end{figure}

\begin{figure}[p]
\centering
\figurenum{3o}
\plottwo{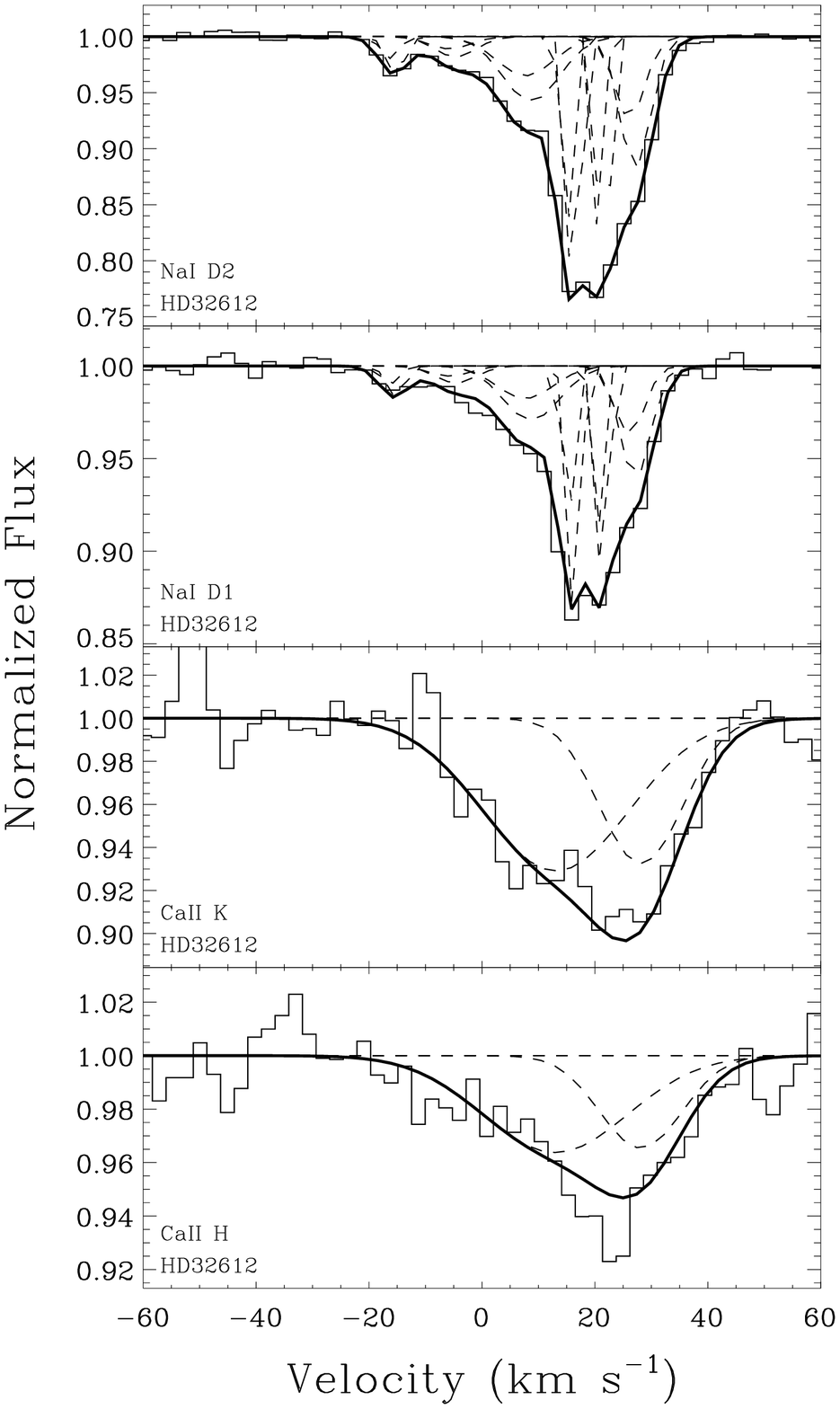}{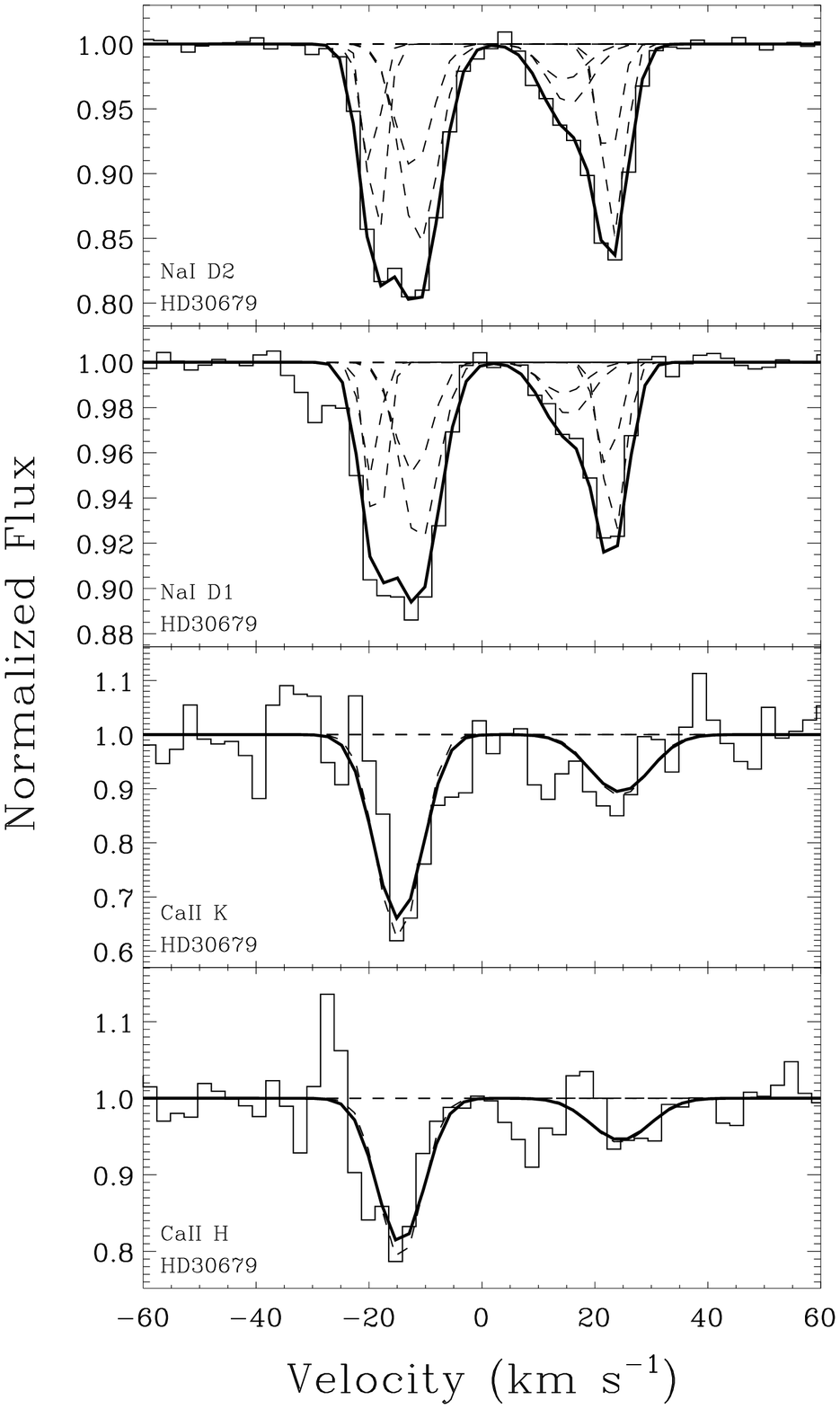}
\caption{\label{470}Same as Figure \ref{123}}
\end{figure}

Our data set contains nine sight lines along which the ISM absorption was measured multiple times at different resolving powers (see Table~\ref{multres}).  In almost all cases, the highest-resolution observation was chosen to represent the ISM absorption along the line of sight.  Figure \ref{twores} shows an example of ISM absorption toward the star HD28497 observed at different spectral resolution.  Spectra obtained at high resolution (TS21; $R \sim 240,000 = 1.3$~km~s$^{-1}$) resulted in a 10 component fit, validated by the F-test, while lower-resolution data (TS23; $R \sim 60,000 = 5$~km~s$^{-1}$) led to a fit consisting of only 6 components.  However, the total column densities were essentially identical (within $\sim$0.3$\sigma$), where the TS23 observations yielded a total \ion{Na}{1} column density of $\log N = 11.948 ^{+0.014}_{-0.016}$, and the TS21 observations, $\log N = 11.936 ^{+0.036}_{-0.060}$.  It is encouraging that we arrive at the same total column density, regardless of the resolution, and this provides confidence that we can safely compare the results of observations over the entire range of spectral resolution.

In our sample of multiply observed sight lines, more than 70\% show agreement in the total column density to within 3$\sigma$.  Of those that are discrepant by $>$3$\sigma$, all but one involve a column density derived from a high-resolution observation that is significantly less than that from a lower-resolution observation.  This may result from differing accuracies in the line-spread functions or in the continuum placement.  Whatever the cause, the impact of such a systematic error is likely to be less in the highest-resolution data.

In terms of the comparison of individual component fits, while the gross profile is similar, not surprisingly, the narrowest features are lost in the low-resolution spectra.  This is clearly seen in our complete sample, in which most of the narrow components are detected in our highest-resolution settings.  However, detectability of narrow ISM components is influenced by several other factors other than spectral resolution, including $S/N$ and degree of blending (i.e., proximity in radial velocity to another absorber).  Indeed, blending can be so severe, even in the highest-resolution spectra, that the weakest components are unable to be resolved in the fit.  Broad features, such as the shallow component near $\sim$5890.4 \AA\ (at 23.5 km~s$^{-1}$, and with a Doppler parameter of 8.1 km~s$^{-1}$) in the bottom panel of Figure \ref{twores} are not necessarily representative of a single ISM structure along that line of sight, but rather a blend of several small absorption features too weak to be resolved into individual components.  It is precisely these narrow, weak components -- perhaps tracing small but dense ISM clouds -- that we hope to identify in the historical solar path, and therefore, while our total column density measurements seem to support the use of a wide range of resolving powers, it is clearly advantageous to use the highest-resolving power possible.

\subsection{Comparison With Previous Observations}

While this work represents the first densely sampled ISM survey in the direction of the historical solar trajectory, some of our targets had been observed before.  Of our sodium observations, 16 of 43 (37\%) had been previously observed, and in calcium, 15 of 43 (35\%).  Most of these were taken as parts of much larger surveys \citep[e.g.,][]{frisch90b,welsh05b}, and the relevant references are listed in Tables~\ref{CAtotecomps} and \ref{NAtotecomps}.  While these observations were taken using various instruments and analyzed by different researchers (including some of the earliest high-resolution ISM observations: \citealt{adams49,burbidge53,munch61}), they nonetheless provide an interesting comparison.  

Overall, in terms of the total column density, there is significant disagreement beyond the 3$\sigma$ level in the majority of cases for \ion{Ca}{2}, while most of the \ion{Na}{1} comparisons agree quite well.  For \ion{Ca}{2}, the most distant targets (i.e., the sight lines with the strongest ISM absorption) tend to agree more closely than the nearest sight lines.  A likely explanation is that the systematic errors (e.g., continuum placement, number of assumed components) overwhelm the statistical errors, particularly for sight lines with weak absorption, which is typical for nearby sight lines and \ion{Ca}{2} observations.  A systematic trend is evident when we compare individual analyses.  In \ion{Ca}{2}, our measurements are systematically higher (by $\sim$0.4 dex for 4 sight lines) than those same sight lines analyzed by \citet{frisch90b}, while our measurements are systematically lower (by $\sim$0.2 dex for 6 sight lines) than those analyzed by \citet{welsh05b}.  

A comparison of individual component fits shows excellent agreement in radial velocity and Doppler parameter for the strongest components.  Determining the total number of absorbers appears to be the largest source of disagreement.  Strong absorbers may or may not be subdivided depending on the spectral resolution and $S/N$ of the data, and the analysis strategy of the researcher.  Given the unavoidable blending along distant sight lines, an accurate inventory of individual absorbers may not be possible, regardless of the quality of the data.  Herein lies the inherent limitation of the component fitting technique, and one which we tried to address by using the apparent column density method detailed in Section~\ref{aodcol} and using total column densities in Section~\ref{totcol}.  However, if one looks at the profile shapes and integrates individual components into similar radial velocities bins, the agreement can be quite good.  Two sight lines (HD34085, \citealt{price01}; HD28497, \citealt{blades97}), both of which were previously observed at high spectral resolution with the UHRF, make for good direct comparisons with our observations.  For HD34085 in \ion{Ca}{2}, we get an identical fit, differing only in how we interpreted the two most closely spaced blended components in the \citet{price01} fit, less than 2 km~s$^{-1}$ apart, as a single absorber in our fit.  Similarly for \ion{Na}{1}, while the full profile is in good agreement, we typically merge multiple neighboring components (within a few km~s$^{-1}$) identified in \citet{price01} into single components, resulting in our fit with 5 components to their fit with 12 components.  However, the column-density-weighted radial velocities between our fits match extremely well.  The same goes for the comparison with HD28497, where we have slightly lower numbers of components for both \ion{Ca}{2} (we have 6 whereas \citealt{blades97} have 10) and \ion{Na}{1} (10 versus 13), but again, the column-density-weighted radial velocities match very well.  

A detailed look at previous observations of the same sight lines highlight two important issues that are critical to this work.  First, given the impact of systematic errors, it is important to have as close to a homogeneous data sample as possible and to have that data be analyzed uniformly by a single researcher.  Second, given unavoidable blending, the results of the traditional technique of fitting a discrete number of components to the absorption profile, need to be augmented by other techniques that can attempt to evaluate the intervening absorption independent of the number of assumed components (see Sections~\ref{aodcol} and \ref{totcol}).

\section{Survey Overview}
\label{survey.overview.s}

The final tally of our component fitting analysis yielded 141 individual ISM absorption components in \ion{Ca}{2} and \ion{Na}{1} toward 43 separate sight lines.  The true number of ISM absorbers is probably greater than this based on our inability to resolve the broad shallow components into individual absorbers (23\% of all absorbers have $b > 6$ km~s$^{-1}$, mostly in \ion{Ca}{2}).  Neutral sodium accounted for $\sim$60\% of all detected absorbers (85 components), and calcium the remaining $\sim$40\% (56 components).  Calcium components tend to be broader (presumably tracing warmer gas) and the $S/N$ lower than for \ion{Na}{1}, which could explain the lower number of components and higher percentage of broad components.  The number of calcium components observed in any sight line with detected absorption ranges from 1--8, with an average of 2.2.  For \ion{Na}{1}, the number of absorbers ranges from 1--10, with an average of 2.8.  In order to compare individual components in both sodium and calcium, we identified paired components, in which the difference in measured interstellar radial velocity is $<$3~km~s$^{-1}$.  Those components that satisfy this requirement (36 in total, 64\% of the \ion{Ca}{2} sample and 42\% of the \ion{Na}{1} sample) are identified in the last column of Tables~\ref{CAallcomps} and \ref{NAallcomps}. 

Of the 43 sight lines chosen for this study, the nearest 11 sight lines ($d \leq 102.2$ pc) contain no detectable ISM absorption.  The observations of these nearest stars are consistent with the low-density interior of the Local Bubble, such that detectable quantities of both ions, particularly \ion{Na}{1}, are rare within approximately 100\,pc \citep{lallement03}.  Stronger transitions of more abundant ions are needed to trace the interstellar clouds within the Local Bubble.  This is possible only with access to high-resolution spectroscopy in the ultraviolet (UV), where the vast majority of such transitions are found \citep{redfield06}.

\subsection{Central Velocity}

Figure \ref{hist} shows the distribution of central velocities of all
absorbers for the entire set of sodium and calcium observations.  Both
the sodium and calcium component velocities range from approximately
--40 km s$^{-1}$ to +40 km s$^{-1}$.  The high number of components
found around 23~km~s$^{-1}$ suggests the presence of a nearby
interstellar cloud.  Such a large nearby cloud would be seen in
absorption toward all stars at increasing distances.  Indeed, we associated this absorption with an absorber at or near the Local Bubble boundary, see Section 3.3.1 and 3.3.2).  There is a
consistent pattern between \ion{Na}{1} and \ion{Ca}{2}, indicating that
to a large degree they are both tracing similar collections of gas.

Figure~\ref{dv} shows the difference in central velocity for our paired component subsample.  While these were defined as paired components based on whether the difference in their velocities was $<$3 km~s$^{-1}$, the distribution of the velocity differences is typically much less and clearly peaks at zero.  However, only 81\% of the components are actually consistent with the zero velocity difference at 3$\sigma$.  While it seems clear that there is often a strong correlation between the sodium and calcium absorption, it is not obvious that they are identically distributed.

\subsection{Doppler Parameter}
\label{dp}

Figure \ref{hist} shows the distribution of measured line width, or Doppler parameter, for each component.  The high Doppler parameter tail in both the sodium and calcium distributions is a reflection of our inability to resolve the weakest components in the interstellar absorption profile.  As a result, multiple weak absorbers that are likely present are blended into broad shallow components.  Although the total column density remains the same, it makes it difficult to assess the true number of absorbers and the true line widths.  For the entire sample, the median value of the Doppler parameter for \ion{Na}{1} is 2.3 km~s$^{-1}$ whereas for \ion{Ca}{2} it is 4.7 km~s$^{-1}$.  However, it is clear from Figure~\ref{hist} that a distinct population of small Doppler parameter measurements differentiate themselves from the large Doppler parameter tail.  For \ion{Na}{1}, 61\% of the components have $b_{\rm Na} \le 2.5$ km~s$^{-1}$, with a mean of 1.5 km~s$^{-1}$.  For \ion{Ca}{2}, 48\% of the components have $b_{\rm Ca} \le 4.8$ km~s$^{-1}$, with a mean of 2.5 km~s$^{-1}$.  These values agree fairly well qualitatively with other all-sky surveys.  \citet{welty94,welty96} find lower median values for \ion{Na}{1} (0.73 km~s$^{-1}$) and \ion{Ca}{2} (1.31 km~s$^{-1}$), but many of their components have had their Doppler parameters fixed in order to converge on a consistent fit of severely blended lines, which may be biasing their sample to lower values.

\subsubsection{Temperature and Turbulent Velocity}
\label{temperature.sss}

Comparison of line widths of ions with different atomic weight that show absorption of the same gas can be used to estimate the
temperature and turbulent velocity of that gas.  Thermal motions are
inversely proportional to the atomic weight while turbulent motions
are independent of atomic weight.  This relationship is described in the following equation that relates the Doppler width parameter ($b$ [in km~s$^{-1}$]) to the temperature ($T$ [in K]) and nonthermal, turbulent velocity ($\xi$ [in km~s$^{-1}$]),
\begin{equation}
b^2 = \frac{2kT}{m} + \xi^2 = 0.016629\frac{T}{A} + \xi^2,
\label{redLinEq1}
\end{equation}
where $k$ is Boltzmann's constant, $m$ is the mass of the ion observed, and $A$ is the atomic weight of the element in atomic mass units.  \citet{redfield04} did an extensive
survey of temperature and turbulent velocity measurements of 50
absorbers within the LISM using line widths of as many as 8 different ions.  Their results showed that all observed line widths could be satisfactorily fit with a self-consistent temperature and turbulent velocity.  The atomic weight of calcium ($A =
40.08$) is significantly higher than that of sodium ($A=22.99$), 
therefore presenting an opportunity in the current survey to disentangle the thermal and
turbulent broadening contributions.  

As discussed above, the mean values of the Doppler parameter for
\ion{Ca}{2} are significantly higher than those for \ion{Na}{1}.  Taken
at face value, this is inconsistent with both ions present in a single
cloud, characterized by a single temperature and turbulent velocity.
This is true for our paired component sample as well.  In
Figure~\ref{bs}, the Doppler parameter is plotted for both ions.  The
solid line indicates the relationship of Doppler parameters if purely
determined by turbulent motion (i.e., $b_{\rm Na} = b_{\rm Ca}$), whereas the dashed line indicates the
scenario in which only thermal motions are included (i.e., $b_{\rm Na} = \sqrt{A_{\rm Ca}/A_{\rm Na}}\, b_{\rm Ca} = 1.320\,b_{\rm Ca}$).  In reality, both
thermal and turbulent motions likely contribute to the line width, and therefore a collection of gas that includes
both ions should fall between these two lines.  For our sample of 36
paired components, only 4 (11\%) lie within this self-consistent
region, which has been noted in other surveys of \ion{Na}{1} and
\ion{Ca}{2} \citep[e.g.,][]{welty96}.  This argues that these two ions are not
well mixed, likely due to differences in the ionization and depletion
structure of these two elements.  However, based on the correlation in
observed radial velocity of the components, it is clear that
\ion{Na}{1} and \ion{Ca}{2} are nonetheless correlated and may simply
sample different regions of a typical interstellar cloud structure.  

In Figure~\ref{hist}, we convert the observed Doppler line width into the maximum temperature of the gas, assuming that there is no contribution from nonthermal, or turbulent, broadening.  If we ignore the high width tails to both distributions and look at the mean values given above in Section~\ref{dp}, the mean maximum temperature for \ion{Na}{1} is $T_{\rm max} = 3100$ K, and for \ion{Ca}{2}, $T_{\rm max} = 15,000$ K.  Clearly, \ion{Na}{1} is sampling a cooler, and likely less turbulent, interstellar material compared to \ion{Ca}{2}.  These maximum temperatures are likely significant overestimates of the true temperature, as there is always some contribution to the line broadening via turbulent motions, but serve as a stringent upper limit.  The assumption that the sound speed can be used as an upper limit to the turbulent velocity, and therefore provide a lower limit to the temperature of the gas has been applied to evaluations of \ion{Ca}{2} line widths \citep[e.g.,][]{welty96}.  This is reasonable for warm partially ionized clouds which typically have subsonic turbulence \citep[e.g.,][]{redfield04sw}, but may not be valid for cold neutral clouds probed by \ion{Na}{1} absorption which typically have supersonic turbulence \citep[e.g.,][]{heiles03,meyer12}.

Instead, here we make some reasonable assumptions about the contribution of the turbulent broadening in order to make a better estimate of the temperature of the gas.  The LISM is a convenient standard with which to evaluate \ion{Ca}{2}.  Not only are the lines widths of many ions detected in the LISM self-consistent along individual sight lines \citep{redfield04}, but the entire collection of LISM clouds also are very similar, and therefore, mean values of line widths of the entire sample are self-consistent, with an average temperature, $T \sim 6900$ K, and turbulent velocity, $\xi \sim 1.67$ km~s$^{-1}$ \citep{redfield04sw}.  Note that this is subsonic turbulence, typical of warm partially ionized clouds.  If we use this same turbulent velocity, for \ion{Ca}{2}, this leads to a mean temperature in our sample of $\sim$8300 K.  This turbulent velocity is inconsistent with the average Doppler width of \ion{Na}{1} observed in our sample, which again argues that \ion{Na}{1} is sampling a distinctly different interstellar environment.  If instead, we use the turbulent velocity observed for the Local Leo Cold Cloud \citep[$\xi \sim 0.23$ km~s$^{-1}$;][]{meyer12}, which is quite small but supersonic as found for other cold neutral clouds \citep[e.g.,][]{heiles03}, it leads to a temperature just slightly less than our maximum temperature, or $\sim$3000 K.  While perhaps correlated, the \ion{Na}{1} and \ion{Ca}{2} appear to be sampling spatially distinct locations.

%A good assumption for a lower limit on the temperature would be the
%sound speed, at least for warm clouds like
%those we identified with \ion{Ca}{2}. \citep{redfield04sw} find that the
%turbulent Mach number for clouds in the LISM is 0.19, so it's
%reasonable to expect some sub-sonic turbulent broadening. However,
%\citep{heiles97} found that cold clouds tend to be supersonic with
%turbulent Mach numbers of around 3. So assuming a a turbulent velocity
%equal to the sound speed might not be a valid lower limit for our cool
%\ion{Na}{1} clouds. The maximum temperature range plotted along the
%top of Figure~\ref{hist}, though greatly overestimating the true temperature
%of the gas, serves as a hard upper limit.

\subsection{Column Density}

Figure \ref{hist} shows the distribution of observed individual
component column densities determined from our fitting procedure.
A distinction here is made between observations with different
instruments, which vary in spectral resolution and sensitivity.  These
both impact our detection limit.  The vertical lines
indicate those limits for each instrument, which are 3$\sigma$ upper
limits using the mean $S/N$ for observations taken with that
instrument.  In general, since ISM lines are quite
narrow, one gains significantly in sensitivity as one goes to higher
spectral resolution.  Our calculations support this trend, other than
for observations of \ion{Na}{1} with TS12, which is significantly less
sensitive than the other instruments used at this wavelength, and hence
the detection limit is higher than would be expected.  The number of
weak absorbers is almost certainly underestimated due to the
observational bias introduced by our sensitivity limits.  As can be
seen in the \ion{Ca}{2} distribution, the vast majority of low column
density absorbers are detected by our most sensitive (and
highest-resolving power) instrument, the UHRF.

The median observed \ion{Ca}{2} column density is $\log
N($\ion{Ca}{2}$) \sim 11.3$, and the median observed \ion{Na}{1}
column density is $\log N($\ion{Na}{1}$) \sim 11.2$.  In
Figure~\ref{hist}, the \ion{Na}{1} column density has also been
translated into a total hydrogen column density.  Even though much of
the sodium is ionized, \citet{ferlet85} demonstrated a correlation exists between \ion{Na}{1} and the total hydrogen column
density $N({\rm HI} + {\rm H}_2)$, over \ion{Na}{1} column densities
consistent with our sample (i.e., $10.0 \leq \log N($\ion{Na}{1}$)
\leq 13.0$).  The mean
\ion{Na}{1} column density thus translates into a mean hydrogen column
density of $\log N({\rm HI} + {\rm H}_2) \sim 19.5$.  

For the purposes of reconstructing the ISM density along the
historical solar trajectory, measurements of the column density of each individual absorbing cloud is of the utmost importance.  In Paper II, we present an analysis that synthesizes our fits of individual absorbers into an assignment of specific interstellar clouds.  However, we can evaluate the column density observed in this direction in even simpler terms which also avoid the bias of component identification.  As discussed in
Section~\ref{multiple.resolution.observations.ss}, we find that the
total column density estimate is unaffected by spectral resolution of
the spectrograph, but the identification of individual absorbers is a
difficult task, often biased by choices made by the researcher.  This
is particularly true given the high level of complexity we observe
even within a narrow observing cone.  We use two techniques that
should be relatively free of the choices we made in making the
individual fits in order to characterize the variation of column
density along the historical solar trajectory.  The first is to use
the apparent column density, and the other is to simply use the total
column density.

\subsubsection{Apparent Column Density}
\label{aodcol}

In order to characterize the column density structure along this line
of sight independent of how individual components are fit to the
absorption profile, we employed the apparent column density method
described by \citet{savage91}.  In this case, the interstellar
absorption is analyzed by converting absorption line profiles into
profiles of apparent optical depth, $\tau_a(v),$ and apparent column
density, $N_a(v)$, per unit velocity.  This very quickly provides a
diagnostic of the column density and velocity structure regardless
of the degree of blending.  Apparent optical depth as a function of
wavelength is obtained from Equation \ref{tau}, where $I_{0}(\lambda)$
is our normalized flux continuum (1 in our case), and
$I_{obs}(\lambda)$ is the absorption line profile for a particular
observation,

\begin{equation}
\tau_{a}(\lambda) = ln\left[\frac{I_{0}(\lambda)}{I_{obs}(\lambda)}\right].
\label{tau}
\end{equation}

Apparent optical depth is different from true optical depth in that
the apparent optical depth has been blurred by the resolution of the
instrument used to take the data.  In our case, despite using a
variety of instruments, they all were sufficiently high resolution
that the instrumental impact on the apparent column density
measurement is minimal.  The apparent optical depth is a good
approximation of the true optical depth when the instrument resolution
is high, the continuum is well defined, and the measurements have a
high signal-to-noise ratio, which is, in general, satisfied by all of
our observations.  When these conditions are met, $\tau_{a}(\lambda)
\approx \tau (\lambda)$, and we can make use of the relation between
true optical depth and column density:

\begin{equation}
\tau(\lambda) = \frac{\pi e^2}{m_e c^2} f\lambda^2 N(\lambda).
\label{taul}
\end{equation}

Total column density can be found from $N = \int N(\lambda)d\lambda$
and expressing total column density as a function of velocity, we can
solve Equation \ref{taul} for $N(v)$ and integrate, giving:
\begin{equation}
N = \frac{m_{e}c}{\pi e^{2} f \lambda} \int ln \frac{I_{0}(v)}{I(v)}dv,
\end{equation}
where $m_{e}$ is the mass of an electron, $f$ is the transition
oscillator strength, $c$ is the speed of light, and $\lambda$ is
wavelength in \r{A}ngstroms.  Apparent column density per unit velocity
as a function of distance is plotted in Figure \ref{aods} for the
sodium and calcium data.  As was discussed above, the sodium column
density has been translated into hydrogen column densities following
the relationship derived by \citet{ferlet85}.

The onset of detected absorption is at a similar distance for both
ions and consistent with the edge of the Local Bubble at $\sim$120 pc.
Shortly thereafter, there is a strong absorption feature at $\sim$23~km
s$^{-1}$ that is consistently seen in absorption profiles at
increasing distances.  This suggests that there is a cloud at or near the
edge of the Local Bubble taking up a significant portion of the sample
area.  Given the peculiar motion of the Sun (13.38 km~s$^{-1}$), it will 
travel this equivalent distance in $\sim$10 Myr.
Unfortunately, \ion{Ca}{2} and \ion{Na}{1} are largely insensitive to
the warm, partially ionized clouds that are known to reside within the
Local Bubble, which require UV spectroscopic observations \citep{frisch11}.

The sodium and calcium apparent column density plots look very
similar, again arguing for a correlation in the spatial distribution of
\ion{Na}{1} and \ion{Ca}{2}.  The broader profiles
of \ion{Ca}{2} are also evident in Figure~\ref{aods}.

\subsubsection{Total Column Density}
\label{totcol}

An additional technique to evaluate the column density along this unique direction is to work with the total observed column density.  Again, this is independent of the strategy of component fitting and appears robust to a wide range of resolving powers.  The total column densities for sight lines in our sample are listed in Tables \ref{CaTotesN} and \ref{NaTotesN}.  Figure~\ref{coldist} displays these values as a function of distance.  Again, insensitive to the low column densities of the LISM clouds within the Local Bubble, our nearest sight lines provide only upper limits.  The edge of the Local Bubble is clearly seen at $\sim$120 pc, with a rapid increase in total column density, which quickly levels off.  Our measurements of the distance to the edge of the Local Bubble are in agreement with \citet{lallement03} who did an all-sky survey of interstellar neutral sodium out to a distance of 500 parsecs.  This sharp rise is most apparent in the sodium column density profile but is also seen in the calcium data.  The error in the distance measurements is from \textit{Hipparcos} parallax measurements \citep{vanleeuwen07}, and for this reason, the error bars tend to grow with distance as parallax becomes more difficult to measure.

The spread in column densities at any given distance is likely an indication that we are observing small-scale structure variations in the ISM across this relatively narrow observing cone.  At the largest distances ($\sim$500 pc), targets at the edge of our observing cone would be separated from the center by a physical distance of 80\,pc.  These spatial variations could explain drops in the total column density for sight lines at larger distances in Figure~\ref{coldist} as well as the apparent disappearance and reappearance of absorbing components in the apparent column density profiles presented in Figure~\ref{aods}.  

In order to evaluate the spatial variations in our sample, the total column density measurements were analyzed to see if a self-consistent series of absorbers within individual distance bins could replicate our observations.  We used distance bins of $<$150 pc, 150--200 pc, 200--250 pc, 250--350 pc, and 350--700 pc.  A minimum curvature surface interpolation is made using the measurements toward targets that fall within these distance bins.  More distant targets provide limits to the column density of preceding distance bins.  Once a spatial map of column density is determined for the closest distance bin, it is subtracted from more distant sight lines and the procedure is repeated for the next distance bin.  Despite this being the most densely sampled high-resolution absorption line survey of an individual sight line, we are still left with only 43 pencil-beam measurements.  Given the large distances of some of our sight lines, the physical spacings can be quite dramatic.  For example, in the farthest distance bin, with a mean distance of 525\,pc, the radius of our field of view ($10^\circ$) is equivalent to 92.6\,pc, whereas for our closest bin with a mean distance of 125\,pc such an angular separation is only 22.0\,pc.  Nonetheless, a self-consistent model of the spatial variation of column density in both \ion{Ca}{2} and \ion{Na}{1} can be made.  The results are shown in Figure~\ref{dslice}.  Overall, both ions show a similar, although not identical, spatial distribution.  Again, this argues for correlated, though not necessarily identical, distributions of \ion{Ca}{2} and \ion{Na}{1}.  Additionally, significant structure is observed, highlighting the complexity of the ISM and need for densely sampled surveys in order to characterize this complexity.

\subsubsection{\ion{Na}{1}/\ion{Ca}{2} Ratio}
\label{naicaii.ss}

Neutral sodium and singly ionized calcium are two of the strongest resonance lines in the optical \citep{redfield06} and have therefore been extensively used to diagnose the physical properties of the ISM over a range of conditions.  Measuring the column density ratio of these two ions, which is relatively simple observationally, holds the promise of evaluating the impact of ionization and depletion, both of which will impact the ratio.  In the diffuse, warm ISM, both \ion{Na}{1} and \ion{Ca}{2} are trace species, with \ion{Na}{2} and \ion{Ca}{3} dominating \citep{slavin02}.  Additionally, the two elements have very different condensation temperatures and therefore, have very different patterns of depletion onto dust, namely calcium is typically depleted in the gas phase by more than two orders of magnitude compared to sodium \citep{savage96}.  For these reasons, the \ion{Na}{1}/\ion{Ca}{2} column density ratio can vary by several orders of magnitude and potentially be a sensitive diagnostic of these phenomena.  In our sample of paired components, we see column density ratios that range more than two orders of magnitude from $N($\ion{Na}{1}$)/N($\ion{Ca}{2}$) = $0.066--12 (see Figure~\ref{nrat}), while even larger variations have been detected along other sight lines \citep{welty96}.  

Based on many studies of \ion{Na}{1} and \ion{Ca}{2} \citep[e.g.,][]{routly52,siluk74,crawford92,bertin93,welty96}, the largest column density ratio ranges tend to be associated with cold ($T \sim 100$ K), dense interstellar clouds, where \ion{Na}{1} is prevalent, \ion{Ca}{2} is rare, and calcium in general is significantly depleted onto dust.  The lowest column density ratios are typically associated with warm ($T \sim 1000$ K), high-velocity interstellar environments \citep[e.g., supernova remnants and shells;][]{siluk74}, where dust has either been destroyed and the constituents returned to the gas phase, or where depletion has not yet initiated.  We certainly see evidence to support this in our sample, in which the largest \ion{Na}{1}/\ion{Ca}{2} values are located just beyond the Local Bubble boundary (see Figures~\ref{nrat} and \ref{aodsrat}), which delineates the warm, rarefied Local Bubble gas from dense molecular gas.  Our dense survey of nearby stars, which have accurate distance measurements, allows us to pinpoint the location of the high column density ratio gas.  

Additional evidence of the physical conditions of gas with different \ion{Na}{1}/\ion{Ca}{2} ratios comes in the form of the Routly-Spitzer Effect \citep{routly52}, which is exemplified in a canonical figure of observed ISM velocity versus column density ratio.  We show this figure for our own sample at the top of Figure~\ref{nrat}.  The highest \ion{Na}{1}/\ion{Ca}{2} column density ratios tend to be at low velocity, whereas the highest velocities tend have low column density ratios.  \citet{siluk74} presented the interpretation that these high peculiar velocities could be signatures of old supernova remnants and that shocks associated with these remnants would preferentially destroy dust grains and enhance the gas phase calcium abundance, thereby lowering the \ion{Na}{1}/\ion{Ca}{2} column density ratio.  Our sample also shows the canonical Routly-Spitzer Effect, and the majority of the low column density ratio measurements are associated with our longest sight lines, which likely traverse a wide range of interstellar environments.  

Finally, in Figure~\ref{nrat}, we also see a tight correlation between the \ion{Na}{1}/\ion{Ca}{2} column density ratio and \ion{Na}{1} column density, which as mentioned above is correlated with hydrogen column density \citep{ferlet85}.  While observational bias could influence this plot, as shown in Figure~\ref{hist}, our detection limits are low enough as to not significantly impact the observed correlation.  Instead, it is likely that the variation in calcium depletion is the cause of the correlation.  The largest sodium column densities are likely indicative of the densest interstellar environments, in which calcium can be depleted by more than 3 orders of magnitude, whereas sodium maintains a modest depletion over a wide range of densities \citep{phillips84}.

While these connections between the \ion{Na}{1}/\ion{Ca}{2} column density ratio and other physical quantities imply a physical association, we know from the comparison of the Doppler parameter shown in Figure~\ref{bs}, that \ion{Na}{1} and \ion{Ca}{2} are not identically distributed.  Because they do not share an identical temperature or turbulent velocity, they cannot be components of the same parcel of gas.  \citet{routly52} made note of the inconsistent differences in the Doppler parameter between sodium and calcium but, given the moderate spectral resolution of the observations, the larger \ion{Ca}{2} line widths could be the result of component blending, which due to the Routly-Spitzer Effect will be more significant for \ion{Ca}{2} than \ion{Na}{1}.  In essence, the low \ion{Na}{1}/\ion{Ca}{2} column density ratio, for which, by necessity, there would be a strong \ion{Ca}{2} signal, is also found at a wide range of velocities, and so the blending of these lines would lead to a large Doppler width.  However, as the spectral resolution of observations has improved, the low column density ratio sight lines have not resolved into a series of blended interstellar components.  More than half of our paired sample were taken from the highest-resolution spectrographs used in this survey, and yet the line width discrepancy, $b_{\rm Ca} > b_{\rm Na}$ still holds.  This makes using the ionization and depletion to derive fundamental physical properties (e.g., electron density, temperature) problematic.  \citet{slavin08}, using a series of photoionization models of the local interstellar medium, also emphasize the difficulty in using this ratio of trace species as a diagnostic of physical properties in warm partially ionized clouds.  In addition, a third possibility needs to be explored to explain the observed \ion{Na}{1}/\ion{Ca}{2} column density ratio, and that is geometry.  Because we do not know how \ion{Na}{1} and \ion{Ca}{2} are distributed in space, the orientation of the absorbing clouds, particularly if they are filamentary, can have a significant impact on the \ion{Na}{1}/\ion{Ca}{2} column density ratio.  Filamentary structures are quite common in the ISM \citep{frisch83,heiles97,redfield08,peek11}, presumably organized by the presence of magnetic fields \citep{jackson03}.  So, for example, at the Local Bubble boundary, where there is likely to be interactions between the hot Local Bubble gas and the cold, dense molecular material, the orientation of structures at this interface may play a significant role in the large range of \ion{Na}{1}/\ion{Ca}{2} column density ratios.

\subsection{Small-Scale Structure}

In this section, we examine the subset of sight lines that are close in angle in order to search for evidence of small-scale structure in the ISM along our observing cone.  There are 5 sight line pairs that are within 1$^{\circ}$ of each other and show ISM absorption: HD30963--HD31089, 0.31$^\circ$; HD30535--HD30679, 0.57$^\circ$; HD27563--HD27436, 0.60$^\circ$; HD28763--HD28497, 0.82$^\circ$; HD28843--HD29248, 0.93$^\circ$.  Of particular note are the first two pairs listed (HD30963--HD31089 and HD30535--HD30679) because they are within 2$^\circ$ of the solar trajectory direction.  A detailed comparison between the sight lines is not discussed here, as interested readers can easily make their own comparisons using the spectra in Figures~\ref{123}--\ref{470}, and Tables~\ref{CAallcomps} and \ref{NAallcomps}, but we do highlight a common occurrence in many of the pairs that argues for the existence of small-scale structure.  Excess absorption is often detected in the closer of the two targets, requiring that the absorption discrepancy between the sight lines occur interior to the nearest star.  For example, HD30963 shows significant \ion{Na}{1} absorption at velocities $<$10 km~s$^{-1}$, whereas HD31089 shows none, which is in contrast to the \ion{Na}{1} absorption near 20 km~s$^{-1}$, which increases by about a factor of 3 between HD30963 and HD31089.  Likewise, HD30535, which is significantly closer than HD30679, shows \ion{Na}{1} absorption near 20 km~s$^{-1}$ that is about twice that of the absorption at the same velocity in the more distant target HD30679.  Finally, the last listed pair (HD28843--HD29248) is also of interest because it is the closest of the pairs, with stellar distances of 145.8\,pc and 207.0\,pc, respectively, and hence the distance errors are small ($<$10\,pc).  This pair also shows excess absorption (near 0 km~s$^{-1}$) in the spectrum of the closer star for both \ion{Na}{1} and \ion{Ca}{2}.  For the three cases discussed above, the maximum size scale of the variation (i.e., the projected length between the two stars at a distance just interior to the nearest star), is 1.3, 3.3, and 2.4\,pc, respectively, although if the variation is assumed to be at the Local Bubble boundary at 120\,pc, the scale of ISM variation is even smaller at only 0.6, 1.2, and 1.9\,pc, respectively.  

Others sight line pairs of interest include those with small physical separations.  These tend to be pairings of the closest targets at essentially identical distances.  There are 13 pairs in which the physical separation is $\le$15 pc.  More than half of these include pairings between the targets within the Local Bubble and therefore, have no detected absorption.  However, there are 5 pairings in which both have detected absorption, including HD28208 and HD28980 with a physical separation of 7.7 pc, and HD31512 and HD32249 with a separation of 9.4 pc.  Again, despite being at practically identical distances, there are significant differences in the ISM absorption profile.  The \ion{Na}{1} column densities differ by a factor of 3 between HD28208 and HD28980, and by a factor of 2 between HD31512 and HD32249.  The \ion{Ca}{2} is similar between the pairs, but still shows variations in observed velocity (e.g., HD28208--HD28980) and component structure (e.g., HD31512--HD32249).  

Variations in the ISM on a small scale (i.e., sub-parsec) have been found in many other observations, including wide binaries \citep{meyer96,watson96}, and high proper motion pulsars \citep{frail94,stanimirovic10}.  Only in the very closest ISM, primarily observed in the UV using dominant ionization stage ions, do we see little evidence for small-scale structure, but instead coherent absorption over large angles \citep{redfield01}.  While a large coherent pattern of absorption is evident in our sample (see Figure~\ref{aods}) a detailed look at practically any pair of sight lines, shows significant discrepancies, indicating a fundamental complexity in the distribution and/or physical properties of the observed ISM.  In the context of reconstructing the ISM along the historical solar trajectory, this small-scale structure makes it very difficult to identify anything other than the largest structures.  However, it motivates even more strongly the need to understand the interaction between stars and the ISM through which they are passing because the abundance of small-scale structures implies that dramatic changes in the properties of our surrounding ISM are a common occurrence.

\section{Conclusions}

We performed the densest ISM survey at high spectral resolution, observing 43 targets within 9 degrees of the historical solar trajectory out to a distance of $\sim$500~pc.  Observations were taken of \ion{Na}{1} and \ion{Ca}{2} at resolving powers ranging from $60,000 < R < 940,000$.  We fit the absorption profiles with 85 \ion{Na}{1} and 56 \ion{Ca}{2} components.

\begin{enumerate}
\item Component analysis is consistent between observations over a
  range of resolving powers, although the highest spectral resolution
  provides the highest sensitivity to narrow and low column density
  absorbers.  The total column density appears insensitive to
  resolution among observations in our study and even to analysis
  strategy among similar observations analyzed by other researchers.  

\item However, the number of components, which has a significant
  impact on measured line widths, velocities, and individual column
  densities, is variable depending on resolving power, $S/N$, and
  researcher.  This argues for a homogeneous data set analyzed
  together, as has been done for this sample.  In addition, we present
  two alternative techniques to assess the ISM profile along the
  direction of the historical solar trajectory: the apparent column
  density and the total column density.

\item The nearest significant ISM absorber has a velocity of $\sim$23 km~s$^{-1}$ and is located at $\sim$120 pc, which
  is consistent with the edge of the Local Bubble.  This interstellar
  material would have been encountered by the Sun $\sim$10 Myr ago.
  However, we know that significant interstellar material resides
  within this volume, including the LIC, which defines the current
  structure of the heliosphere.  These optical transitions are simply
  not sensitive enough to probe the very closest material, and
  therefore, to measure the ISM that the Sun encountered within the
  last few million years, we need spectroscopic observations in the
  UV.

%\item The predominance of the feature $\sim$23 km~s$^{-1}$ could be a
%detection of Local Bubble Wall motion. A similar-size data set has
%been taken of ISM absoprtion in the Sun's future trajectory, if we
%see a similar number of positive velocity ISM detections at a
%distance close to where the LB wall is expected to reside, this
%could suggest an expanding Local Bubble. 

\item \ion{Na}{1} and \ion{Ca}{2} do not seem to be identically
  distributed in interstellar clouds, as indicated by the
  inconsistency of the line widths, namely that $b_{\rm Ca} > b_{\rm
    Na}$, despite the fact that calcium has a higher atomic weight.
  Nonetheless, the two ions do appear to be correlated, given the
  similarity in velocity structure and therefore, are likely
  constituents in related but distinct structures within a single ISM
  cloud complex, with \ion{Ca}{2} tracing warm ($\sim$8000 K) gas and \ion{Na}{1}
  tracing cold ($<$3000 K) gas.

\item The \ion{Na}{1}/\ion{Ca}{2} column density ratio shows the
  largest variation at the Local Bubble boundary, possibly due to the
  complexity of physical conditions near that interface (e.g.,
  temperature, ionization, depletion) or variations in orientation of
  the dissimilar distributions of \ion{Na}{1} and \ion{Ca}{2}.  Given
  that these two ions are not identically distributed, it makes using
  the column density ratio problematic in evaluating ionization and
  depletion effects.  Indeed, the non-coincident spatial distribution
  of the \ion{Na}{1} and \ion{Ca}{2} in these clouds, and their
  possibly independent orientation along the line of sight, may also
  have a large role in determining the column density ratio that is
  observed. %The detected components are probably associated somehow,
            %though not occupying the same volume of space. 

\item Significant spatial variation is detected, although individual
  absorbers are consistently detected in practically all sight lines
  (e.g., the 23 km~s$^{-1}$ component).  A corollary to this
  conclusion is that small-scale structure is observed to be
  ubiquitous, again arguing for complexity in cloud structure or
  distribution.  While this makes a reconstruction of the ISM more
  difficult, it makes it clear that dramatic variations in
  interstellar density are a common occurrence and that understanding
  the interaction between the ISM and the Sun, together with the
  heliosphere and the planets, is important.

\end{enumerate}

The connection between the ISM and ultimately planets, via a dynamic
heliosphere is captivating, but also complex.  The work presented here
lays the foundation for evaluating this relationship by presenting
observations and analysis of the interstellar medium along the
historical solar trajectory, the material that the Sun is likely to
have interacted within the last $\sim$40 million years.  There are
many additional questions and assumptions that must be addressed
before one is able to use these observations to make predictions
regarding the influence on planets.  For example, an estimate of the volume density is critical, but difficult to obtain.  We will explore several options in Paper II, including calculating the density using the column density and distance differences between the sight line where the absorption is first detected and the sight line just preceding in distance in which the absorption is not detected.  Alternatively, we can make order of magnitude estimates from assuming the spatial extent is comparable to the radial extent, or simply using canonical cloud sizes derived from other work.  
Volume densities will then allow us to
calculate heliospheric compression and the implications for phenomena that may
affect planets (e.g., the cosmic ray flux). These topics are discussed
more in depth in Paper~II.  While this work is clearly tied to the Sun
and heliosphere, these same issues are at work for other
planet-bearing stars and their own surrounding interstellar
environments.  Indeed, the vast majority of known exoplanets reside
within $\sim$500 pc, and therefore are encountering interstellar
clouds not all that different from those studied here.  If an
ISM-planet connection can be established, it may have important
consequences for evaluating the long-term evolution of the atmospheres
of these planets.

\label{conclusion}

\acknowledgments
We are grateful to the staff at the McDonald Observatory, in particular David Doss, and at the Anglo-Australian Observatory, in particular Stuart Ryder.  Their assistance was essential to the success of the observations that were acquired for this work.  K.W. would like to thank S.R. for his support and guidance throughout this master's thesis project.

\clearpage

{\it Facilities:} \facility{Smith (TS12,TS21,TS23)} \facility{AAT (UHRF)}

\bibliographystyle{apj}

%\bibliography{kwAPJpaper.bib}

\begin{thebibliography}{92}
\expandafter\ifx\csname natexlab\endcsname\relax\def\natexlab#1{#1}\fi

\bibitem[{{Adams}(1949)}]{adams49}
{Adams}, W.~S. 1949, \apj, 109, 354

\bibitem[{{Albert} {et~al.}(1993){Albert}, {Blades}, {Morton}, {Lockman},
  {Proulx}, \& {Ferrarese}}]{albert93}
{Albert}, C.~E., {Blades}, J.~C., {Morton}, D.~C., {Lockman}, F.~J., {Proulx},
  M., \& {Ferrarese}, L. 1993, \apjs, 88, 81

\bibitem[{{Begelman} \& {Rees}(1976)}]{begelman76}
{Begelman}, M.~C., \& {Rees}, M.~J. 1976, \nat, 261, 298

\bibitem[{{Beintema}(1975)}]{beintema75}
{Beintema}, D. 1975, PhD thesis, , Univ.~Groningen, (1975)

\bibitem[{{Bertin} {et~al.}(1993){Bertin}, {Lallement}, {Ferlet}, \&
  {Vidal-Madjar}}]{bertin93}
{Bertin}, P., {Lallement}, R., {Ferlet}, R., \& {Vidal-Madjar}, A. 1993, \aap,
  278, 549

\bibitem[{Bevington(2003)}]{bevington}
Bevington, P.~R. 2003, Data Reduction and Error Analysis for the Physical
  Sciences (McGraw-Hill)

\bibitem[{{Blades} {et~al.}(1997){Blades}, {Sahu}, {He}, {Crawford}, {Barlow},
  \& {Diego}}]{blades97}
{Blades}, J.~C., {Sahu}, M.~S., {He}, L., {Crawford}, I.~A., {Barlow}, M.~J.,
  \& {Diego}, F. 1997, \apj, 478, 648

\bibitem[{{Burbidge} \& {Burbidge}(1953)}]{burbidge53}
{Burbidge}, E.~M., \& {Burbidge}, G.~R. 1953, \apj, 117, 465

\bibitem[{{Crawford}(1992)}]{crawford92}
{Crawford}, I.~A. 1992, \mnras, 259, 47

\bibitem[{{Decker} {et~al.}(2005){Decker}, {Krimigis}, {Roelof}, {Hill},
  {Armstrong}, {Gloeckler}, {Hamilton}, \& {Lanzerotti}}]{decker05}
{Decker}, R.~B., {Krimigis}, S.~M., {Roelof}, E.~C., {Hill}, M.~E.,
  {Armstrong}, T.~P., {Gloeckler}, G., {Hamilton}, D.~C., \& {Lanzerotti},
  L.~J. 2005, Science, 309, 2020

\bibitem[{{Dehnen} \& {Binney}(1998)}]{dehnen98}
{Dehnen}, W., \& {Binney}, J.~J. 1998, \mnras, 298, 387

\bibitem[{{Diego} {et~al.}(1995)}]{diego95}
{Diego}, F., {et~al.} 1995, \mnras, 272, 323

\bibitem[{{Fahr}(1968)}]{fahr68}
{Fahr}, H.~J. 1968, \apss, 2, 474

\bibitem[{{Ferlet} {et~al.}(1985){Ferlet}, {Vidal-Madjar}, \& {Gry}}]{ferlet85}
{Ferlet}, R., {Vidal-Madjar}, A., \& {Gry}, C. 1985, \apj, 298, 838

\bibitem[{{Frail} {et~al.}(1994){Frail}, {Weisberg}, {Cordes}, \&
  {Mathers}}]{frail94}
{Frail}, D.~A., {Weisberg}, J.~M., {Cordes}, J.~M., \& {Mathers}, C. 1994,
  \apj, 436, 144

\bibitem[{{Frisch} \& {York}(1986)}]{frisch86}
{Frisch}, P., \& {York}, D.~G. 1986, in The Galaxy and the Solar System, ed.
  R.~{Smoluchowski}, J.~M. {Bahcall}, \& M.~S. {Matthews}, 83--100

\bibitem[{{Frisch}(2006)}]{frisch06book}
{Frisch}, P.~C. 2006, in {Astrophysics and Space Science Library, Vol.~338,
  Solar Journey: The Significance of our Galactic Environment for the
  Heliosphere and Earth}, ed. P.~C. {Frisch} (Dordrecht: Springer)

\bibitem[{{Frisch} {et~al.}(2011){Frisch}, {Redfield}, \& {Slavin}}]{frisch11}
{Frisch}, P.~C., {Redfield}, S., \& {Slavin}, J.~D. 2011, \araa, 49, 237

\bibitem[{{Frisch} {et~al.}(1990){Frisch}, {Sembach}, \& {York}}]{frisch90b}
{Frisch}, P.~C., {Sembach}, K., \& {York}, D.~G. 1990, \apj, 364, 540

\bibitem[{{Frisch} \& {York}(1983)}]{frisch83}
{Frisch}, P.~C., \& {York}, D.~G. 1983, \apjl, 271, L59

\bibitem[{{G{\'e}nova} \& {Beckman}(2003)}]{genova03}
{G{\'e}nova}, R., \& {Beckman}, J.~E. 2003, \apjs, 145, 355

\bibitem[{{Giacalone} \& {Jokipii}(1999)}]{giacalone99}
{Giacalone}, J., \& {Jokipii}, J.~R. 1999, \apj, 520, 204

\bibitem[{{Grenier} {et~al.}(1999){Grenier}, {Baylac}, {Rolland}, {Burnage},
  {Arenou}, {Briot}, {Delmas}, {Duflot}, {Genty}, {G{\'o}mez}, {Halbwachs},
  {Marouard}, {Oblak}, \& {Sellier}}]{gren}
{Grenier}, S., {Baylac}, M., {Rolland}, L., {Burnage}, R., {Arenou}, F.,
  {Briot}, D., {Delmas}, F., {Duflot}, M., {Genty}, V., {G{\'o}mez}, A.~E.,
  {Halbwachs}, J., {Marouard}, M., {Oblak}, E., \& {Sellier}, A. 1999, \aaps,
  137, 451

\bibitem[{{Habing}(1969)}]{habing69}
{Habing}, H.~J. 1969, \bain, 20, 177

\bibitem[{{Hearty} {et~al.}(2000){Hearty}, {Fern{\'a}ndez}, {Alcal{\'a}},
  {Covino}, \& {Neuh{\"a}user}}]{hearty00}
{Hearty}, T., {Fern{\'a}ndez}, M., {Alcal{\'a}}, J.~M., {Covino}, E., \&
  {Neuh{\"a}user}, R. 2000, \aap, 357, 681

\bibitem[{{Heiles}(1997)}]{heiles97}
{Heiles}, C. 1997, \apj, 481, 193

\bibitem[{{Heiles} \& {Troland}(2003)}]{heiles03}
{Heiles}, C., \& {Troland}, T.~H. 2003, \apj, 586, 1067

\bibitem[{{Hobbs}(1969)}]{hobbs69}
{Hobbs}, L.~M. 1969, \apj, 157, 165

\bibitem[{{Hobbs}(1974)}]{hobbs74}
---. 1974, \apj, 191, 381

\bibitem[{{Hobbs}(1978)}]{hobbs78}
---. 1978, \apjs, 38, 129

\bibitem[{{Hobbs}(1984)}]{hobbs84}
---. 1984, \apjs, 56, 315

\bibitem[{{Holweger} {et~al.}(1999){Holweger}, {Hempel}, \&
  {Kamp}}]{holweger99}
{Holweger}, H., {Hempel}, M., \& {Kamp}, I. 1999, \aap, 350, 603

\bibitem[{{Hoyle} \& {Lyttleton}(1939)}]{hoyle39}
{Hoyle}, F., \& {Lyttleton}, R.~A. 1939, in Proceedings of the Cambridge
  Philosophical Society, Vol.~35, Proceedings of the Cambridge Philosophical
  Society, 405

\bibitem[{{Jackson} {et~al.}(2003){Jackson}, {Werner}, \&
  {Gautier}}]{jackson03}
{Jackson}, T., {Werner}, M., \& {Gautier}, III, T.~N. 2003, \apjs, 149, 365

\bibitem[{{Lallement} {et~al.}(1993){Lallement}, {Bertin}, {Chassefiere}, \&
  {Scott}}]{lallement93}
{Lallement}, R., {Bertin}, P., {Chassefiere}, E., \& {Scott}, N. 1993, \aap,
  271, 734

\bibitem[{{Lallement} {et~al.}(2003){Lallement}, {Welsh}, {Vergely}, {Crifo},
  \& {Sfeir}}]{lallement03}
{Lallement}, R., {Welsh}, B.~Y., {Vergely}, J.~L., {Crifo}, F., \& {Sfeir}, D.
  2003, \aap, 411, 447

\bibitem[{{Linsky} \& {Wood}(1996)}]{linsky96}
{Linsky}, J.~L., \& {Wood}, B.~E. 1996, \apj, 463, 254

\bibitem[{{Magnani} {et~al.}(1985){Magnani}, {Blitz}, \& {Mundy}}]{magnani85}
{Magnani}, L., {Blitz}, L., \& {Mundy}, L. 1985, \apj, 295, 402

\bibitem[{{Makaganiuk} {et~al.}(2011){Makaganiuk}, {Kochukhov}, {Piskunov},
  {Jeffers}, {Johns-Krull}, {Keller}, {Rodenhuis}, {Snik}, {Stempels}, \&
  {Valenti}}]{makaganiuk11}
{Makaganiuk}, V., {Kochukhov}, O., {Piskunov}, N., {Jeffers}, S.~V.,
  {Johns-Krull}, C.~M., {Keller}, C.~U., {Rodenhuis}, M., {Snik}, F.,
  {Stempels}, H.~C., \& {Valenti}, J.~A. 2011, \aap, 529, A160

\bibitem[{{Meyer} \& {Blades}(1996)}]{meyer96}
{Meyer}, D.~M., \& {Blades}, J.~C. 1996, \apjl, 464, L179

\bibitem[{{Meyer} {et~al.}(2006){Meyer}, {Lauroesch}, {Heiles}, {Peek}, \&
  {Engelhorn}}]{meyer06}
{Meyer}, D.~M., {Lauroesch}, J.~T., {Heiles}, C., {Peek}, J.~E.~G., \&
  {Engelhorn}, K. 2006, \apjl, 650, L67

\bibitem[{{Meyer} {et~al.}(2012){Meyer}, {Lauroesch}, {Peek}, \&
  {Heiles}}]{meyer12}
{Meyer}, D.~M., {Lauroesch}, J.~T., {Peek}, J.~E.~G., \& {Heiles}, C. 2012,
  \apj, 752, 119

\bibitem[{{Morton}(2003)}]{morton03}
{Morton}, D.~C. 2003, \apjs, 149, 205

\bibitem[{{M{\"u}ller} {et~al.}(2006){M{\"u}ller}, {Frisch}, {Florinski}, \&
  {Zank}}]{muller06}
{M{\"u}ller}, H.-R., {Frisch}, P.~C., {Florinski}, V., \& {Zank}, G.~P. 2006,
  \apj, 647, 1491

\bibitem[{{M{\"u}nch} \& {Zirin}(1961)}]{munch61}
{M{\"u}nch}, G., \& {Zirin}, H. 1961, \apj, 133, 11

\bibitem[{{Nordstr{\"o}m} {et~al.}(2004){Nordstr{\"o}m}, {Mayor}, {Andersen},
  {Holmberg}, {Pont}, {J{\o}rgensen}, {Olsen}, {Udry}, \& {Mowlavi}}]{nord}
{Nordstr{\"o}m}, B., {Mayor}, M., {Andersen}, J., {Holmberg}, J., {Pont}, F.,
  {J{\o}rgensen}, B.~R., {Olsen}, E.~H., {Udry}, S., \& {Mowlavi}, N. 2004,
  \aap, 418, 989

\bibitem[{{O'dell}(2001)}]{odell01}
{O'dell}, C.~R. 2001, \araa, 39, 99

\bibitem[{{Pavlov} {et~al.}(2005){Pavlov}, {Toon}, {Pavlov}, {Bally}, \&
  {Pollard}}]{pavlov05}
{Pavlov}, A.~A., {Toon}, O.~B., {Pavlov}, A.~K., {Bally}, J., \& {Pollard}, D.
  2005, \grl, 32, L03705

\bibitem[{{Peek} {et~al.}(2011){Peek}, {Heiles}, {Peek}, {Meyer}, \&
  {Lauroesch}}]{peek11}
{Peek}, J.~E.~G., {Heiles}, C., {Peek}, K.~M.~G., {Meyer}, D.~M., \&
  {Lauroesch}, J.~T. 2011, \apj, 735, 129

\bibitem[{{Penprase}(1993)}]{penprase93}
{Penprase}, B.~E. 1993, \apjs, 88, 433

\bibitem[{{Phillips} {et~al.}(1984){Phillips}, {Pettini}, \&
  {Gondhalekar}}]{phillips84}
{Phillips}, A.~P., {Pettini}, M., \& {Gondhalekar}, P.~M. 1984, \mnras, 206,
  337

\bibitem[{{Press} {et~al.}(2002){Press}, {Teukolsky}, {Vetterling}, \&
  {Flannery}}]{press02}
{Press}, W.~H., {Teukolsky}, S.~A., {Vetterling}, W.~T., \& {Flannery}, B.~P.
  2002, {Numerical recipes in C++ : the art of scientific computing}

\bibitem[{{Price} {et~al.}(2001){Price}, {Crawford}, {Barlow}, \&
  {Howarth}}]{price01}
{Price}, R.~J., {Crawford}, I.~A., {Barlow}, M.~J., \& {Howarth}, I.~D. 2001,
  \mnras, 328, 555

\bibitem[{{Redfield}(2006)}]{redfield06}
{Redfield}, S. 2006, in Astronomical Society of the Pacific Conference Series,
  Vol. 352, New Horizons in Astronomy: Frank N. Bash Symposium, ed.
  {S.~J.~Kannappan, S.~Redfield, J.~E.~Kessler-Silacci, M.~Landriau, \&
  N.~Drory}, 79

\bibitem[{{Redfield} \& {Linsky}(2000)}]{redfield00}
{Redfield}, S., \& {Linsky}, J.~L. 2000, \apj, 534, 825

\bibitem[{{Redfield} \& {Linsky}(2001)}]{redfield01}
---. 2001, \apj, 551, 413

\bibitem[{{Redfield} \& {Linsky}(2002)}]{redfield02}
---. 2002, \apjs, 139, 439

\bibitem[{{Redfield} \& {Linsky}(2004{\natexlab{a}})}]{redfield04sw}
---. 2004{\natexlab{a}}, \apj, 602, 776

\bibitem[{{Redfield} \& {Linsky}(2004{\natexlab{b}})}]{redfield04}
---. 2004{\natexlab{b}}, \apj, 613, 1004

\bibitem[{{Redfield} \& {Linsky}(2008)}]{redfield08}
---. 2008, \apj, 673, 283

\bibitem[{{Routly} \& {Spitzer}(1952)}]{routly52}
{Routly}, P.~M., \& {Spitzer}, Jr., L. 1952, \apj, 115, 227

\bibitem[{{Savage} \& {Sembach}(1991)}]{savage91}
{Savage}, B.~D., \& {Sembach}, K.~R. 1991, \apj, 379, 245

\bibitem[{{Savage} \& {Sembach}(1996)}]{savage96}
---. 1996, \araa, 34, 279

\bibitem[{{Scherer} {et~al.}(2006){Scherer}, {Fichtner}, {Borrmann}, {Beer},
  {Desorgher}, {Fl{\"u}kiger}, {Fahr}, {Ferreira}, {Langner}, {Potgieter},
  {Heber}, {Masarik}, {Shaviv}, \& {Veizer}}]{scherer06}
{Scherer}, K., {Fichtner}, H., {Borrmann}, T., {Beer}, J., {Desorgher}, L.,
  {Fl{\"u}kiger}, E., {Fahr}, H.-J., {Ferreira}, S.~E.~S., {Langner}, U.~W.,
  {Potgieter}, M.~S., {Heber}, B., {Masarik}, J., {Shaviv}, N., \& {Veizer}, J.
  2006, \ssr, 127, 327

\bibitem[{{Sch{\"o}nrich} {et~al.}(2010){Sch{\"o}nrich}, {Binney}, \&
  {Dehnen}}]{schonrich10}
{Sch{\"o}nrich}, R., {Binney}, J., \& {Dehnen}, W. 2010, \mnras, 403, 1829

\bibitem[{{Shapley}(1921)}]{shapley21}
{Shapley}, H. 1921, Journal of Geology, 29, 502

\bibitem[{{Shaviv}(2003)}]{shaviv03}
{Shaviv}, N.~J. 2003, \na, 8, 39

\bibitem[{{Shorlin} {et~al.}(2002){Shorlin}, {Wade}, {Donati}, {Landstreet},
  {Petit}, {Sigut}, \& {Strasser}}]{shorlin}
{Shorlin}, S.~L.~S., {Wade}, G.~A., {Donati}, J., {Landstreet}, J.~D., {Petit},
  P., {Sigut}, T.~A.~A., \& {Strasser}, S. 2002, \aap, 392, 637

\bibitem[{{Shull} {et~al.}(1977){Shull}, {York}, \& {Hobbs}}]{shull77}
{Shull}, J.~M., {York}, D.~G., \& {Hobbs}, L.~M. 1977, \apjl, 211, L139

\bibitem[{{Siluk} \& {Silk}(1974)}]{siluk74}
{Siluk}, R.~S., \& {Silk}, J. 1974, \apj, 192, 51

\bibitem[{{Slavin} \& {Frisch}(2002)}]{slavin02}
{Slavin}, J.~D., \& {Frisch}, P.~C. 2002, \apj, 565, 364

\bibitem[{{Slavin} \& {Frisch}(2008)}]{slavin08}
---. 2008, \aap, 491, 53

\bibitem[{{Smith} \& {Scalo}(2009)}]{smith09}
{Smith}, D.~S., \& {Scalo}, J.~M. 2009, Astrobiology, 9, 673

\bibitem[{{Stanimirovi{\'c}} {et~al.}(2010){Stanimirovi{\'c}}, {Weisberg},
  {Pei}, {Tuttle}, \& {Green}}]{stanimirovic10}
{Stanimirovi{\'c}}, S., {Weisberg}, J.~M., {Pei}, Z., {Tuttle}, K., \& {Green},
  J.~T. 2010, \apj, 720, 415

\bibitem[{{Stawikowski} \& {Glebocki}(1994)}]{staw}
{Stawikowski}, A., \& {Glebocki}, R. 1994, \actaa, 44, 33

\bibitem[{{Talbot} {et~al.}(1976){Talbot}, {Butler}, \& {Newman}}]{talbot76}
{Talbot}, R.~J., {Butler}, D.~M., \& {Newman}, M.~J. 1976, \nat, 262, 561

\bibitem[{{Thaddeus}(1986)}]{thaddeus86}
{Thaddeus}, P. 1986, The Galaxy and the Solar System, 61

\bibitem[{{Tody}(1993)}]{tody93}
{Tody}, D. 1993, in Astronomical Society of the Pacific Conference Series,
  Vol.~52, Astronomical Data Analysis Software and Systems II, ed.
  {R.~J.~Hanisch, R.~J.~V.~Brissenden, \& J.~Barnes}, 173--+

\bibitem[{{Tull}(1972)}]{tull72}
{Tull}, R.~G. 1972, in Proc. ESO/CERN Conference on Auxiliary Instrumentation
  for Large Telescopes (Geneva: ESO), 259

\bibitem[{{Tull} {et~al.}(1995){Tull}, {MacQueen}, {Sneden}, \&
  {Lambert}}]{tull95}
{Tull}, R.~G., {MacQueen}, P.~J., {Sneden}, C., \& {Lambert}, D.~L. 1995,
  \pasp, 107, 251

\bibitem[{{Usoskin} {et~al.}(2005){Usoskin}, {Alanko-Huotari}, {Kovaltsov}, \&
  {Mursula}}]{usoskin05}
{Usoskin}, I.~G., {Alanko-Huotari}, K., {Kovaltsov}, G.~A., \& {Mursula}, K.
  2005, Journal of Geophysical Research (Space Physics), 110, 12108

\bibitem[{{Vallerga} {et~al.}(1993){Vallerga}, {Vedder}, {Craig}, \&
  {Welsh}}]{vallerga93}
{Vallerga}, J.~V., {Vedder}, P.~W., {Craig}, N., \& {Welsh}, B.~Y. 1993, \apj,
  411, 729

\bibitem[{{van Leeuwen}(2007)}]{vanleeuwen07}
{van Leeuwen}, F. 2007, \aap, 474, 653

\bibitem[{{Watson} \& {Meyer}(1996)}]{watson96}
{Watson}, J.~K., \& {Meyer}, D.~M. 1996, \apjl, 473, L127

\bibitem[{{Welsh} {et~al.}(1994){Welsh}, {Craig}, {Vedder}, \&
  {Vallerga}}]{welsh94}
{Welsh}, B.~Y., {Craig}, N., {Vedder}, P.~W., \& {Vallerga}, J.~V. 1994, \apj,
  437, 638

\bibitem[{{Welsh} {et~al.}(2005){Welsh}, {Sallmen}, \& {Jelinsky}}]{welsh05b}
{Welsh}, B.~Y., {Sallmen}, S., \& {Jelinsky}, S. 2005, \aap, 440, 547

\bibitem[{{Welsh} \& {Shelton}(2009)}]{welsh09}
{Welsh}, B.~Y., \& {Shelton}, R.~L. 2009, \apss, 323, 1

\bibitem[{{Welsh} {et~al.}(1991){Welsh}, {Vedder}, {Vallerga}, \&
  {Craig}}]{welsh91}
{Welsh}, B.~Y., {Vedder}, P.~W., {Vallerga}, J.~V., \& {Craig}, N. 1991, \apj,
  381, 462

\bibitem[{{Welty} {et~al.}(1994){Welty}, {Hobbs}, \& {Kulkarni}}]{welty94}
{Welty}, D.~E., {Hobbs}, L.~M., \& {Kulkarni}, V.~P. 1994, \apj, 436, 152

\bibitem[{{Welty} {et~al.}(1996){Welty}, {Morton}, \& {Hobbs}}]{welty96}
{Welty}, D.~E., {Morton}, D.~C., \& {Hobbs}, L.~M. 1996, \apjs, 106, 533

\bibitem[{{Wood} {et~al.}(2005){Wood}, {M{\"u}ller}, {Zank}, {Linsky}, \&
  {Redfield}}]{wood05let}
{Wood}, B.~E., {M{\"u}ller}, H.-R., {Zank}, G., {Linsky}, J., \& {Redfield}, S.
  2005, \apjl, 628, L143

\bibitem[{{Wyman} \& {Redfield}(2013)}]{wyman13}
{Wyman}, K., \& {Redfield}, S. 2013, in preparation

\bibitem[{{Zank} \& {Frisch}(1999)}]{zankfrisch99}
{Zank}, G.~P., \& {Frisch}, P.~C. 1999, \apj, 518, 965

\end{thebibliography}

\clearpage

\clearpage

%\addtocounter{figure}{1}
%

%\clearpage


\clearpage

%\null
%\vfill

\setcounter{figure}{3}

\begin{figure}
\centering
\epsscale{1.}
\plotone{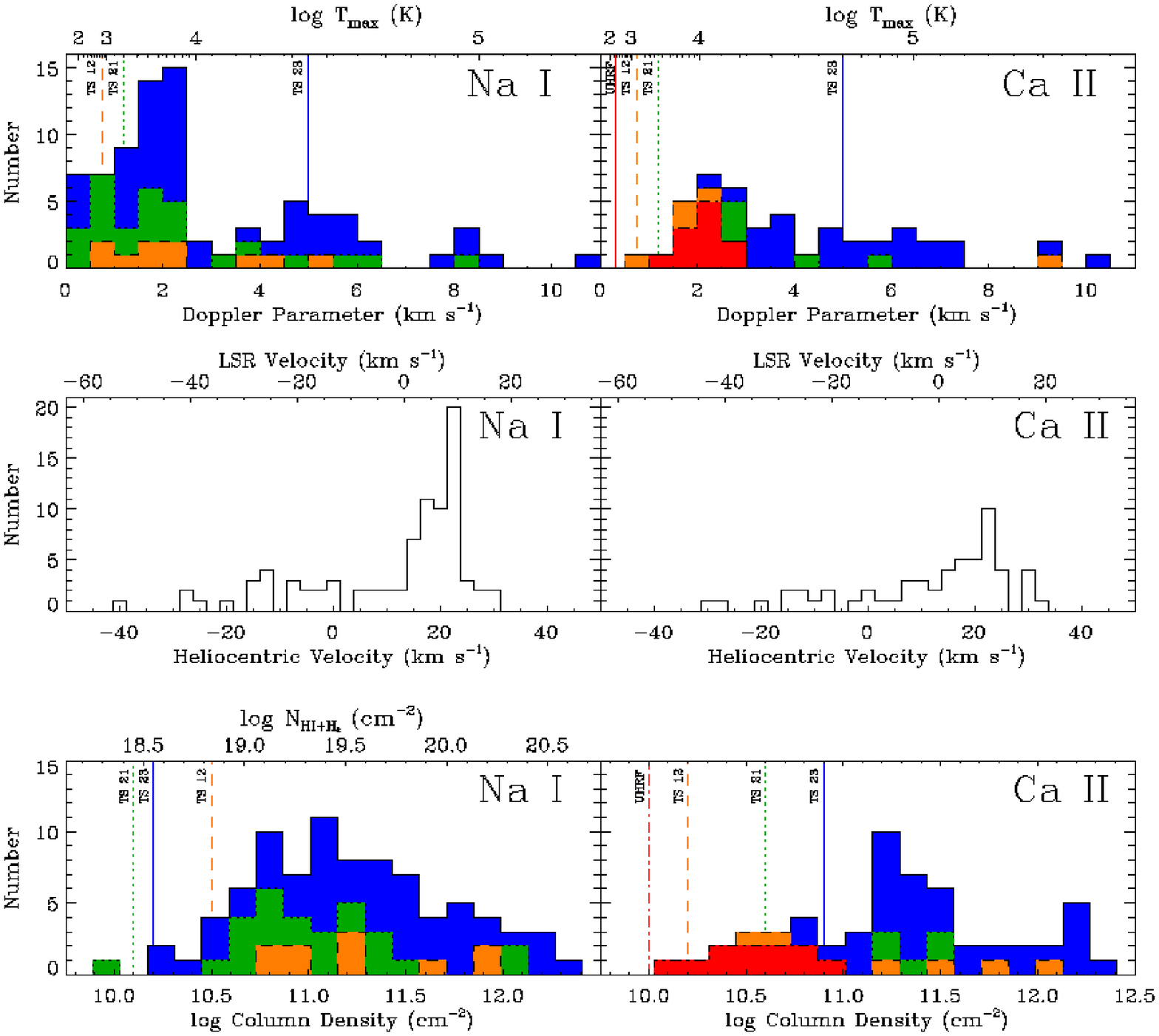}
\caption{See caption on next page.}
\label{hist}
\end{figure}

\clearpage 

%\raggedright

Fig. 4.--- Distribution of ISM absorption fit parameters for both \ion{Na}{1} and \ion{Ca}{2}: Doppler parameter, or line width, ($b$) at the top; radial velocity ($v$) in the middle; and column density ($N$) at the bottom.  The line width and column density distributions are color coded to differentiate between the various spectrographs used and their different resolving powers.  The Doppler parameter distribution includes the spectral resolution limits of the various instruments.  In addition, the top axis of the Doppler parameter distribution is a conversion of the line width into a maximum temperature, assuming no nonthermal, or turbulent, broadening.  The \ion{Na}{1} lines are consistently narrower than the \ion{Ca}{2} lines, despite the lower atomic weight of sodium and larger sensitivity to thermal broadening.  The velocity distribution is consistent between \ion{Na}{1} and \ion{Ca}{2}, and shows a preponderance of absorption at $\sim$20 km~s$^{-1}$.  The velocity distribution is not color coded due to the lack of any radial velocity bias induced by the differing spectral resolution.  The column density distribution is shown together with an estimate of the sensitivity of each instrument.  While dependent on spectral resolution, another significant factor is the efficiency of the spectrograph, which does not necessarily scale with the spectral resolution.  The \ion{Ca}{2} distribution shows a clear correlation with instrument, with the lowest column density measurements ($N($\ion{Ca}{2}$) < 11$) being made with the highest-resolution instrument (UHRF).  The \citet{ferlet85} correlation between $N($\ion{Na}{1}$)$ and $N($\ion{H}{1}$ + $H$_2)$ is used in the top axis of the \ion{Na}{1} distribution.

\clearpage

\begin{figure}
\centering
\epsscale{0.85}
\plotone{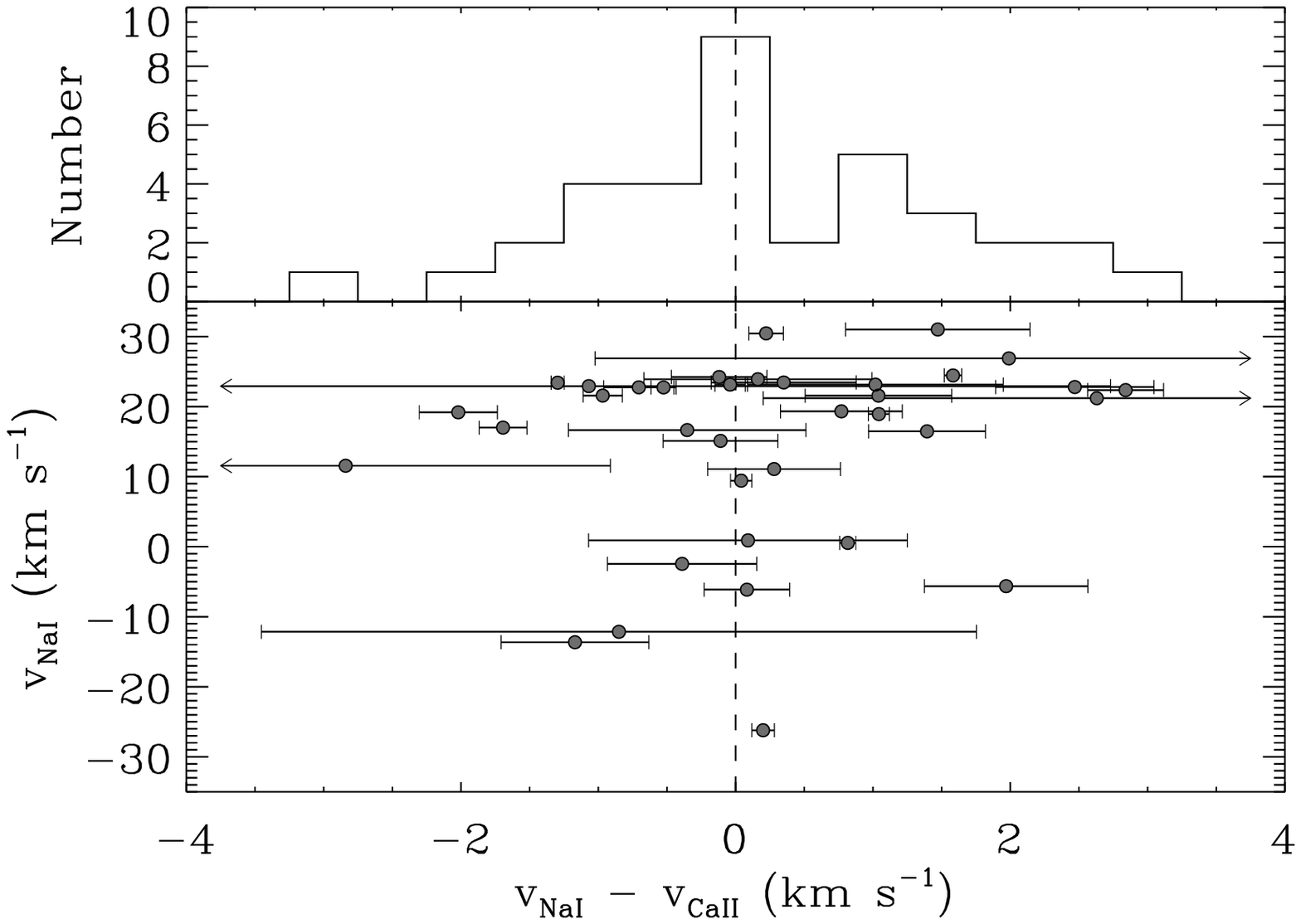}
\caption{Comparison of central velocities between \ion{Na}{1} and \ion{Ca}{2} for those identified as paired components (i.e., $\Delta v < 3$ km~s$^{-1}$).  The distribution clearly peaks at zero although only 81\% (29 out of 36) agree with zero within 3$\sigma$.
}
\label{dv}
\end{figure}

\begin{figure}
\centering
\epsscale{0.65}
\plotone{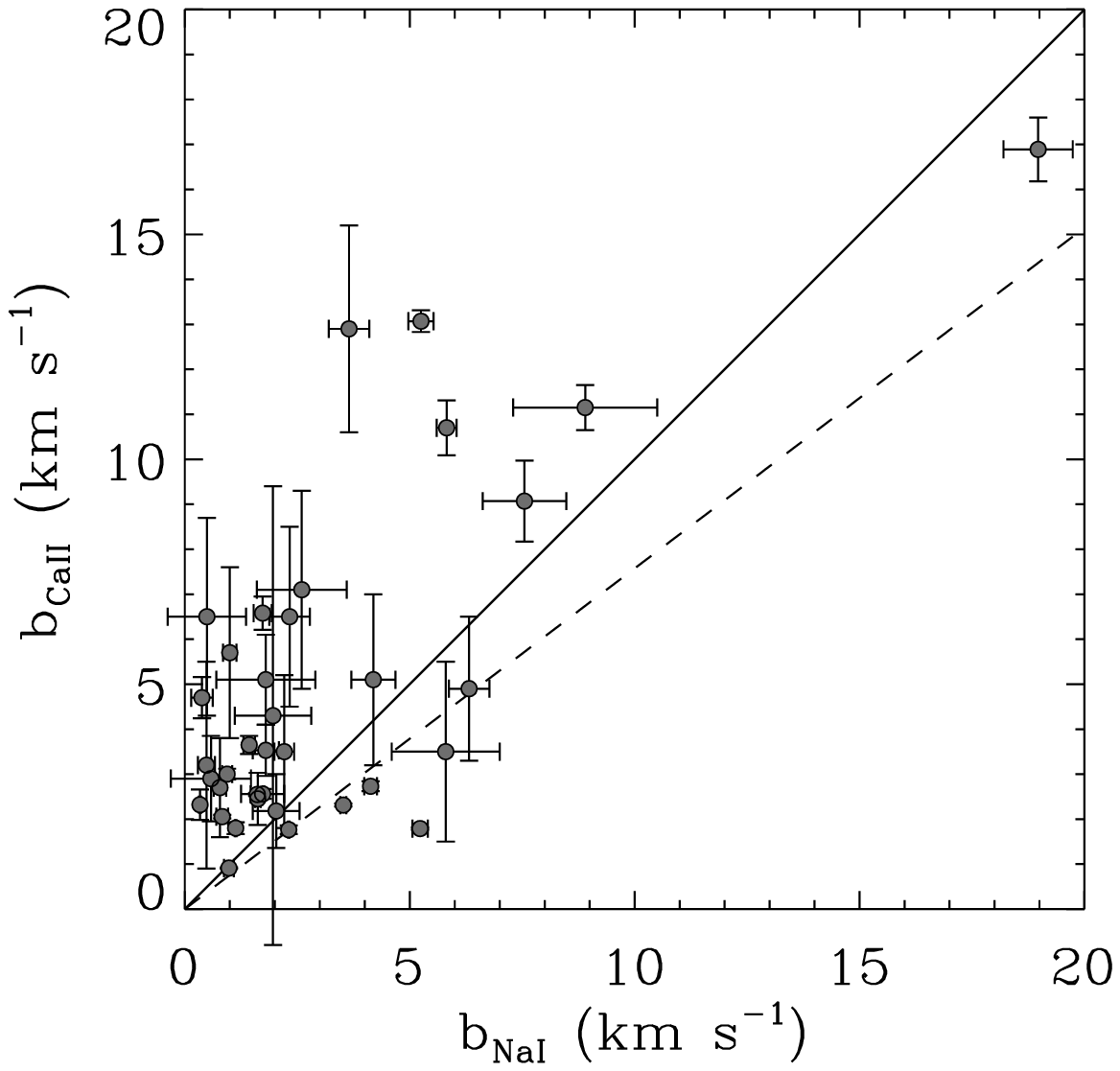}
\caption{Comparison of Doppler parameters between \ion{Na}{1} and \ion{Ca}{2} for those identified as paired components (i.e., $\Delta v < 3$ km~s$^{-1}$).  The solid line indicates a purely turbulence-broadened profile, while the dashed line indicates a purely thermally broadened profile.  Very few paired components fall between these two regimes, indicating that \ion{Na}{1} and \ion{Ca}{2} are not identically distributed.  Instead, while possibly correlated, the two ions are largely segregated.}
\label{bs}
\end{figure}

\begin{figure}
\centering
\epsscale{0.95}
\plottwo{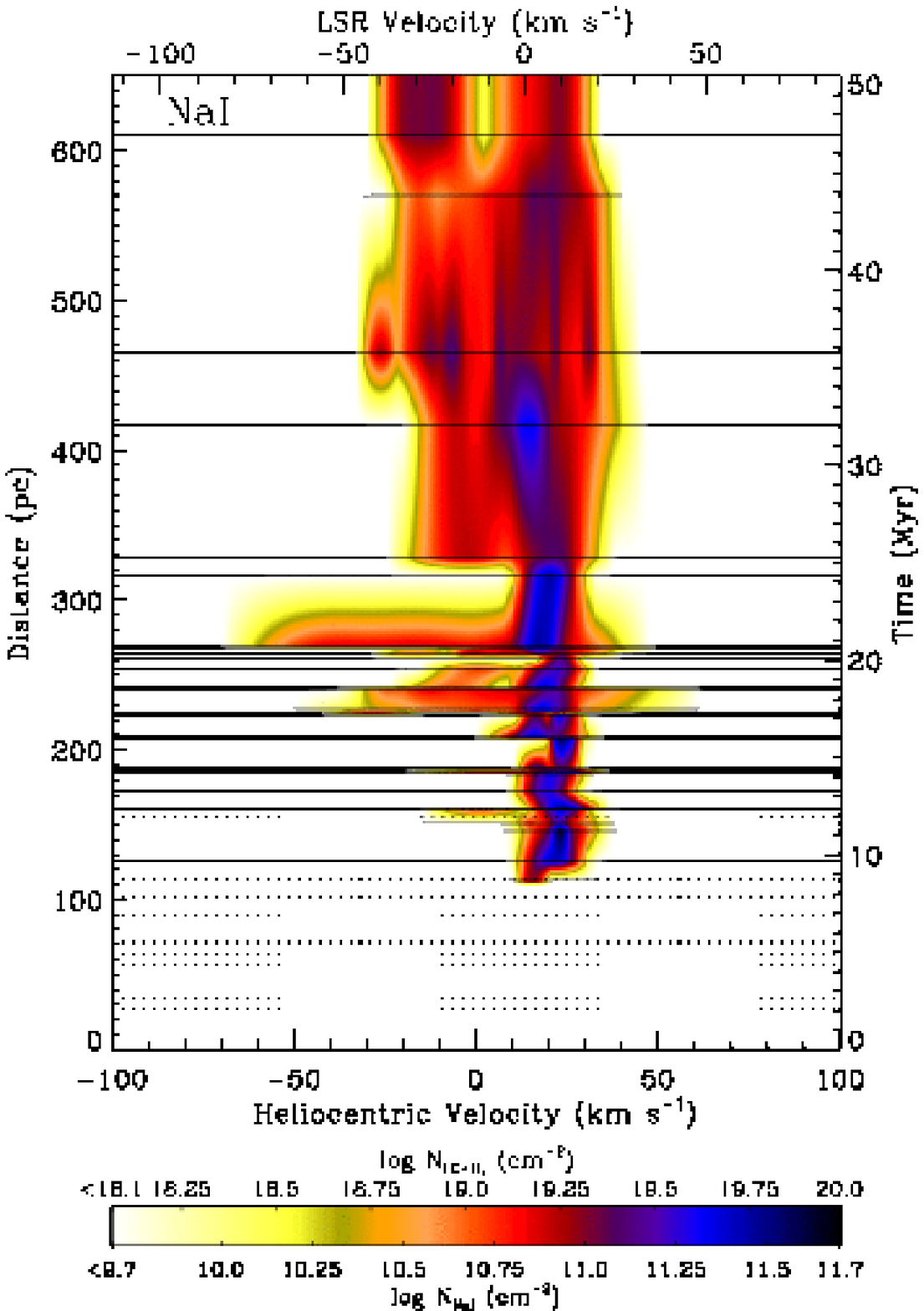}{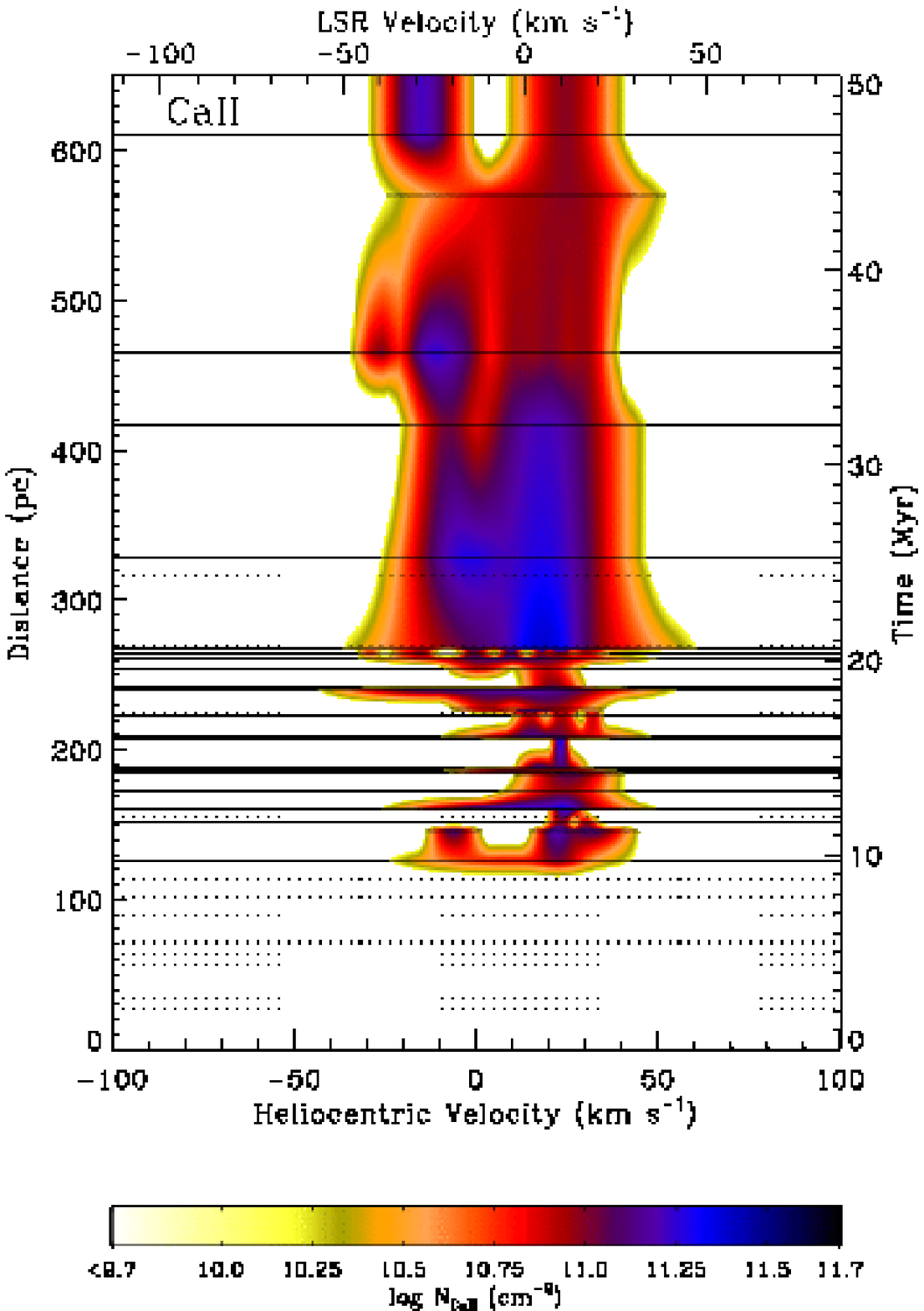}
\caption{{\it left:} Apparent column density profile determined from the sodium data plotted at the distance to each star.  The stellar distance is marked by black horizontal bars.  Sight lines for which only upper limits were made are indicated by dotted lines.  The distance axis is also displayed as a function of time, based on current values of the solar motion \citep{dehnen98}.  The \ion{Na}{1} column density ($N($\ion{Na}{1}$)$) is converted to hydrogen column density ($N($\ion{H}{1}$+$H$_2)$) based on the correlation found by \citet{ferlet85}.  {\it right:} Same for the calcium data.}
\label{aods}
\end{figure}

\begin{figure}
\centering
\epsscale{.80}
\plotone{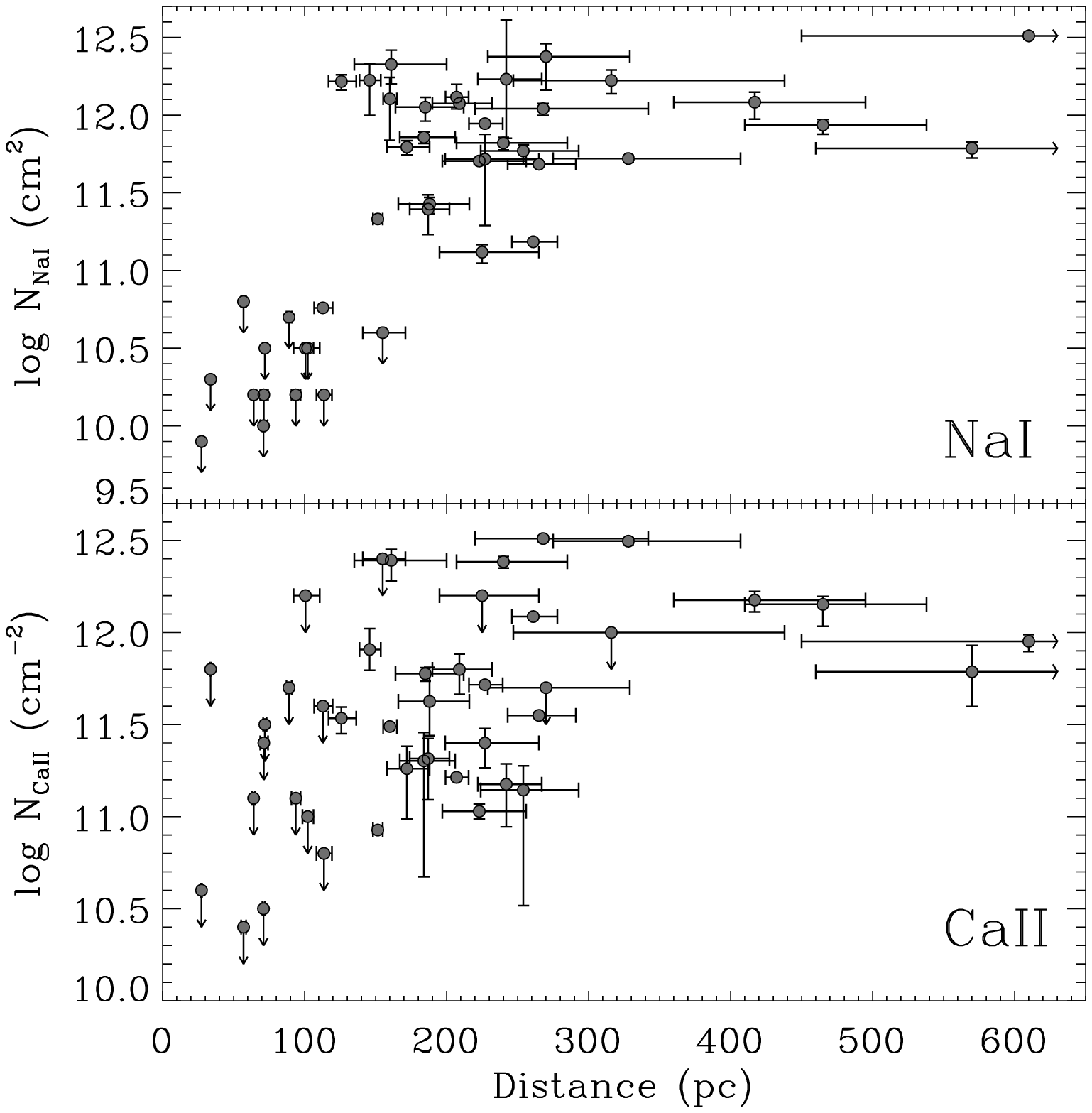}
\caption{Total column density as a function of distance.  Calcium (bottom panel) has significantly more scatter in the measurements of column density than the sodium data (top panel), likely due to the more dramatic effects of depletion on the calcium abundance in the gas phase.  The first detection of cold, dense ISM lies at $\sim$120 pc, consistent with measurements of the Local Bubble boundary made by \cite{lallement03}.  Distances and errors are from the new analysis of \textit{Hipparcos} parallax measurements \citep{vanleeuwen07}.  Data are shown for 43 stars between $\sim$25 and $\sim$600 pc, with a median distance between stars of 5 pc.  A dozen ISM absorption measurements inside the Local Bubble resulted in column density upper limits.\label{coldist}}
\end{figure}
\begin{figure}
\centering
\epsscale{1}
\plottwo{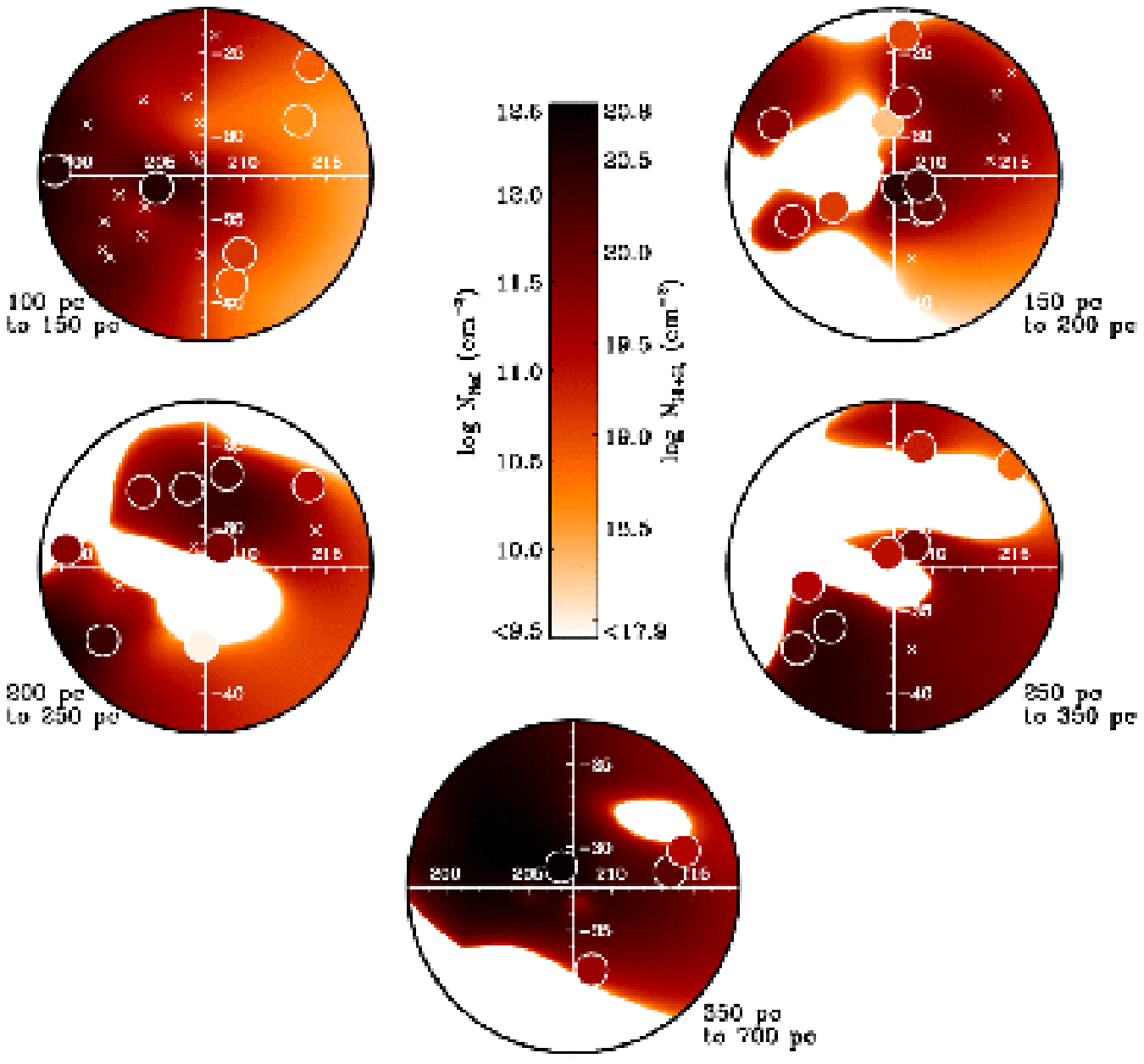}{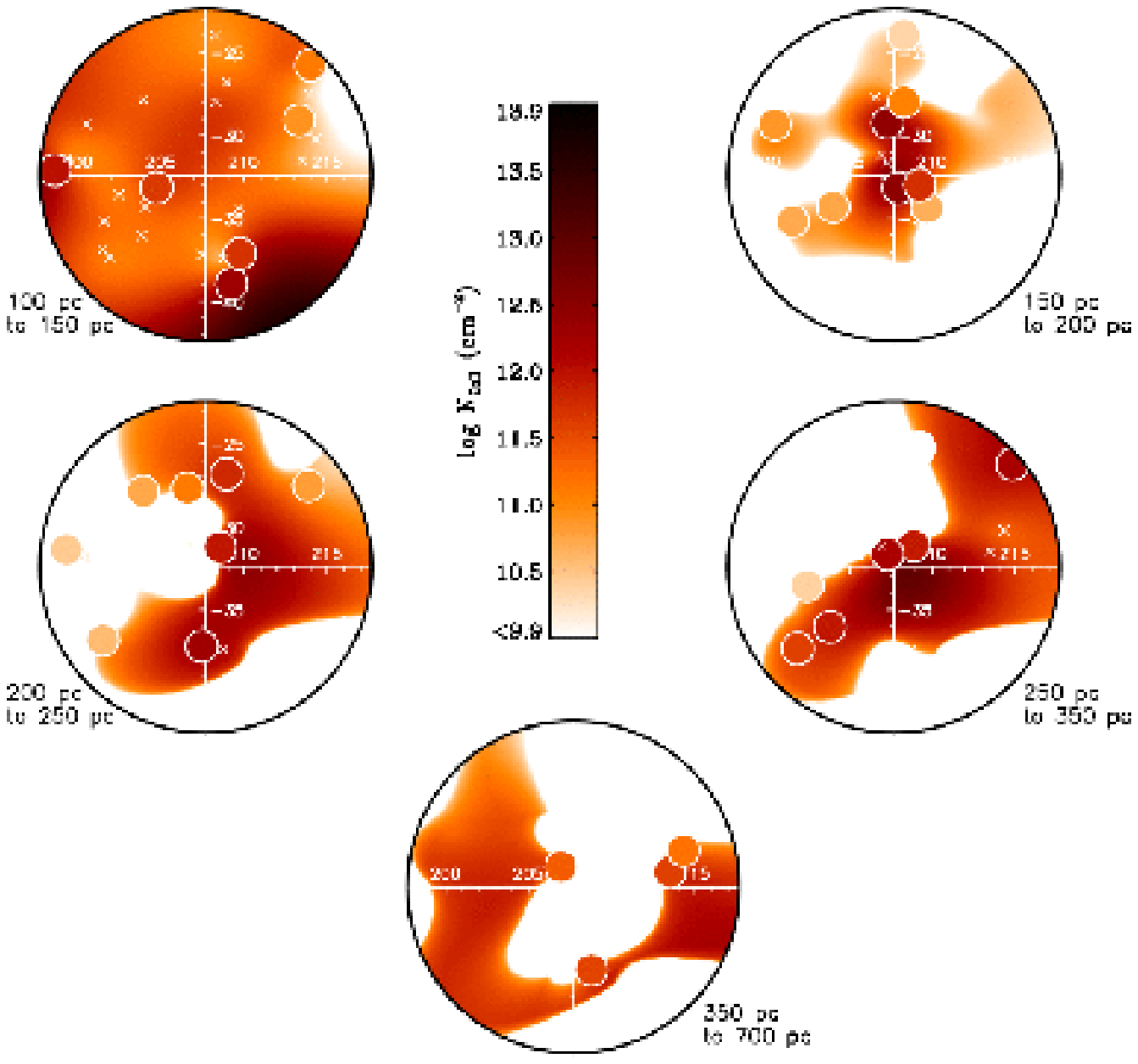}
\caption{{\it left:} Spatial maps of sodium column density measurements in successive distance bins.  The field of view has a radius of 10$^\circ$.  Circles indicate observed sight lines, and the color coding indicates the observed column density.  The crosses are sight lines with constraining upper limits.  The color coding throughout the field of view is an interpolation of the observed sight lines using a minimum curvature surface.  Subsequent distance bins have previous column density maps subtracted, and so provide a contour map of ISM in each distance bin.  The physical distances of $10^\circ$ at the mean distance of each bin, from the nearest bin to the farthest, are 22.0\,pc, 30.0\,pc, 39.7\,pc, 52.9\,pc, and 92.6\,pc, respectively.  {\it right:} Same for the calcium data.  While gross similarities exist between the ions, there are also significant differences, further emphasizing that these two ions are not spatially coincident. }
\label{dslice}
\end{figure}

\begin{figure}
\centering
\plotone{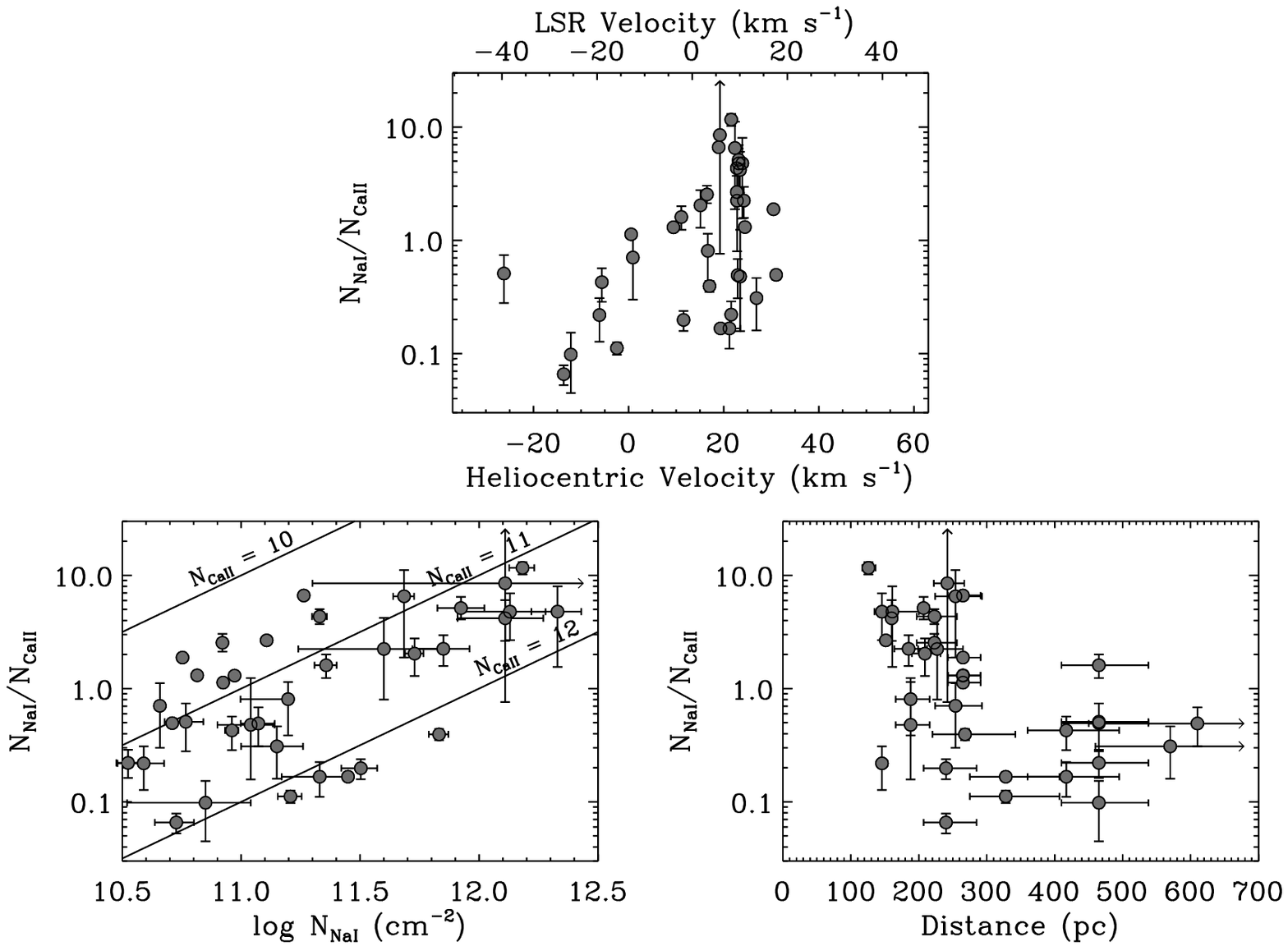}
\caption{Measurements of the column density ratio between \ion{Na}{1} and \ion{Ca}{2} for those identified as paired components (i.e., $\Delta v < 3$ km~s$^{-1}$).  The top plot is the canonical Routly-Spitzer Effect, showing a large spread in velocity for components with small \ion{Na}{1}/\ion{Ca}{2} column density ratios, whereas those with high ratios have a small variation in velocity.  The bottom left shows the column density ratio as a function of \ion{Na}{1} column density.  The solid lines indicate lines of constant \ion{Ca}{2} column density ($\log N_{\rm CaII} = 10$, 11, 12).  The correlation likely indicates the increase in calcium depletion in the densest ISM clouds.  The bottom right shows the column density ratio as a function of stellar distance.  The largest ratios are confined to the Local Bubble boundary region beyond $\sim$120 pc.  
}
\label{nrat}
\end{figure}

\begin{figure}
\centering
\epsscale{.6}
\plotone{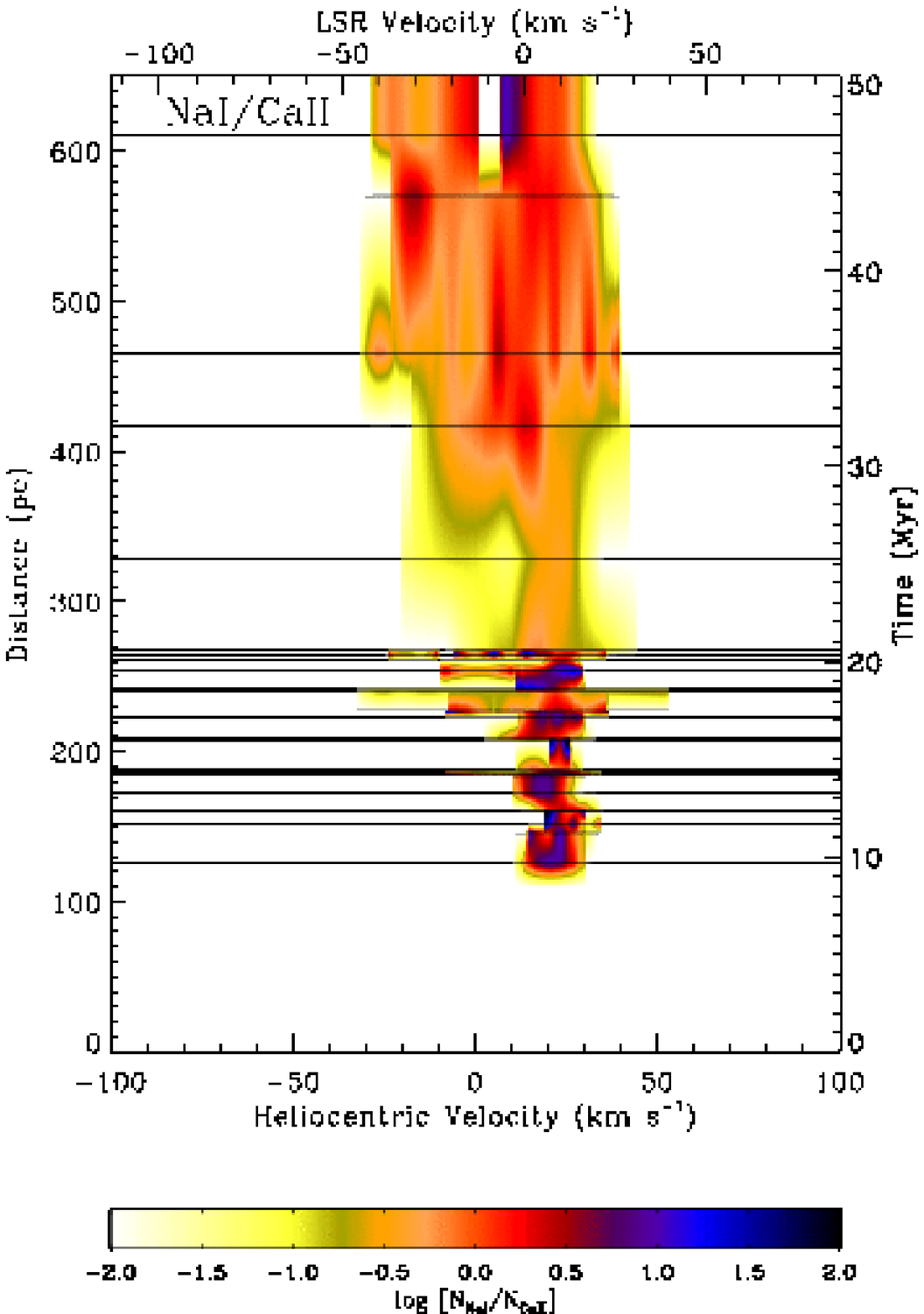}
\caption{The \ion{Na}{1}/\ion{Ca}{2} column density ratio based on the apparent column density profiles shown in Figure~\ref{aods}.  The high column density ratios are confined to the nearest sight lines and have a very narrow spread in velocity (i.e., the Routly-Spitzer Effect).  }
\label{aodsrat}
\end{figure}

\clearpage

\begin{deluxetable}{lllcllcclc}
\tablewidth{0pt}
\tabletypesize{\scriptsize}
\tablecolumns{10}
\tablecaption{Stellar Parameters for Targets along the Past Solar Trajectory\tablenotemark{a}}
\tablehead{ HD & Other & Spectral & $m_{\rm V}$ & $v_{\rm R}$ & $v \sin i$ & $l$ & $b$ & Distance\tablenotemark{b} & $\Delta\theta$\tablenotemark{c} \\
\# & Name & Type & (mag) & (km s$^{-1}$) & (km s$^{-1}$) & ($\circ$) & ($\circ$) & (pc) & ($\circ$) }
\startdata
33111 & $\beta$ Eri & A3III  & 2.8 &--9.2 & 195 & 205.34 & --25.32 & 27.40$^{+0.32}_{-0.31}$ & 7.38 \\
30743 & HR1545 & F5V & 6.3 &--3.0 & 7.0\tablenotemark{d}& 212.08& --33.12& 33.81$^{+0.57}_{-0.55}$ & 3.75 \\
%26574 & $o^1$ Eri & F2II-III & 4.0&11.0& 115 & 199.32 & --38.39 & 37.31$^{+0.45}_{-0.44}$ & 9.07 \\
33904 & $\mu$ Lep & B9IV & 3.3 &27.7 & 18 & 217.25 & --28.91 & 57.0$^{+1.8}_{-1.7}$ & 8.92 \\
27861 & $\xi$ Eri&A2V & 	5.2 &--11.0 & 179 & 197.82 & --34.30 & 64.1$^{+1.0}_{-1.0}$ & 8.46 \\
33802 & $\iota$ Lep & B8V & 4.5 &25.0 & 190 & 212.67 & --27.32 & 71.07$^{+0.82}_{-0.80}$ & 6.67 \\
29573 & HR1483 & A0V & 5.0 &6.5 &  27 & 208.96 & --34.85 & 71.3$^{+2.9}_{-2.6}$ & 2.66 \\
31109 & $\omega$ Eri & A9IV & 4.4 &--6.0 & 186 & 203.75 & --28.78 & 72.0$^{+1.3}_{-1.2}$ & 4.97 \\
32045 & S Eri & F0V & 4.8 &--8.9 & 212 & 211.94 & --30.34 & 89.0$^{+1.9}_{-1.8}$ & 4.17 \\
30127 & HR1513 & A1V & 5.5 &5.0 & 195 & 216.99 & --36.20 & 93.8$^{+3.4}_{-3.1}$ & 8.55 \\
%32964 & EN Eri & B9V & 5.1 &30.9 &  30 & 204.78 & --25.35 & 94.7$^{+3.2}_{-3.0}$ & 7.51 \\
27925 & HIP20521 & A9IV/V & 8.0 & & & 209.24 & --38.96 & 100.6$^{+10.0}_{-8.4}$ & 6.67 \\
34863 & $\nu$ Lep & B7V & 5.3 &16.0 & 285 & 214.00 & --25.79 & 102.2$^{+4.0}_{-3.7}$ & 8.61 \\
28763 & HR1438 & A3V & 6.3 &--11.9\tablenotemark{e} & 102 & 209.75 & --37.13 & 112.9$^{+6.9}_{-6.1}$ & 5.01 \\
32996 & HR1661 & B9.5/A0IV & 6.0 &25.8\tablenotemark{e} &  26 & 213.32 & --29.10 & 113.6$^{+5.7}_{-5.2}$ & 5.85 \\
%30020 & DW Eri & F4III  & 6.8 &40.0 &  65 & 205.96 & --32.37 & 124$^{+26}_{-18}$ & 1.47 \\
29554 & HIP21640 & B9 & 7.7 & & & 204.73 & --33.15 & 125.9$^{+10.5}_{-9.0}$ & 2.60 \\
28843 & DZ Eri & B9III  & 	5.8 &18.4 & & 198.62 & --32.10 & 145.8$^{+7.8}_{-7.1}$ & 7.68 \\
34503 & $\tau$ Ori & B5III  & 3.6 &  20.1 &  40 & 208.28 & --23.96 & 151.5$^{+3.5}_{-3.4}$ & 8.47 \\
31625 & HIP23011 & A5 & 6.9 &  27.7 &  77 & 207.34 & --29.25 & 155$^{+16}_{-14}$ & 3.18 \\
30211 & $\mu$ Eri & B5IV & 4.0 &   8.5 & 150 & 200.53 & --29.34 & 160.0$^{+5.0}_{-4.7}$ & 6.87 \\
30050 & RZ Eri & A & 7.9 &  32.0 & 11\tablenotemark{g} & 208.00 & --33.16 & 161$^{+39}_{-26}$ & 0.79 \\
29851 & HIP21837 & A2IV/V & 6.7 & & & 209.70 & --34.37 & 172$^{+16}_{-14}$ & 2.57 \\
28208 & HIP20747 & B9 & 7.3 &&     & 201.61 & --35.18 & 184$^{+22}_{-17}$ & 5.77 \\
30332 & HIP22169 & B9 & 7.7 & & & 209.25 & --33.00 & 185$^{+27}_{-21}$ & 1.43 \\
28980 & HIP21258 & A0 & 6.6 & & & 204.05 & --34.31 & 187$^{+15}_{-13}$ & 3.59 \\
32468 & HIP23493 & A0 & 6.8 & & & 208.30 & --28.00 & 188$^{+28}_{-22}$ & 4.44 \\
%29173 & HR1460 & 	A1 & 6.4 &15\tablenotemark{f} &  26 & 205.87 & --34.63 & 198$^{+261}_{-72}$ & 2.69 \\
29248 & $\nu$ Eri & 	B2III & 3.9 &  14.9 &  20 & 199.31 & --31.38 & 207.0$^{+8.5}_{-7.8}$ & 7.20 \\
33224 & HR1671 & B8V & 5.8 &  27.0 & 155 & 208.95 & --26.83 & 209$^{+23}_{-19}$ & 5.68 \\
33949 & $\kappa$ Lep & B7V & 4.4 &  18.0 & 120 & 213.88 & --27.55 & 223$^{+33}_{-26}$ & 7.23 \\
28377 & HIP20849 & A9IV & 7.5 & & & 207.45 & --37.20 & 225$^{+40}_{-30}$ & 4.79 \\
32249 & $\psi$ Eri & B3V & 4.8 &  25.4 &  30 & 206.59 & --27.69 & 227$^{+13}_{-11}$ & 4.82 \\
31512 & 62 Eri & B6V & 5.5 &  24.0 & 80 & 203.93 & --27.88 & 227$^{+38}_{-28}$ & 5.58 \\
30963 & HIP22588 & B9 & 7.3 & & & 208.57 & --31.24 & 240$^{+45}_{-33}$ & 1.38 \\
27563 & EM Eri & B5III  & 	5.8 & 11.2 & 35 & 201.50 & --36.83 & 242$^{+25}_{-20}$ & 6.75 \\
29009 & EH Eri & B9 & 5.7 & 1.7 &  55 & 202.47 & --33.55 & 254$^{+39}_{-30}$ & 4.53 \\
34816 & $\lambda$ Lep & B0.5IV & 	4.3 &  20.2 &  25 & 214.83 & --26.24 & 261$^{+17}_{-15}$ & 8.75 \\
34085 & $\beta$ Ori & B8I & 0.1 &  20.7 &  40 & 209.24 & --25.25 & 265$^{+26}_{-22}$ & 7.29 \\
31089 & HIP22669 & B9V & 7.8 & & & 208.90 & --31.12 & 268$^{+74}_{-48}$ & 1.64 \\
27436 & HIP20179 & A0 & 7.0 & & & 201.90 & --37.34 & 270$^{+59}_{-41}$ & 6.85 \\
%26994 & HIP19856 & B7III & 6.9 & & & 211.14 & --41.78 & 289$^{+44}_{-34}$ & 9.76 \\
28262 & HIP20787 & A0 & 8.6 & & & 203.85 & --36.03 & 316$^{+122}_{-69}$ & 4.82 \\
30535 & HIP22304 & A0 & 7.2 & & & 207.32 & --31.65 & 328$^{+79}_{-53}$ & 0.83 \\
31726 & HR1595 & B2V & 6.1 & 11.4 &   5 & 213.50 & --31.51 & 417$^{+78}_{-57}$ & 5.00 \\
28497 & DU Eri & B2V & 5.6 &  22.0 & 295 & 208.78 & --37.40 & 465$^{+73}_{-55}$ & 5.07 \\
32612 & HR1640 & 	B2IV & 6.4 & 16.0 &  65 & 214.33 & --30.21 & 570$^{+170}_{-110}$ & 6.08 \\
30679 & HIP22406 & A2 & 7.7 & & & 206.93 & --31.19 & 610$^{+330}_{-160}$ & 1.38 \\
%30076 & DX Eri & B2V & 5.8 & 15.1 & 160 & 205.72 & --32.12 & 1490$^{+\infty}_{-750}$ & 1.70 \\
\enddata
\label{stars}
\tablenotetext{a}{All values from SIMBAD unless otherwise noted.}
\tablenotetext{b}{Distances calculated from {\it Hipparcos} parallaxes.}
\tablenotetext{c}{Angular distance from direction of the past solar trajectory, $l=207.70^{\circ}$ and $b=-32.41^{\circ}$ \citep{dehnen98}.}
\tablenotetext{d}{\cite{nord}}
\tablenotetext{e}{\cite{gren}}
\tablenotetext{f}{\cite{shorlin}}
\tablenotetext{g}{\cite{staw}}
%vsini typically from 2002ApJ, 573, 359 Abt et al.
%  also Royer at al. 2002 A&A 393, 897
%radial velocity typically from Wilson general catalogue of stellar RV (1953)
%  also Wielen et al. 1999  and 2000
%  also nordstroem et al. 2004 A&A 418 989
\end{deluxetable}  	

\clearpage

\begin{deluxetable}{llllllcrr}
\tablewidth{0pt}
\tabletypesize{\scriptsize}
\tablecaption{Observational Parameters for Stars along the Past Solar Trajectory}
\tablehead{ HD & Other & Date & Telescope\tablenotemark{a} & Instrument\tablenotemark{b} & Ion & $v_{\rm atm}$ & Exposure & S/N  \\
\# & Name & & & & & km s$^{-1}$ & (s) & }
\startdata
33111 & $\beta$ Eri   & 2003 Dec 3  & McD2.7 & TS12 & \ion{Ca}{2} & & 4800 & 27 \\
33111 & $\beta$ Eri   & 2004 Oct 19 & McD2.7 & TS21 & \ion{Na}{1}  & +21.2 & 1200 & 385 \\
33111 & $\beta$ Eri   & 2004 Dec 1  & McD2.7 & TS23 & \ion{Ca}{2}, \ion{Na}{1} & +2.7 &  300 & 52, 465 \\
33111 & $\beta$ Eri   & 2005 Feb 27 & AAT3.9 & UHRF & \ion{Na}{1}  & --29.7 & 1200 & 82 \\
33111 & $\beta$  Eri  & 2005 Feb 28 & AAT3.9 & UHRF & \ion{Ca}{2} & & 1200 & 42 \\
30743 & HR1545        & 2005 Feb 14 & McD2.7 & TS21 & \ion{Na}{1}  & --28.5 & 1200 & 46 \\ 
30743 & HR1545        & 2005 Feb 14 & McD2.7 & TS21 & \ion{Ca}{2} & & 2400 & 13 \\
%26574 & $o^1$ Eri     & 2004 Dec 3  & McD2.7 & TS23 & \ion{Ca}{2}, \ion{Na}{1} & --5.4 & 3600 & 0, 8 \\
%26574 & $o^1$ Eri     & 2004 Dec 6  & McD2.7 & TS23 & \ion{Ca}{2}, \ion{Na}{1} & --6.8 &  900 & 46, 442 \\
33904 & $\mu$ Lep     & 2003 Oct 5  & McD2.7 & TS12 & \ion{Ca}{2} & & 3600 & 17 \\
33904 & $\mu$ Lep     & 2003 Dec 2  & McD2.7 & TS12 & \ion{Na}{1}  & +3.9 & 3600 & 30 \\
33904 & $\mu$ Lep     & 2004 Oct 19 & McD2.7 & TS21 & \ion{Ca}{2} & & 1200 & 117 \\
33904 & $\mu$ Lep     & 2005 Oct 17 & AAT3.9 & UHRF & \ion{Ca}{2} & & 2400 & 50 \\
27861 & $\xi$ Eri     & 2005 Mar 16 & McD2.7 & TS21 & \ion{Ca}{2} & & 1800 & 49 \\
27861 & $\xi$ Eri     & 2005 Mar 16 & McD2.7 & TS21 & \ion{Na}{1}  & --26.6 & 1200 & 72 \\
33802 & $\iota$ Lep   & 2005 Jan 28 & McD2.7 & TS12 & \ion{Ca}{2} & & 4800 & 15 \\
33802 & $\iota$ Lep   & 2005 Feb 14 & McD2.7 & TS21 & \ion{Na}{1}  & --27.4 & 1200 & 108 \\
33802 & $\iota$ Lep   & 2005 Feb 14 & McD2.7 & TS21 & \ion{Ca}{2} & & 1200 & 72 \\
33802 & $\iota$ Lep   & 2005 Feb 28 & AAT3.9 & UHRF & \ion{Ca}{2} & & 1800 & 37 \\
29573 & HR1483        & 2003 Oct 2  & McD2.7 & TS12 & \ion{Ca}{2} & & 3600 & 4 \\
29573 & HR1483        & 2004 Oct 20 & McD2.7 & TS21 & \ion{Na}{1}  & +19.2 & 1200 & 114 \\
29573 & HR1483        & 2004 Oct 20 & McD2.7 & TS21 & \ion{Ca}{2} & & 1200 & 42 \\
31109 & $\omega$ Eri  & 2004 Dec 1  & McD2.7 & TS23 & \ion{Ca}{2}, \ion{Na}{1} & +0.7 &  900 & 23, 438 \\
32045 & S Eri         & 2004 Oct 20 & McD2.7 & TS21 & \ion{Ca}{2} & & 1200 & 54 \\
32045 & S Eri         & 2004 Oct 20 & McD2.7 & TS21 & \ion{Na}{1}  & +21.1 &  600 & 86 \\
30127 & HR1513        & 2004 Oct 20 & McD2.7 & TS21 & \ion{Ca}{2} & & 1800 & 37 \\
30127 & HR1513        & 2004 Oct 20 & McD2.7 & TS21 & \ion{Na}{1}  & +20.3 & 1800 & 109 \\
%32964 & EN Eri        & 2004 Dec 1  & McD2.7 & TS23 & \ion{Ca}{2}, \ion{Na}{1} & +2.5 &  900 & 29, 248 \\
27925 & HIP20521      & 2004 Dec 6  & McD2.7 & TS23 & \ion{Ca}{2}, \ion{Na}{1} & --4.9 & 1800 & 6, 83 \\
34863 & $\nu$ Lep     & 2005 Mar 16 & McD2.7 & TS21 & \ion{Na}{1}  & --29.3 & 1200 & 56 \\
34863 & $\nu$ Lep     & 2005 Mar 16 & McD2.7 & TS21 & \ion{Ca}{2} & & 1800 & 47 \\
28763 & HR1438        & 2004 Oct 20 & McD2.7 & TS21 & \ion{Ca}{2} & & 1800 & 24 \\
28763 & HR1438        & 2004 Oct 20 & McD2.7 & TS21 & \ion{Na}{1}  & +18.7 & 1800 & 78 \\
32996 & HR1661        & 2004 Dec 2  & McD2.7 & TS23 & \ion{Ca}{2}, \ion{Na}{1} & +2.5 & 1800 & 92, 315 \\
29554 & HIP21640      & 2004 Nov 30 & McD2.7 & TS23 & \ion{Ca}{2}, \ion{Na}{1} & --0.3 & 1800 & 60, 206 \\
28843 & DZ Eri        & 2005 Mar 14 & McD2.7 & TS21 & \ion{Ca}{2} & & 2700 & 36 \\
28843 & DZ Eri        & 2005 Mar 14 & McD2.7 & TS21 & \ion{Na}{1}  & --27.2 & 1800 & 32 \\
34503 & $\tau$ Ori    & 2003 Dec 3  & McD2.7 & TS12 & \ion{Ca}{2} & & 4800 & 38 \\
34503 & $\tau$ Ori    & 2004 Oct 19 & McD2.7 & TS21 & \ion{Na}{1}  & +22.3 & 1200 & 265 \\
31625 & HIP23011      & 2004 Dec 6  & McD2.7 & TS23 & \ion{Ca}{2}, \ion{Na}{1} & --1.1 & 1800 & 4, 113 \\
30211 & $\mu$ Eri     & 2003 Dec 3  & McD2.7 & TS12 & \ion{Ca}{2} & & 3600 & 18 \\
30211 & $\mu$ Eri     & 2004 Oct 18 & McD2.7 & TS21 & \ion{Na}{1}  & +19.3 & 1200 & 131 \\
30211 & $\mu$ Eri     & 2004 Oct 18 & McD2.7 & TS21 & \ion{Ca}{2} & &  600 & 41 \\
30211 & $\mu$ Eri     & 2004 Nov 30 & McD2.7 & TS23 & \ion{Ca}{2}, \ion{Na}{1} & +0.1 &  900 & 184, 522 \\
%29173 & HR1460        & 2004 Dec 2  & McD2.7 & TS23 & \ion{Ca}{2}, \ion{Na}{1} & --1.7 & 1800 & 46, 193 \\
%30020 & DW Eri        & 2004 Oct 20 & McD2.7 & TS21 & \ion{Na}{1}  & +19.2 & 1200 & 57 \\
%30020 & DW Eri        & 2004 Oct 20 & McD2.7 & TS21 & \ion{Ca}{2} & & 1800 & 14 \\
%30020 & DW Eri        & 2004 Dec 2  & McD2.7 & TS23 & \ion{Ca}{2}, \ion{Na}{1} & --0.7 &  900 & 0, 96 \\
30050 & RZ Eri        & 2004 Dec 1  & McD2.7 & TS23 & \ion{Ca}{2}, \ion{Na}{1} & --0.0 & 1800 & 16, 156 \\
29851 & HIP21837      & 2004 Dec 2  & McD2.7 & TS23 & \ion{Ca}{2}, \ion{Na}{1} & --0.7 & 1800 & 45, 238 \\
28208 & HIP20747      & 2004 Dec 6  & McD2.7 & TS23 & \ion{Ca}{2}, \ion{Na}{1} & --5.1 & 1800 & 15, 96 \\
30332 & HIP22169      & 2004 Nov 30 & McD2.7 & TS23 & \ion{Ca}{2}, \ion{Na}{1} & +0.9 &  600 & 0, 3 \\
30332 & HIP22169      & 2004 Dec 1  & McD2.7 & TS23 & \ion{Ca}{2}, \ion{Na}{1} & +0.4 & 1800 & 40, 136 \\
28980 & HIP21258      & 2004 Dec 2  & McD2.7 & TS23 & \ion{Ca}{2}, \ion{Na}{1} & --2.0 & 1800 & 46, 181 \\
32468 & HIP23493      & 2005 Feb 14 & McD2.7 & TS21 & \ion{Ca}{2} & & 1800 & 21 \\
32468 & HIP23493      & 2005 Feb 14 & McD2.7 & TS21 & \ion{Na}{1}  & --27.5 & 1200 & 28 \\
29248 & $\nu$ Eri     & 2003 Dec 1  & McD2.7 & TS12 & \ion{Ca}{2} & & 3600 & 36 \\
29248 & $\nu$ Eri     & 2003 Dec 2  & McD2.7 & TS12 & \ion{Na}{1}  & --1.5 & 3600 & 25 \\
33224 & HR1671        & 2004 Nov 30 & McD2.7 & TS23 & \ion{Ca}{2}, \ion{Na}{1} & +3.5 &  900 & 38, 186 \\
33949 & $\kappa$ Lep  & 2004 Dec 3  & McD2.7 & TS23 & \ion{Ca}{2}, \ion{Na}{1} & +2.8 & 1800 & 0, 7 \\
33949 & $\kappa$ Lep  & 2005 Feb 14 & McD2.7 & TS21 & \ion{Ca}{2} & & 1200 & 68 \\
33949 & $\kappa$ Lep  & 2005 Feb 14 & McD2.7 & TS21 & \ion{Na}{1}  & --27.4 & 1200 & 93 \\
33949 & $\kappa$ Lep  & 2005 Feb 27 & AAT3.9 & UHRF & \ion{Na}{1}  & --29.7 & 2400 & 30 \\
33949 & $\kappa$ Lep  & 2005 Feb 28 & AAT3.9 & UHRF & \ion{Ca}{2} & & 1800 & 37 \\
28377 & HIP20849      & 2004 Dec 6  & McD2.7 & TS23 & \ion{Ca}{2}, \ion{Na}{1} & --4.5 & 1800 & 8, 101 \\
32249 & $\psi$ Eri    & 2003 Dec 1  & McD2.7 & TS12 & \ion{Ca}{2} & & 4800 & 19 \\
32249 & $\psi$ Eri    & 2003 Dec 2  & McD2.7 & TS12 & \ion{Na}{1}  & +1.9 & 4800 & 19 \\
32249 & $\psi$ Eri    & 2004 Nov 30 & McD2.7 & TS23 & \ion{Ca}{2}, \ion{Na}{1} & +2.5 &  900 & 164, 398 \\
31512 & 62 Eri        & 2004 Nov 30 & McD2.7 & TS23 & \ion{Ca}{2}, \ion{Na}{1} & +1.7 &  900 & 71, 240 \\
30963 & HIP22588      & 2004 Nov 30 & McD2.7 & TS23 & \ion{Ca}{2}, \ion{Na}{1} & +1.5 & 1800 & 66, 241 \\
27563 & EM Eri        & 2004 Dec 1  & McD2.7 & TS23 & \ion{Ca}{2}, \ion{Na}{1} & --3.2 &  900 & 74, 273 \\
29009 & EH Eri        & 2004 Dec 1  & McD2.7 & TS23 & \ion{Ca}{2}, \ion{Na}{1} & --1.6 &  900 & 71, 275 \\
34816 & $\lambda$ Lep & 2003 Dec 1  & McD2.7 & TS12 & \ion{Ca}{2} & & 4800 & 16 \\
34816 & $\lambda$ Lep & 2003 Dec 2  & McD2.7 & TS12 & \ion{Na}{1}  & +4.6 & 4800 & 25 \\
34085 & $\beta$ Ori   & 2003 Dec 1  & McD2.7 & TS12 & \ion{Ca}{2} & & 2400 & 119 \\
34085 & $\beta$ Ori   & 2003 Dec 2  & McD2.7 & TS12 & \ion{Na}{1}  & +3.7 & 2400 & 143 \\
34085 & $\beta$ Ori   & 2004 Dec 2  & McD2.7 & TS23 & \ion{Ca}{2}, \ion{Na}{1} & +3.3 &   10 & 105, 431 \\
34085 & $\beta$ Ori   & 2005 Feb 27 & AAT3.9 & UHRF & \ion{Na}{1}  & --29.7 &  900 & 316 \\
34085 & $\beta$ Ori   & 2005 Feb 28 & AAT3.9 & UHRF & \ion{Ca}{2} & &  900 & 246 \\
34085 & $\beta$ Ori   & 2005 Mar 22 & McD2.7 & TS12 & \ion{Ca}{2} & & 2400 & 129 \\
34085 & $\beta$ Ori   & 2006 Feb 17 & McD2.7 & TS21 & \ion{Ca}{2} & & 1200 & 107 \\
34085 & $\beta$ Ori   & 2006 Feb 17 & McD2.7 & TS21 & \ion{Na}{1}  & --27.4 & 600  & 346 \\
31089 & HIP22669      & 2004 Dec 1  & McD2.7 & TS23 & \ion{Ca}{2}, \ion{Na}{1} & +1.1 & 1800 & 41, 143 \\
27436 & HIP20179      & 2004 Dec 6  & McD2.7 & TS23 & \ion{Ca}{2}, \ion{Na}{1} & --5.9 & 1800 & 15, 102 \\
%26994 & HIP19856      & 2004 Dec 6  & McD2.7 & TS23 & \ion{Ca}{2}, \ion{Na}{1} & --5.7 & 1800 & 46, 142 \\
28262 & HIP20787      & 2004 Dec 6  & McD2.7 & TS23 & \ion{Ca}{2}, \ion{Na}{1} & --4.9 & 2700 & 9, 65 \\
30535 & HIP22304      & 2004 Dec 1  & McD2.7 & TS23 & \ion{Ca}{2}, \ion{Na}{1} & +0.4 & 1800 & 29, 178 \\
31726 & HR1595        & 2004 Dec 1  & McD2.7 & TS23 & \ion{Ca}{2}, \ion{Na}{1} & +2.0 & 1800 & 122, 301 \\
28497 & DU Eri        & 2004 Oct 19 & McD2.7 & TS21 & \ion{Ca}{2} & & 1200 & 52 \\
28497 & DU Eri        & 2004 Oct 19 & McD2.7 & TS21 & \ion{Na}{1}  & +18.8 & 1200 & 98 \\
28497 & DU Eri        & 2004 Nov 30 & McD2.7 & TS23 & \ion{Ca}{2}, \ion{Na}{1} & --1.2 &  900 & 70, 228 \\
32612 & HR1640        & 2004 Dec 2  & McD2.7 & TS23 & \ion{Ca}{2}, \ion{Na}{1} & +2.3 & 1800 & 86, 284 \\
30679 & HIP22406      & 2004 Dec 2  & McD2.7 & TS23 & \ion{Ca}{2}, \ion{Na}{1} & +0.1 & 1800 & 17, 190 \\
%30076 & DX Eri        & 2003 Dec 1  & McD2.7 & TS12 & \ion{Ca}{2} & & 7200 & 8 \\
%30076 & DX Eri        & 2003 Dec 2  & McD2.7 & TS12 & \ion{Na}{1}  & --0.2 & 3600 & 6 \\
%30076 & DX Eri        & 2004 Nov 30 & McD2.7 & TS23 & \ion{Ca}{2}, \ion{Na}{1} & +0.4 &  900 & 101, 300 \\
\enddata
\label{obsParams}
\tablenotetext{a}{McD2.7: the Harlan J. Smith 2.7m Telescope at McDonald Observatory; AAT3.9: the Anglo-Australian 3.9m Telescope at the Anglo-Australian Observatory.}
\tablenotetext{b}{TS12: Coud\'{e} double-pass Spectrometer ($R \sim $400,000); TS21: Cross-Dispersed Echelle Spectrometer (2D Coud\'{e}) Focus 1 ($R \sim $240,000); TS23: Cross-Dispersed Echelle Spectrometer (2D Coud\'{e}) Focus 3 ($R \sim $60,000); UHRF: Ultra-High-Resolution Facility ($R \sim $940,000).}
\end{deluxetable}  

\clearpage

\begin{deluxetable}{llclcccc}
\tablewidth{0pt}
\tabletypesize{\scriptsize}
\tablecaption{\ion{Ca}{2} ISM Fit Parameters for Targets along the Past Solar Trajectory\label{CAallcomps}}
\tablehead{ HD & Other & Component & Instrument &$v$ & $b$ & $\log N$ & Component \# of \\
\#  & Name & \# & & (km s$^{-1}$) & (km s$^{-1}$) & (cm$^{-2}$) & Pair in \ion{Na}{1} }
\startdata
29554 & HIP21640 & 1 & TS23 & 7.8 $\pm$ 4.1 & 19.3 $\pm$ 4.5 & 11.32 $^{+0.11}_{-0.12}$ & \\
29554 & HIP21640 & 2 & TS23 & 22.548 $\pm$ 0.091 & 3.65 $\pm$ 0.20 & 11.117 $^{+0.011}_{-0.011}$ & 2 \\
28843 & DZ Eri & 1 & TS21 & --6.212 $\pm$ 0.21 & 2.90 $\pm$ 0.95 & 11.25 $^{+0.13}_{-0.19}$ & 1 \\
28843 & DZ Eri & 2 & TS21 & 22.13 $\pm$ 0.93 & 2.7 $\pm$ 1.1 & 11.451 $^{+0.073}_{-0.089}$ & 3 \\
28843 & DZ Eri & 3 & TS21 & 31.2 $\pm$ 2.4 & 5.6 $\pm$ 2.6 & 11.5 $^{+0.3}_{-1.2}$ & \\
34503 & $\tau$ Ori & 1 & TS12 & 23.266 $\pm$ 0.091 & 1.80 $\pm$ 0.13 & 10.681 $\pm$0.021 & 2 \\
34503 & $\tau$ Ori & 2 & TS12 & 30.23 $\pm$ 0.11 & 1.64 $\pm$ 0.14 & 10.562 $\pm$0.028 & \\

30211 & $\mu$ Eri & 1 & TS12 & 24.715 $\pm$ 0.037 & 2.055 $\pm$ 0.041 & 11.489 $\pm$0.0094 & 1 \\
30050 & RZ Eri & 1 & TS23 & 11.0 $\pm$ 2.2 & 14.5 $\pm$ 3.0 & 12.305 $^{+0.061}_{-0.071}$ & \\
30050 & RZ Eri & 2 & TS23 & 23.75 $\pm$ 0.83 & 3.5 $\pm$ 1.7 & 11.65 $^{+0.21}_{-0.44}$ & 1 \\
29851 & HIP21837 & 1 & TS23 & 23.36 $\pm$ 0.36 & 6.2 $\pm$ 3.9 & 11.26 $^{+0.14}_{-0.20}$ & \\
28208 & HIP20747 & 1 & TS23 & 23.8 $\pm$ 1.6 & 7.3 $\pm$ 1.1 & 11.30 $^{+0.19}_{-0.33}$ & \\
30332 & HIP22169 & 1 & TS23 & 8.4 $\pm$ 1.5 & 6.01 $\pm$ 0.68 & 11.448 $^{+0.032}_{-0.033}$ & \\
30332 & HIP22169 & 2 & TS23 & 24.35 $\pm$ 0.22 & 6.5 $\pm$ 2.0 & 11.499 $^{+0.058}_{-0.067}$ & 2 \\
28980 & HIP21258 & 1 & TS23 & 13.15 $\pm$ 0.95 & 10.2 $\pm$ 2.6 & 11.31 $^{+0.12}_{-0.17}$ & \\
32468 & HIP23493 & 1 & TS21 & 17.01 $\pm$ 0.86 & 2.548 $\pm$ 0.032 & 11.29 $^{+0.15}_{-0.20}$ & 1 \\
32468 & HIP23493 & 2 & TS21 & 23.10 $\pm$ 0.47 & 4.3 $\pm$ 5.1 & 11.36 $\pm$0.41 & 2 \\
29248 & $\nu$ Eri & 1 & TS12 & 23.190 $\pm$ 0.015 & 0.912 $\pm$ 0.016 & 11.213 $\pm$0.0057 & 1 \\
33224 & HR1671 & 1 & TS23 & 15.22 $\pm$ 0.40 & 5.7 $\pm$ 1.9 & 11.42 $^{+0.13}_{-0.19}$ & 1 \\
33224 & HR1671 & 2 & TS23 & 20.6 $\pm$ 1.4 & 11.7 $\pm$ 1.6 & 11.57 $^{+0.13}_{-0.14}$ & \\
33949 & $\kappa$ Lep & 1 & UHRF & 15.08 $\pm$ 0.42 & 2.45 $\pm$ 0.58 & 10.516 $\pm$0.076 & 1 \\
33949 & $\kappa$ Lep & 2 & UHRF & 23.46 $\pm$ 0.25 & 2.32 $\pm$ 0.34 & 10.693 $\pm$0.058 & 2 \\
33949 & $\kappa$ Lep & 3 & UHRF & 32.00 $\pm$ 0.34 & 1.38 $\pm$ 0.67 & 10.391 $\pm$0.081 & \\
32249 & $\psi$ Eri & 1 & TS12 & 14.03 $\pm$ 0.15 & 9.19 $\pm$ 0.22 & 11.715 $\pm$0.0091 & \\
31512 & 62 Eri & 1 & TS23 & 2.9 $\pm$ 1.2 & 3.82 $\pm$ 0.90 & 10.87 $^{+0.10}_{-0.12}$ & \\
31512 & 62 Eri & 2 & TS23 & 20.33 $\pm$ 0.45 & 6.5 $\pm$ 2.2 & 11.25 $^{+0.12}_{-0.16}$ & 4 \\
30963 & HIP22588 & 1 & TS23 & --12.47 $\pm$ 0.092 & 13.07 $\pm$ 0.24 & 11.909 $^{+0.029}_{-0.031}$ & 2 \\
30963 & HIP22588 & 2 & TS23 & 14.4 $\pm$ 1.9 & 16.89 $\pm$ 0.71 & 12.206 $^{+0.042}_{-0.047}$ & 3 \\
27563 & EM Eri & 1 & TS23 & 21.21 $\pm$ 0.18 & 3.2 $\pm$ 2.3 & 11.18 $^{+0.13}_{-0.18}$ & 2 \\
29009 & EH Eri & 1 & TS23 & 0.81 $\pm$ 0.59 & 3.5 $\pm$ 2.0 & 10.81 $^{+0.20}_{-0.37}$ & 1 \\
29009 & EH Eri & 2 & TS23 & 19.52 $\pm$ 0.18 & 3.53 $\pm$ 0.57 & 10.87 $^{+0.23}_{-0.53}$ & 3 \\
34816 & $\lambda$ Lep & 1 & TS12 & 8.61 $\pm$ 0.21 & 17.65 $\pm$ 0.28 & 12.086 $\pm$0.0062 & \\
34085 & $\beta$ Ori & 1 & UHRF & --29.20 $\pm$ 0.061 & 1.645 $\pm$ 0.076 & 10.157 $\pm$0.019 & \\
34085 & $\beta$ Ori & 2 & UHRF & --19.48 $\pm$ 0.45 & 2.27 $\pm$ 0.37 & 10.241 $\pm$0.089 & \\
34085 & $\beta$ Ori & 3 & UHRF & --15.32 $\pm$ 0.071 & 2.201 $\pm$ 0.085 & 10.834 $\pm$0.036 & \\
34085 & $\beta$ Ori & 4 & UHRF & --0.271 $\pm$ 0.024 & 2.312 $\pm$ 0.032 & 10.871 $\pm$0.0052 & 1 \\
34085 & $\beta$ Ori & 5 & UHRF & 9.389 $\pm$ 0.019 & 1.790 $\pm$ 0.026 & 10.857 $\pm$0.0061 & 2 \\
34085 & $\beta$ Ori & 6 & UHRF & 17.898 $\pm$ 0.076 & 1.765 $\pm$ 0.095 & 10.441 $\pm$0.021 & 3 \\
34085 & $\beta$ Ori & 7 & UHRF & 22.867 $\pm$ 0.060 & 2.56 $\pm$ 0.11 & 10.698 $\pm$0.014 & 4 \\
34085 & $\beta$ Ori & 8 & UHRF & 30.238 $\pm$ 0.076 & 2.73 $\pm$ 0.11 & 10.479 $\pm$0.015 & 5 \\
31089 & HIP22669 & 1 & TS23 & 11.98 $\pm$ 0.57 & 20.3 $\pm$ 4.6 & 12.180 $^{+0.017}_{-0.017}$ & \\
31089 & HIP22669 & 2 & TS23 & 18.70 $\pm$ 0.17 & 6.58 $\pm$ 0.37 & 12.236 $^{+0.026}_{-0.027}$ & 1 \\
%26994 & HIP19856 & 1 & TS23 & --4.70 $\pm$ 0.25 & 4.02 $\pm$ 0.43 & 11.68 $^{+0.10}_{-0.13}$ & \\
%26994 & HIP19856 & 2 & TS23 & 5.542 $\pm$ 0.047 & 5.35 $\pm$ 0.34 & 11.913 $^{+0.025}_{-0.027}$ & \\
%26994 & HIP19856 & 3 & TS23 & 2.60 $\pm$ 0.97 & 15.2 $\pm$ 1.9 & 11.78 $^{+0.11}_{-0.15}$ & \\
30535 & HIP22304 & 1 & TS23 & --2.06 $\pm$ 0.19 & 9.07 $\pm$ 0.90 & 12.159 $^{+0.021}_{-0.022}$ & 1 \\
30535 & HIP22304 & 2 & TS23 & 18.56 $\pm$ 0.41 & 10.70 $\pm$ 0.61 & 12.227 $^{+0.032}_{-0.034}$ & 2 \\
31726 & HR1595 & 1 & TS23 & --7.60 $\pm$ 0.57 & 5.1 $\pm$ 1.9 & 11.33 $^{+0.12}_{-0.17}$ & 1 \\
31726 & HR1595 & 2 & TS23 & 18.57 $\pm$ 0.78 & 11.15 $\pm$ 0.50 & 12.108 $^{+0.055}_{-0.063}$ & 4 \\
28497 & DU Eri & 1 & TS23 & --26.41 $\pm$ 0.019 & 2.18 $\pm$ 0.82 & 11.06 $^{+0.15}_{-0.23}$ & 1 \\
28497 & DU Eri & 2 & TS23 & --11.3 $\pm$ 2.6 & 5.1 $\pm$ 1.0 & 11.858 $^{+0.045}_{-0.051}$ & 3 \\
28497 & DU Eri & 3 & TS23 & --10.2 $\pm$ 4.9 & 3.10 $\pm$ 0.69 & 11.27 $^{+0.26}_{-0.75}$ & \\
28497 & DU Eri & 4 & TS23 & 10.81 $\pm$ 0.20 & 4.9 $\pm$ 1.6 & 11.151 $^{+0.086}_{-0.099}$ & 7 \\
28497 & DU Eri & 5 & TS23 & 20.53 $\pm$ 0.46 & 4.70 $\pm$ 0.46 & 11.18 $^{+0.11}_{-0.12}$ & 8 \\
28497 & DU Eri & 6 & TS23 & 29.55 $\pm$ 0.67 & 3.00 $\pm$ 0.12 & 11.015 $^{+0.039}_{-0.040}$ & 10 \\
%30076 & 1 & TS23 & --20.40 $\pm$ 0.12 & 1.38 $\pm$ 0.88 & 11.31 $^{+0.29}_{-0.48}$ & \\
%30076 & 2 & TS23 & --14.01 $\pm$ 0.86 & 5.0 $\pm$ 1.1 & 11.70 $^{+0.10}_{-0.10}$ & \\
%30076 & 3 & TS23 & 0.36 $\pm$ 0.77 & 4.4 $\pm$ 1.2 & 11.051 $^{+0.048}_{-0.051}$ & \\
%30076 & 4 & TS23 & 13.50 $\pm$ 0.38 & 3.20 $\pm$ 0.89 & 11.330 $^{+0.072}_{-0.080}$ & \\
%30076 & 5 & TS23 & 20.43 $\pm$ 0.72 & 1.10 $\pm$ 0.24 & 10.74 $^{+0.12}_{-0.16}$ & \\
%30076 & 6 & TS23 & 26.82 $\pm$ 0.25 & 7.45 $\pm$ 0.73 & 11.703 $^{+0.032}_{-0.032}$ & \\
%30076 & 7 & TS23 & 46.73 $\pm$ 0.79 & 5.85 $\pm$ 0.94 & 11.291 $^{+0.023}_{-0.024}$ & \\
32612 & HR1640 & 1 & TS23 & 4.1 $\pm$ 9.7 & 13.4 $\pm$ 4.9 & 11.19 $^{+0.29}_{-0.73}$ & \\
32612 & HR1640 & 2 & TS23 & 24.9 $\pm$ 3.0 & 12.9 $\pm$ 2.3 & 11.66 $^{+0.15}_{-0.21}$ & 6 \\
%32612 & HR1640 & 3 & TS23 & 63.3 $\pm$ 1.3 & 2.9 $\pm$ 1.1 & 10.466 $^{+0.079}_{-0.097}$ & \\
30679 & HIP22406 & 1 & TS23 & --14.5 $\pm$ 1.5 & 4.80 $\pm$ 0.43 & 11.816 $^{+0.021}_{-0.022}$ & \\
30679 & HIP22406 & 2 & TS23 & 24.0 $\pm$ 3.8 & 7.1 $\pm$ 2.2 & 11.38 $^{+0.13}_{-0.18}$ & 4 \\
\enddata
\tablecomments{This table only lists those sight lines with detected \ion{Ca}{2} absorption. The upper limits for sight lines with no detected absorption are given in Table~\ref{CAtotecomps}.}
\end{deluxetable}

\clearpage

\begin{deluxetable}{llclcl}
\tablewidth{0pt}
\tabletypesize{\scriptsize}
\tablecaption{\ion{Ca}{2} ISM Fit Parameters Total Column Density Per Sight Line\label{CAtotecomps}}
\tablehead{ HD & Other & Components &Instrument& $\log N$ & Other \\
\# & Name & & & (cm$^{-2}$) & References }
\startdata
33111 & $\beta$ Eri &&UHRF& $<$10.6 & 1 \\
30743 & HR1545 &&TS21& $<$11.8 & \nodata \\
%26574 & $o^1$ Eri &&TS23& $<$11.1\\
33904 & $\mu$ Lep &&UHRF& $<$10.4 & \nodata \\
27861 & $\xi$ Eri&&TS21& $<$11.1 & \nodata \\
33802 & $\iota$ Lep &&TS21& $<$10.5 & 1 \\
29573 & HR1483 &&TS21& $<$11.4 & 2 \\
31109 & $\omega$ Eri &&TS23& $<$11.5 & \nodata \\
32045 & S Eri &&TS21& $<$11.7 & \nodata \\
30127 & HR1513 &&TS21& $<$11.1 & \nodata \\
27925 & HIP20521 &&TS23& $<$12.2 & \nodata \\
34863 & $\nu$ Lep &&TS21& $<$11.0 & 1 \\
28763 & HR1438 &&TS21& $<$11.6 & \nodata \\
32996 & HR1661 &&TS23& $<$10.8 & \nodata \\
29554 & HIP21640 & 2 & TS23 & 11.534 $^{+0.061}_{-0.084}$ & \nodata \\
28843 & DZ Eri & 3 & TS21 & 11.91 $\pm$0.11 & \nodata \\
34503 & $\tau$ Ori & 2 & TS12 & 10.927 $\pm$0.017 & 1 \\
31625 & HIP23011 &&TS23& $<$12.4 & \nodata \\
30211 & $\mu$ Eri & 1 & TS12 & 11.489 $\pm$0.0094 & 3 \\
30050 & RZ Eri & 2 & TS23 & 12.39 $^{+0.06}_{-0.11}$ & \nodata \\
29851 & HIP21837 & 1 & TS23 & 11.26 $^{+0.12}_{-0.27}$ & \nodata \\
28208 & HIP20747 & 1 & TS23 & 11.30 $^{+0.15}_{-0.63}$ & \nodata \\
30332 & HIP22169 & 2 & TS23 & 11.775 $^{+0.033}_{-0.041}$ & \nodata \\
28980 & HIP21258 & 1 & TS23 & 11.31 $^{+0.10}_{-0.22}$ & \nodata \\
32468 & HIP23493 & 2 & TS21 & 11.63 $\pm$0.19 & \nodata \\
29248 & $\nu$ Eri & 1 & TS12 & 11.213 $\pm$0.0057 & 3,4,5,6,7 \\
33224 & HR1671 & 2 & TS23 & 11.80 $^{+0.08}_{-0.14}$ & \nodata \\
33949 & $\kappa$ Lep & 3 & UHRF & 11.028 $\pm$0.040 & 1 \\
28377 & HIP20849 &&TS23& $<$12.2 & \nodata \\
32249 & $\psi$ Eri & 1 & TS12 & 11.715 $\pm$0.0091 & 1,3 \\
31512 & 62 Eri & 2 & TS23 & 11.40 $^{+0.08}_{-0.14}$ & \nodata \\
30963 & HIP22588 & 2 & TS23 & 12.384 $^{+0.029}_{-0.034}$ & \nodata \\
27563 & EM Eri & 1 & TS23 & 11.18 $^{+0.11}_{-0.23}$ & 3 \\
29009 & EH Eri & 2 & TS23 & 11.14 $^{+0.13}_{-0.63}$ & \nodata \\
34816 & $\lambda$ Lep & 1 & TS12 & 12.086 $\pm$0.0062 & 1,8\\
34085 & $\beta$ Ori & 8 & UHRF & 11.549 $\pm$0.011 & 7,9,10,11\\
31089 & HIP22669 & 2 & TS23 & 12.510 $^{+0.015}_{-0.017}$ & \nodata \\
27436 & HIP20179 &&TS23& $<$11.7 & \nodata \\
%26994 & HIP19856 & 3 & TS23 & 12.277 $^{+0.043}_{-0.064}$ \\
28262 & HIP20787 &&TS23& $<$12.0 & \nodata \\
30535 & HIP22304 & 2 & TS23 & 12.496 $^{+0.019}_{-0.022}$ & \nodata \\
31726 & HR1595 & 2 & TS23 & 12.175 $^{+0.047}_{-0.064}$ & 3,12\\
%30076 & 7 & TS23 & 12.253 $^{+0.043}_{-0.069}$ \\
28497 & DU Eri & 6 & TS23 & 12.15 $^{+0.04}_{-0.12}$ & 3,7,13,14,15,\\
32612 & HR1640 & 3 & TS23 & 11.79 $^{+0.14}_{-0.19}$ & 12 \\
30679 & HIP22406 & 2 & TS23 & 11.952 $^{+0.036}_{-0.055}$ & \nodata \\
\enddata
\tablerefs{(1) \citealt{frisch90b}; (2) \citealt{holweger99}; (3) \citealt{welsh05b}; (4) \citealt{adams49}; (5) \citealt{burbidge53}; (6) \citealt{munch61}; (7) \citealt{hobbs84}; (8) \citealt{habing69}; (9) \citealt{hobbs74}; (10) \citealt{beintema75}; (11) \citealt{price01}; (12) \citealt{albert93}; (13) \citealt{shull77}; (14) \citealt{penprase93}; (15) \citealt{blades97}.}
\label{CaTotesN}
\end{deluxetable}

\clearpage

\begin{deluxetable}{llclcccc}
\tablewidth{0pt}
\tabletypesize{\scriptsize}
\tablecaption{\ion{Na}{1} ISM Fit Parameters for Targets along the Past Solar Trajectory\label{NAallcomps}}
\tablehead{ HD & Other & Component & Instrument & $v$ & $b$ & $\log N$ & Component \# of \\
\#  & Name & \# & &(km s$^{-1}$) & (km s$^{-1}$) & (cm$^{-2}$) & Pair in \ion{Ca}{2} }
\startdata
28763 & HR1438 & 1 & TS21 & 15.491 $\pm$ 0.050 & 1.302 $\pm$ 0.040 & 10.759 $^{+0.0041}_{-0.0042}$ & \\
%30020 & 1 & TS21 & 13.088 $\pm$ 0.065 & 0.492 $\pm$ 0.096 & 10.741 $^{+0.0045}_{-0.0046}$ & \\
%30020 & 2 & TS21 & 20.810 $\pm$ 0.074 & 1.00 $\pm$ 0.36 & 10.906 $^{+0.053}_{-0.060}$ & \\
29554 & HIP21640 & 1 & TS23 & 18.639 $\pm$ 0.045 & 0.482 $\pm$ 0.057 & 11.089 $^{+0.028}_{-0.030}$ & \\
29554 & HIP21640 & 2 & TS23 & 21.58 $\pm$ 0.11 & 1.43 $\pm$ 0.13 & 12.183 $^{+0.049}_{-0.056}$ & 2 \\
28843 & DZ Eri & 1 & TS21 & --6.13 $\pm$ 0.23 & 0.58 $\pm$ 0.89 & 10.590 $^{+0.086}_{-0.11}$ & 1 \\
28843 & DZ Eri & 2 & TS21 & 23.08 $\pm$ 0.21 & 4.78 $\pm$ 0.96 & 11.466 $^{+0.079}_{-0.096}$ & \\
28843 & DZ Eri & 3 & TS21 & 23.148 $\pm$ 0.049 & 0.78 $\pm$ 0.14 & 12.13 $^{+0.15}_{-0.22}$ & 2 \\
34503 & $\tau$ Ori & 1 & TS21 & 14.783 $\pm$ 0.045 & 0.63 $\pm$ 0.37 & 9.9560 $^{+0.037}_{-0.040}$ & \\
34503 & $\tau$ Ori & 2 & TS21 & 22.741 $\pm$ 0.0067 & 1.130 $\pm$ 0.018 & 11.107 $^{+0.0058}_{-0.0059}$ & 1 \\
34503 & $\tau$ Ori & 3 & TS21 & 22.487 $\pm$ 0.035 & 5.95 $\pm$ 0.14 & 10.889 $^{+0.021}_{-0.022}$ & \\
30211 & $\mu$ Eri & 1 & TS21 & 23.419 $\pm$ 0.030 & 0.83 $\pm$ 0.13 & 12.11 $^{+0.16}_{-0.20}$ & 1 \\
30050 & RZ Eri & 1 & TS23 & 23.912 $\pm$ 0.024 & 2.21 $\pm$ 0.22 & 12.33 $^{+0.10}_{-0.11}$ & 2 \\
29851 & HIP21837 & 1 & TS23 & 19.362 $\pm$ 0.051 & 1.53 $\pm$ 0.21 & 11.794 $^{+0.044}_{-0.049}$ & \\
28208 & HIP20747 & 1 & TS23 & 20.603 $\pm$ 0.048 & 2.05 $\pm$ 0.21 & 11.857 $^{+0.035}_{-0.038}$ & \\
30332 & HIP22169 & 1 & TS23 & 5.79 $\pm$ 0.50 & 8.3 $\pm$ 1.0 & 11.134 $^{+0.044}_{-0.049}$ & \\
30332 & HIP22169 & 2 & TS23 & 24.23 $\pm$ 0.27 & 2.33 $\pm$ 0.45 & 11.85 $^{+0.11}_{-0.13}$ & 2 \\
30332 & HIP22169 & 3 & TS23 & 27.49 $\pm$ 0.21 & 0.318 $\pm$ 0.093 & 11.441 $^{+0.060}_{-0.069}$ & \\
28980 & HIP21258 & 1 & TS23 & 20.29 $\pm$ 0.14 & 1.91 $\pm$ 0.70 & 11.40 $^{+0.10}_{-0.14}$ & \\
32468 & HIP23493 & 1 & TS21 & 16.657 $\pm$ 0.090 & 1.61 $\pm$ 0.15 & 11.197 $^{+0.019}_{-0.020}$ & 1 \\
32468 & HIP23493 & 2 & TS21 & 23.45 $\pm$ 0.24 & 1.96 $\pm$ 0.85 & 11.04 $^{+0.10}_{-0.14}$ & 2 \\
29248 & $\nu$ Eri & 1 & TS12 & 23.15 $\pm$ 0.11 & 0.98 $\pm$ 0.11 & 11.924 $\pm$ 0.099& 1 \\
29248 & $\nu$ Eri & 2 & TS12 & 25.06 $\pm$ 0.36 & 1.31 $\pm$ 0.24 & 11.67 $\pm$ 0.13& \\
33224 & HR1671 & 1 & TS23 & 15.11 $\pm$ 0.12 & 1.00 $\pm$ 0.15 & 11.729 $^{+0.037}_{-0.040}$ & 1 \\
33224 & HR1671 & 2 & TS23 & 17.512 $\pm$ 0.067 & 4.84 $\pm$ 0.12 & 11.815 $^{+0.0033}_{-0.0033}$ & \\
33949 & $\kappa$ Lep & 1 & TS21 & 16.474 $\pm$ 0.067 & 1.622 $\pm$ 0.064 & 10.921 $^{+0.015}_{-0.016}$ & 1 \\
33949 & $\kappa$ Lep & 2 & TS21 & 22.755 $\pm$ 0.052 & 0.337 $\pm$ 0.025 & 11.330 $^{+0.030}_{-0.032}$ & 2 \\
33949 & $\kappa$ Lep & 3 & TS21 & 23.772 $\pm$ 0.067 & 2.10 $\pm$ 0.13 & 11.320 $^{+0.029}_{-0.031}$ & \\
28377 & HIP20849 & 1 & TS23 & --28.11 $\pm$ 0.70 & 4.51 $\pm$ 0.36 & 10.853 $^{+0.093}_{-0.12}$ & \\
28377 & HIP20849 & 2 & TS23 & 13.40 $\pm$ 0.46 & 2.30 $\pm$ 0.36 & 10.777 $^{+0.021}_{-0.022}$ & \\
32249 & $\psi$ Eri & 1 & TS12 & 18.484 $\pm$ 0.017 & 1.541 $\pm$ 0.031 & 11.869 $\pm$0.0052 & \\
32249 & $\psi$ Eri & 2 & TS12 & 22.697 $\pm$ 0.045 & 0.861 $\pm$ 0.081 & 11.152 $\pm$0.017 & \\
31512 & 62 Eri & 1 & TS23 & --13.45 $\pm$ 0.25 & 1.62 $\pm$ 0.80 & 10.29 $^{+0.11}_{-0.16}$ & \\
31512 & 62 Eri & 2 & TS23 & --0.370 $\pm$ 0.25 & 2.43 $\pm$ 0.84 & 10.462 $^{+0.055}_{-0.063}$ & \\
31512 & 62 Eri & 3 & TS23 & 19.7 $\pm$ 1.3 & 8.41 $\pm$ 2.0 & 10.90 $^{+0.15}_{-0.24}$ & \\
31512 & 62 Eri & 4 & TS23 & 22.80 $\pm$ 0.36 & 0.49 $\pm$ 0.87 & 11.60 $^{+0.26}_{-0.36}$ & 2 \\
30963 & HIP22588 & 1 & TS23 & --24.2 $\pm$ 1.8 & 1.5 $\pm$ 1.1 & 10.268 $^{+0.050}_{-0.050}$ & \\
30963 & HIP22588 & 2 & TS23 & --13.64 $\pm$ 0.53 & 5.25 $\pm$ 0.28 & 10.727 $^{+0.074}_{-0.090}$ & 1 \\
30963 & HIP22588 & 3 & TS23 & 11.56 $\pm$ 0.33 & 18.97 $\pm$ 0.77 & 11.503 $^{+0.069}_{-0.082}$ & 2 \\
30963 & HIP22588 & 4 & TS23 & 21.411 $\pm$ 0.026 & 1.56 $\pm$ 0.16 & 11.433 $^{+0.038}_{-0.041}$ & \\
27563 & EM Eri & 1 & TS23 & --6.36 $\pm$ 0.51 & 5.8 $\pm$ 3.7 & 10.54 $^{+0.14}_{-0.22}$ & \\
27563 & EM Eri & 2 & TS23 & 18.893 $\pm$ 0.094 & 2.53 $\pm$ 0.67 & 11.59 $^{+0.27}_{-0.77}$ & \\
27563 & EM Eri & 3 & TS23 & 19.19 $\pm$ 0.22 & 0.48 $\pm$ 0.19 & 12.11 $\pm$0.81 & 1 \\
29009 & EH Eri & 1 & TS23 & 0.9 $\pm$ 1.0 & 5.8 $\pm$ 1.2 & 10.658 $^{+0.015}_{-0.016}$ & 1 \\
29009 & EH Eri & 2 & TS23 & 18.3 $\pm$ 1.0 & 1.49 $\pm$ 0.72 & 10.76 $^{+0.25}_{-0.67}$ & \\
29009 & EH Eri & 3 & TS23 & 22.36 $\pm$ 0.21 & 1.80 $\pm$ 0.29 & 11.685 $^{+0.042}_{-0.046}$ & 2 \\
34816 & $\lambda$ Lep & 1 & TS12 & 23.208 $\pm$ 0.066 & 2.14 $\pm$ 0.11 & 11.184 $\pm$0.016 & \\
34085 & $\beta$ Ori & 1 & TS12 & 0.546 $\pm$ 0.054 & 3.524 $\pm$ 0.066 & 10.924 $\pm$0.0095 & 4 \\
34085 & $\beta$ Ori & 2 & TS12 & 9.429 $\pm$ 0.075 & 5.23 $\pm$ 0.17 & 10.972 $\pm$0.011 & 5 \\
34085 & $\beta$ Ori & 3 & TS12 & 18.941 $\pm$ 0.011 & 2.310 $\pm$ 0.019 & 11.263 $\pm$0.0030 & 6 \\
34085 & $\beta$ Ori & 4 & TS12 & 24.450 $\pm$ 0.021 & 1.73 $\pm$ 0.48 & 10.815 $\pm$0.0099 & 7 \\
34085 & $\beta$ Ori & 5 & TS12 & 30.46 $\pm$ 0.10 & 4.13 $\pm$ 0.14 & 10.753 $\pm$0.011 & 8 \\
31089 & HIP22669 & 1 & TS23 & 17.006 $\pm$ 0.039 & 1.73 $\pm$ 0.20 & 11.832 $^{+0.039}_{-0.043}$ & 2 \\
31089 & HIP22669 & 2 & TS23 & 17.4 $\pm$ 2.0 & 11.67 $\pm$ 0.30 & 11.495 $^{+0.081}_{-0.10}$ & \\
31089 & HIP22669 & 3 & TS23 & 22.58 $\pm$ 0.37 & 2.40 $\pm$ 0.75 & 11.04 $^{+0.10}_{-0.13}$ & \\
27436 & HIP20179 & 1 & TS23 & --39.1 $\pm$ 2.9 & 11.8 $\pm$ 2.3 & 11.201 $^{+0.093}_{-0.10}$ & \\
27436 & HIP20179 & 2 & TS23 & --12.3 $\pm$ 5.9 & 17. $\pm$ 10. & 11.46 $^{+0.38}_{-1.1}$ & \\
27436 & HIP20179 & 3 & TS23 & 17.3 $\pm$ 7.5 & 11.0 $\pm$ 6.5 & 11.30 $^{+0.36}_{-0.68}$ & \\
27436 & HIP20179 & 4 & TS23 & 17.375 $\pm$ 0.055 & 2.39 $\pm$ 0.40 & 12.24 $^{+0.10}_{-0.11}$ & \\
28262 & HIP20787 & 1 & TS23 & 20.085 $\pm$ 0.048 & 2.37 $\pm$ 0.21 & 12.223 $^{+0.074}_{-0.077}$ & \\
30535 & HIP22304 & 1 & TS23 & --2.45 $\pm$ 0.51 & 7.55 $\pm$ 0.93 & 11.207 $^{+0.047}_{-0.053}$ & 1 \\
30535 & HIP22304 & 2 & TS23 & 19.33 $\pm$ 0.17 & 5.82 $\pm$ 0.22 & 11.449 $^{+0.0095}_{-0.0097}$ & 2 \\
30535 & HIP22304 & 3 & TS23 & 23.22 $\pm$ 0.20 & 1.18 $\pm$ 0.37 & 10.918 $^{+0.084}_{-0.10}$ & \\
31726 & HR1595 & 1 & TS23 & --5.63 $\pm$ 0.17 & 4.19 $\pm$ 0.49 & 10.961 $^{+0.027}_{-0.029}$ & 1 \\
31726 & HR1595 & 2 & TS23 & 8.36 $\pm$ 0.64 & 5.09 $\pm$ 0.44 & 11.400 $^{+0.045}_{-0.050}$ & \\
31726 & HR1595 & 3 & TS23 & 14.543 $\pm$ 0.012 & 1.04 $\pm$ 0.51 & 11.82 $^{+0.12}_{-0.17}$ & \\
31726 & HR1595 & 4 & TS23 & 21.2 $\pm$ 2.3 & 8.9 $\pm$ 1.6 & 11.33 $^{+0.12}_{-0.16}$ & 2 \\
%26994 & 1 & TS23 & --5.44 $\pm$ 0.24 & 1.71 $\pm$ 0.49 & 11.393 $^{+0.036}_{-0.039}$ & \\
%26994 & 2 & TS23 & 3.22 $\pm$ 0.33 & 0.60 $\pm$ 0.12 & 11.97 $\pm$0.38 & \\
% 26994 & 3 & TS23 & 5.00 $\pm$ 0.26 & 4.13 $\pm$ 0.34 & 12.092 $^{+0.061}_{-0.072}$ & \\
28497 & DU Eri & 1 & TS21 & --26.21 $\pm$ 0.080 & 2.03 $\pm$ 0.52 & 10.767 $^{+0.074}_{-0.089}$ & 1 \\
28497 & DU Eri & 2 & TS21 & --15.3 $\pm$ 1.6 & 3.75 $\pm$ 0.71 & 10.74 $^{+0.31}_{-0.50}$ & \\
28497 & DU Eri & 3 & TS21 & --12.15 $\pm$ 0.13 & 1.8 $\pm$ 1.1 & 10.85 $^{+0.19}_{-0.33}$ & 2 \\
28497 & DU Eri & 4 & TS21 & --6.909 $\pm$ 0.057 & 2.49 $\pm$ 0.51 & 11.23 $^{+0.11}_{-0.14}$ & \\
28497 & DU Eri & 5 & TS21 & --1.2 $\pm$ 1.7 & 3.34 $\pm$ 0.55 & 10.62 $^{+0.16}_{-0.27}$ & \\
28497 & DU Eri & 6 & TS21 & 6.494 $\pm$ 0.054 & 0.21 $\pm$ 0.34 & 10.66 $^{+0.14}_{-0.22}$ & \\
28497 & DU Eri & 7 & TS21 & 11.09 $\pm$ 0.44 & 6.32 $\pm$ 0.45 & 11.357 $^{+0.044}_{-0.049}$ & 4 \\
28497 & DU Eri & 8 & TS21 & 21.57 $\pm$ 0.27 & 0.38 $\pm$ 0.24 & 10.524 $^{+0.044}_{-0.049}$ & 5 \\
28497 & DU Eri & 9 & TS21 & 23.5 $\pm$ 1.0 & 8.1 $\pm$ 1.2 & 11.045 $^{+0.062}_{-0.073}$ & \\
28497 & DU Eri & 10 & TS21 & 31.022 $\pm$ 0.045 & 0.94 $\pm$ 0.11 & 10.710 $^{+0.0070}_{-0.0071}$ & 6 \\
32612 & HR1640 & 1 & TS23 & --15.5 $\pm$ 1.1 & 2.4 $\pm$ 2.4 & 10.42 $^{+0.11}_{-0.11}$ & \\
32612 & HR1640 & 2 & TS23 & --4.1 $\pm$ 1.8 & 4.57 $\pm$ 0.97 & 10.52 $^{+0.10}_{-0.11}$ & \\
32612 & HR1640 & 3 & TS23 & 8.01 $\pm$ 0.35 & 6.12 $\pm$ 0.86 & 11.099 $^{+0.064}_{-0.073}$ & \\
32612 & HR1640 & 4 & TS23 & 15.86 $\pm$ 0.20 & 1.65 $\pm$ 0.44 & 11.175 $^{+0.068}_{-0.072}$ & \\
32612 & HR1640 & 5 & TS23 & 21.14 $\pm$ 0.45 & 1.64 $\pm$ 0.46 & 11.13 $^{+0.13}_{-0.18}$ & 2 \\
32612 & HR1640 & 6 & TS23 & 26.89 $\pm$ 0.28 & 3.65 $\pm$ 0.45 & 11.15 $^{+0.11}_{-0.15}$ & \\
%30076 & 1 & TS23 & --20.29 $\pm$ 0.18 & 1.56 $\pm$ 0.18 & 10.980 $^{+0.057}_{-0.065}$ & \\
%30076 & 2 & TS23 & --14.91 $\pm$ 0.45 & 4.96 $\pm$ 0.71 & 11.259 $^{+0.067}_{-0.079}$ & \\
%30076 & 3 & TS23 & --0.3 $\pm$ 3.1 & 4.9 $\pm$ 2.8 & 10.34 $^{+0.24}_{-0.32}$ & \\
%30076 & 4 & TS23 & 9.69 $\pm$ 0.94 & 3.3 $\pm$ 1.9 & 10.751 $^{+0.086}_{-0.095}$ & \\
%30076 & 5 & TS23 & 22.22 $\pm$ 0.12 & 0.616 $\pm$ 0.051 & 12.218 $^{+0.037}_{-0.041}$ & \\
%30076 & 6 & TS23 & 24.61 $\pm$ 0.40 & 4.86 $\pm$ 0.41 & 12.089 $^{+0.024}_{-0.024}$ & \\
30679 & HIP22406 & 1 & TS23 & --19.04 $\pm$ 0.16 & 2.4 $\pm$ 1.2 & 11.059 $^{+0.096}_{-0.12}$ & \\
30679 & HIP22406 & 2 & TS23 & --11.49 $\pm$ 0.37 & 4.54 $\pm$ 0.43 & 11.335 $^{+0.019}_{-0.019}$ & \\
30679 & HIP22406 & 3 & TS23 & 15.37 $\pm$ 0.82 & 5.25 $\pm$ 0.31 & 10.853 $^{+0.030}_{-0.032}$ & \\
30679 & HIP22406 & 4 & TS23 & 22.931 $\pm$ 0.022 & 2.6 $\pm$ 1.0 & 11.072 $^{+0.069}_{-0.074}$ & 2 \\
\enddata
\end{deluxetable}

\clearpage

\begin{deluxetable}{llclcl}
\tablewidth{0pt}
\tabletypesize{\scriptsize}
\tablecaption{\ion{Na}{1} ISM Fit Parameters Total Column Density Per Sight Line\label{NAtotecomps}}
\tablehead{ HD & Other & Number of  &Instrument& $\log N$ & Other \\
\# & Name & Components && (cm$^{-2}$) & References }
\startdata
33111 & $\beta$ Eri &&	UHRF & $<$9.9 & 1 \\
30743 & HR1545 &	&TS21& $<$10.3 & \nodata \\
%26574 & $o^1$ Eri &	&TS23& $<$10.8  \\
33904 & $\mu$ Lep &	&TS12& $<$10.8 & 2 \\
27861 & $\xi$ Eri&	&TS21& $<$10.2 & \nodata \\
33802 & $\iota$ Lep &	&TS21& $<$10.0 & \nodata \\
29573 & HR1483 &	&TS21& $<$10.2 & 3 \\
31109 & $\omega$ Eri &	&TS23& $<$10.5 & \nodata \\
32045 & S Eri &	&TS21& $<$10.7 & \nodata \\
30127 & HR1513 &	&TS21& $<$10.2 & \nodata \\
%32964 &	&TS23& $<$10.4\\
27925 & HIP20521 &	&TS23& $<$10.5 & \nodata \\
34863 & $\nu$ Lep &	&TS21& $<$10.5 & \nodata \\
28763 & HR1438 & 1 & TS21 & 10.759 $^{+0.0041}_{-0.0042}$ & 4,5 \\
32996 & HR1661 &	&TS23& $<$10.2 & \nodata \\
%29173 &	&TS23& $<$10.2\\
%30020 & 2 & TS21 & 11.132 $^{+0.030}_{-0.036}$ \\
29554 & HIP21640 & 2 & TS23 & 12.216 $^{+0.043}_{-0.055}$ & \nodata \\
28843 & DZ Eri & 3 & TS21 & 12.22 $^{+0.11}_{-0.23}$ & \nodata \\
34503 & $\tau$ Ori & 3 & TS21 & 11.331 $^{+0.0084}_{-0.0090}$ & 6 \\
31625 & HIP23011 &&TS23& $<$10.6 & \nodata \\
30211 & $\mu$ Eri & 1 & TS21 & 12.11 $^{+0.14}_{-0.27}$ & 1,4,7 \\
30050 & RZ Eri & 1 & TS23 & 12.326 $^{+0.091}_{-0.13}$ & \nodata \\
29851 & HIP21837 & 1 & TS23 & 11.794 $^{+0.042}_{-0.052}$ & 5 \\
28208 & HIP20747 & 1 & TS23 & 11.857 $^{+0.034}_{-0.040}$ & \nodata \\
30332 & HIP22169 & 3 & TS23 & 12.051 $^{+0.064}_{-0.090}$ & \nodata \\
28980 & HIP21258 & 1 & TS23 & 11.395 $^{+0.093}_{-0.16}$ & \nodata \\
32468 & HIP23493 & 2 & TS21 & 11.428 $^{+0.042}_{-0.062}$ & \nodata \\
29248 & $\nu$ Eri & 2 & TS12 & 12.116 $^{+0.082}_{-0.076}$ & 4,8\\
33224 & HR1671 & 2 & TS23 & 12.075 $^{+0.016}_{-0.019}$ & \nodata \\
33949 & $\kappa$ Lep & 3 & TS21 & 11.704 $^{+0.017}_{-0.019}$ & 1 \\
28377 & HIP20849 & 2 & TS23 & 11.118 $^{+0.048}_{-0.070}$ & \nodata \\
32249 & $\psi$ Eri & 2 & TS12 & 11.945 $\pm$0.0052 & 4 \\
31512 & 62 Eri & 4 & TS23 & 11.72 $^{+0.16}_{-0.43}$ & 7 \\
30963 & HIP22588 & 4 & TS23 & 11.820 $^{+0.035}_{-0.046}$ & \nodata \\
27563 & EM Eri & 3 & TS23 & 12.23 $\pm$0.38 & 4 \\
29009 & EH Eri & 3 & TS23 & 11.769 $^{+0.040}_{-0.083}$ & \nodata \\
34816 & $\lambda$ Lep & 1 & TS12 & 11.184 $\pm$0.016 & 8 \\
34085 & $\beta$ Ori & 5 & TS12 & 11.684 $\pm$0.0050 & 9,10,11 \\
31089 & HIP22669 & 3 & TS23 & 12.041 $^{+0.034}_{-0.043}$ & \nodata \\
27436 & HIP20179 & 4 & TS23 & 12.376 $^{+0.083}_{-0.22}$ & \nodata \\
%26994 & HIP19856 & 3 & TS23 & 12.38 $\pm$0.13 \\
28262 & HIP20787 & 1 & TS23 & 12.223 $^{+0.068}_{-0.085}$ & \nodata \\
30535 & HIP22304 & 3 & TS23 & 11.720 $^{+0.020}_{-0.024}$ & \nodata \\
31726 & HR1595 & 4 & TS23 & 12.083 $^{+0.065}_{-0.11}$ & 4 \\
28497 & DU Eri & 10 & TS21 & 11.936 $^{+0.036}_{-0.059}$ & 4,5,8,12,13 \\
32612 & HR1640 & 6 & TS23 & 11.786 $^{+0.042}_{-0.062}$ & \nodata \\
%30076 & 6 & TS23 & 12.510 $^{+0.021}_{-0.024}$ \\
30679 & HIP22406 & 4 & TS23 & 11.716 $^{+0.027}_{-0.034}$ & \nodata \\
\enddata
\tablerefs{(1) \citealt{welsh94}; (2) \citealt{welsh91}; (3) \citealt{holweger99}; (4) \citealt{welsh05b}; (5) \citealt{penprase93}; (6) \citealt{frisch90b}; (7) \citealt{genova03}; (8) \citealt{hobbs78}; (9) \citealt{hobbs69}; (10) \citealt{hobbs74}; (11) \citealt{price01}; (12) \citealt{shull77}; (13) \citealt{blades97}.}
\label{NaTotesN}
\end{deluxetable}

\clearpage

\begin{deluxetable}{lllclc}
\tablewidth{0pt}
\tabletypesize{\scriptsize}
\tablecaption{Multiple-Resolution Fits\label{multres}}
\tablehead{ HD & Other & Ion & Components &Instrument& $\log N$ \\
\# & Name & & & & (cm$^{-2}$) }
\startdata
%30020 & \ion{Na}{1} & 2 & TS23 & 11.18 $^{+0.09}_{-0.22}$ \\
%30020 & \ion{Na}{1} & 2 & TS21 & 11.13 $^{+0.030}_{-0.038}$ \\
30211 & $\mu$ Eri & \ion{Na}{1} & 1 & TS21 & 12.11 $^{+0.14}_{-0.27}$ \\
30211 & $\mu$ Eri & \ion{Na}{1} & 2 & TS23 & 12.01 $^{+0.14}_{-0.48}$ \\
33949 & $\kappa$ Lep & \ion{Na}{1} & 3 & TS21 & 11.704 $^{+0.017}_{-0.019}$ \\
33949 & $\kappa$ Lep & \ion{Na}{1} & 3 & UHRF & 11.515 $\pm$0.028 \\
32249 & $\psi$ Eri & \ion{Na}{1} & 2 & TS12 & 11.945 $\pm$0.0052 \\
32249 & $\psi$ Eri & \ion{Na}{1} & 3 & TS23 & 12.142 $^{+0.050}_{-0.066}$ \\
34085 & $\beta$ Ori & \ion{Na}{1} & 5 & TS12 & 11.684 $\pm$0.0050 \\
34085 & $\beta$ Ori & \ion{Na}{1} & 4 & TS23 & 11.671 $^{+0.026}_{-0.035}$ \\
34085 & $\beta$ Ori & \ion{Na}{1} & 4 & UHRF & 11.524 $\pm$0.0016 \\
34085 & $\beta$ Ori & \ion{Na}{1} & 6 & TS21 & 11.664 $^{+0.019}_{-0.027}$ \\
28497 & DU Eri & \ion{Na}{1} & 10 & TS21 & 11.936 $^{+0.036}_{-0.060}$ \\
28497 & DU Eri & \ion{Na}{1} & 6 & TS23 & 11.948 $^{+0.014}_{-0.016}$ \\
30211 & $\mu$ Eri & \ion{Ca}{2} & 1 & TS12 & 11.489 $\pm$0.0094 \\
30211 & $\mu$ Eri & \ion{Ca}{2} & 2 & TS21 & 11.73 $\pm$0.11 \\
30211 & $\mu$ Eri & \ion{Ca}{2} & 2 & TS23 & 11.738 $^{+0.019}_{-0.021}$ \\
32249 & $\psi$ Eri & \ion{Ca}{2} & 1 & TS12 & 11.715 $\pm$0.0091 \\
32249 & $\psi$ Eri & \ion{Ca}{2} & 3 & TS23 & 12.07 $^{+0.10}_{-0.23}$ \\
34085 & $\beta$ Ori & \ion{Ca}{2} & 5 & TS12 & 11.507 $\pm$0.0030 \\
34085 & $\beta$ Ori & \ion{Ca}{2} & 5 & TS23 & 11.48 $^{+0.07}_{-0.13}$ \\
34085 & $\beta$ Ori & \ion{Ca}{2} & 8 & UHRF & 11.549 $\pm$0.011 \\
34085 & $\beta$ Ori & \ion{Ca}{2} & 7 & TS12 & 11.517 $\pm$0.0042 \\
28497 & DU Eri & \ion{Ca}{2} & 5 & TS21 & 12.202 $^{+0.034}_{-0.047}$ \\
28497 & DU Eri & \ion{Ca}{2} & 6 & TS23 & 12.152 $^{+0.043}_{-0.12}$ \\
\enddata
\end{deluxetable}

\end{document}